\documentclass[aps,amsmath,onecolumn, showpacs,superscriptaddress,pre,10.5pt,nofootinbib]{revtex4-1}
\usepackage{amssymb}
\usepackage{amsmath}
\usepackage{graphicx}
\usepackage{array}
\usepackage{physics}
\usepackage{mathtools}
\usepackage{amssymb}
\usepackage{verbatim}
\usepackage{color}
\usepackage{bm}
\usepackage{amsthm}

\usepackage{color}
\usepackage[dvipsnames]{xcolor}
\usepackage[latin1]{inputenc}
\definecolor{Blue}{rgb}{0.00, 0.00, 1.00}
\definecolor{Red}{rgb}{1.00, 0.00, 0.00}

\usepackage[pdfpagemode=UseNone,
		bookmarksopen=false,
		colorlinks=true,
		urlcolor=blue,
		citecolor=red,
		citebordercolor=blue,
		linkcolor=blue, 
		urlcolor=black, 
		filecolor=red]{hyperref}
		
\newcommand{\bea}{\begin{eqnarray}}
\newcommand{\eea}{\end{eqnarray}}
\newcommand{\be}{\begin{equation}}
\newcommand{\ee}{\end{equation}}
\newcommand{\bee}{\begin{equation*}}
\newcommand{\eee}{\end{equation*}}

\colorlet{Mycolor1}{green!10!orange!90!}

\def\Xint#1{\mathchoice
   {\XXint\displaystyle\textstyle{#1}}%
   {\XXint\textstyle\scriptstyle{#1}}%
   {\XXint\scriptstyle\scriptscriptstyle{#1}}%
   {\XXint\scriptscriptstyle\scriptscriptstyle{#1}}%
   \!\int}
\def\XXint#1#2#3{{\setbox0=\hbox{$#1{#2#3}{\int}$}
     \vcenter{\hbox{$#2#3$}}\kern-.5\wd0}}

\def\dashint{\Xint-}

\begin{document}

\title{Non-crossing Brownian paths and Dyson Brownian motion under a moving boundary}

\author{Tristan Gauti\'e}
\affiliation{Laboratoire de Physique de l'Ecole Normale Sup\'erieure,
PSL University, CNRS, Sorbonne Universit\'e, Universit\'e de Paris,
24 rue Lhomond, 75231 Paris, France.}
\author{Pierre Le Doussal}
\affiliation{Laboratoire de Physique de l'Ecole Normale Sup\'erieure,
PSL University, CNRS, Sorbonne Universit\'e, Universit\'e de Paris,
24 rue Lhomond, 75231 Paris, France.}
\author{Satya N. \surname{Majumdar}}
\affiliation{LPTMS, CNRS, Univ. Paris-Sud, Universit\'e Paris-Saclay, 91405 Orsay, France}
\author{Gr\'egory \surname{Schehr}}
\affiliation{LPTMS, CNRS, Univ. Paris-Sud, Universit\'e Paris-Saclay, 91405 Orsay, France}

\begin{abstract} 
We compute analytically the probability $S(t)$ that a set of $N$ Brownian paths do not cross 
each other and stay below a moving boundary $g(\tau)= W \sqrt{\tau}$ up to time $t$. 
We show that for large $t$ it decays as a power law $S(t) \sim t^{- \beta(N,W)}$.
The decay exponent $\beta(N,W)$ is obtained as the ground state energy of
a quantum system of $N$ non-interacting fermions in a harmonic well in the presence of an infinite hard wall
at position $W$. Explicit expressions for $\beta(N,W)$ are obtained in various limits of $N$ and $W$, in particular 
for large $N$ and large $W$. We obtain the joint distribution of the positions of the walkers
in the presence of the moving barrier $g(\tau)  =W \sqrt{\tau}$ at large time. We extend our results to the
case of $N$ Dyson Brownian motions (corresponding to the Gaussian Unitary Ensemble) in the presence of the same moving boundary
$g(\tau)=W\sqrt{\tau}$. For $W=0$ we show that the system provides a realization of
a Laguerre biorthogonal ensemble in random matrix theory. We obtain explicitly
the average density near the barrier, as well as in the bulk far away from the barrier. 
Finally we apply our results to $N$ non-crossing Brownian bridges on the interval $[0,T]$ under a time-dependent barrier
$g_B(\tau)= W \sqrt{\tau(1- \frac{\tau}{T})}$.
\end{abstract}


\maketitle

{%
	\hypersetup{linkcolor=black}
	\tableofcontents
}

\section{Introduction}

Computing exactly the survival probability of a single Brownian motion in one dimension in the presence of a generic time dependent moving boundary (absorbing) remains a challenging problem despite a large body of studies in
probability theory \cite{Doob1949, Brei1966, Uch1980, Sal1988, Nov1981,AK13, AS15}, 
statistics \cite{Kolm1933, Smir1936, ChiBou2012} and physics \cite{Chandra1943,Maj1999,Maj2005,BrayMajSchehr2013,Red2001,BraySmith2007,BraySmith2007deux}.
For a single Brownian particle a lot of asymptotic results for the survival probability are known
for a variety of time dependent boundaries. Consider a Brownian motion
$x(\tau)$ in the presence of a moving boundary at $x=g(\tau)$ and define the survival
probability as
\bea \label{first_def_S}
S(t) = {\rm Pr} \left[ x(\tau) < g(\tau), \, \forall \tau \in [t_0,t] \right] \;,
\eea
i.e. the probability that the particle remains below the boundary during the
time interval $[t_0,t]$ for some arbitrary initial time $t_0$. The asymptotic decay of
$S(t)$ depends on how $g(\tau)$ behaves for large $\tau$, and several cases can be distinguished.

\noindent(i) Fixed barrier $g(\tau)=W>0$: It is well known for a Brownian starting at $x(0)=0$, e.g.
using the method of images (see for example \cite{Red2001}), that
\be
S(t) \sim t^{-\frac{1}{2}} \quad , \quad t \to +\infty \;.
\ee
(ii) Slow barrier: For a function $g(\tau)$ with slow enough growth at infinity, Uchiyama \cite{Uch1980} showed that this universal $\frac{1}{2}$ decay exponent is still valid. More precisely, this holds for $g(\tau)$ continuous and either convex or concave under the condition \cite{Uch1980} 
\be 
\label{eq:ConditionUchiyama}
\int _ { 1 } ^ { \infty } | g ( \tau ) | \tau ^ { - 3 / 2 } \, {d} \tau < \infty \;.
\ee
The exponent $1/2$ thus holds in particular for any barrier $g(\tau) = W \tau^\alpha$ with $W>0$ and $\alpha <\frac{1}{2}$.

\noindent (iii) Fast barrier: On the other hand, for a fast barrier $g(\tau) = W \tau^\alpha$ with $\alpha >1/2$, the survival probability does not vanish for large times \cite{Red2001}: 
\be
S(t) \to S_\infty >0 \quad , \quad t \to +\infty
\ee
where in some cases $S_\infty$ can be calculated explicitly \cite{Nov1981}.

A particularly interesting case is the marginally fast square-root barrier $g(\tau) = W\sqrt{\tau}$.
The barrier then grows like the standard deviation of the process. 
In this case, the survival probability decays as a power-law with a non-universal exponent depending continuously
on $W$ \cite{RedKrap1996, Uch1980, Brei1966, Red2001, Tur1992}:
\be  
S(t) \sim t^{-\beta(W)}  \quad , \quad t \to + \infty  \;.
\ee 
The survival exponent $\beta(W)$ can be computed for a general value of $W$ as
the smallest root of the following equation
\be
D_{2\beta(W)} (-W) = 0    
\ee
where $D_\nu(x)$ is the parabolic cylinder function of index $\nu$. In addition to these
results there are results for the case of {\it two} absorbing time dependent boundaries
(cage model) with similar non-trivial exponents in the critical case \cite{BraySmith2007,Nov1981,RedKrap1996,Tur1992}.

It is natural to generalize this problem to the case of $N$ interacting walkers.
For instance one class of interacting walkers which has been much studied
is the so-called vicious walkers problem \cite{Fish1984}. This problem was first introduced by de Gennes in the context of $N$ polymer fibers bound between two parallel plates. He showed that this problem is related to the quantum mechanics of $N$ non-interacting fermions. The same vicious walkers problem was studied by Fisher and coworkers \cite{Fish1984, HF84} in describing the dynamics of domain walls between different incommensurate phases. Since then, the vicious walkers problem has been studied extensively both in physics and mathematics.

\begin{figure}[t]
\includegraphics[width = 0.5 \linewidth]{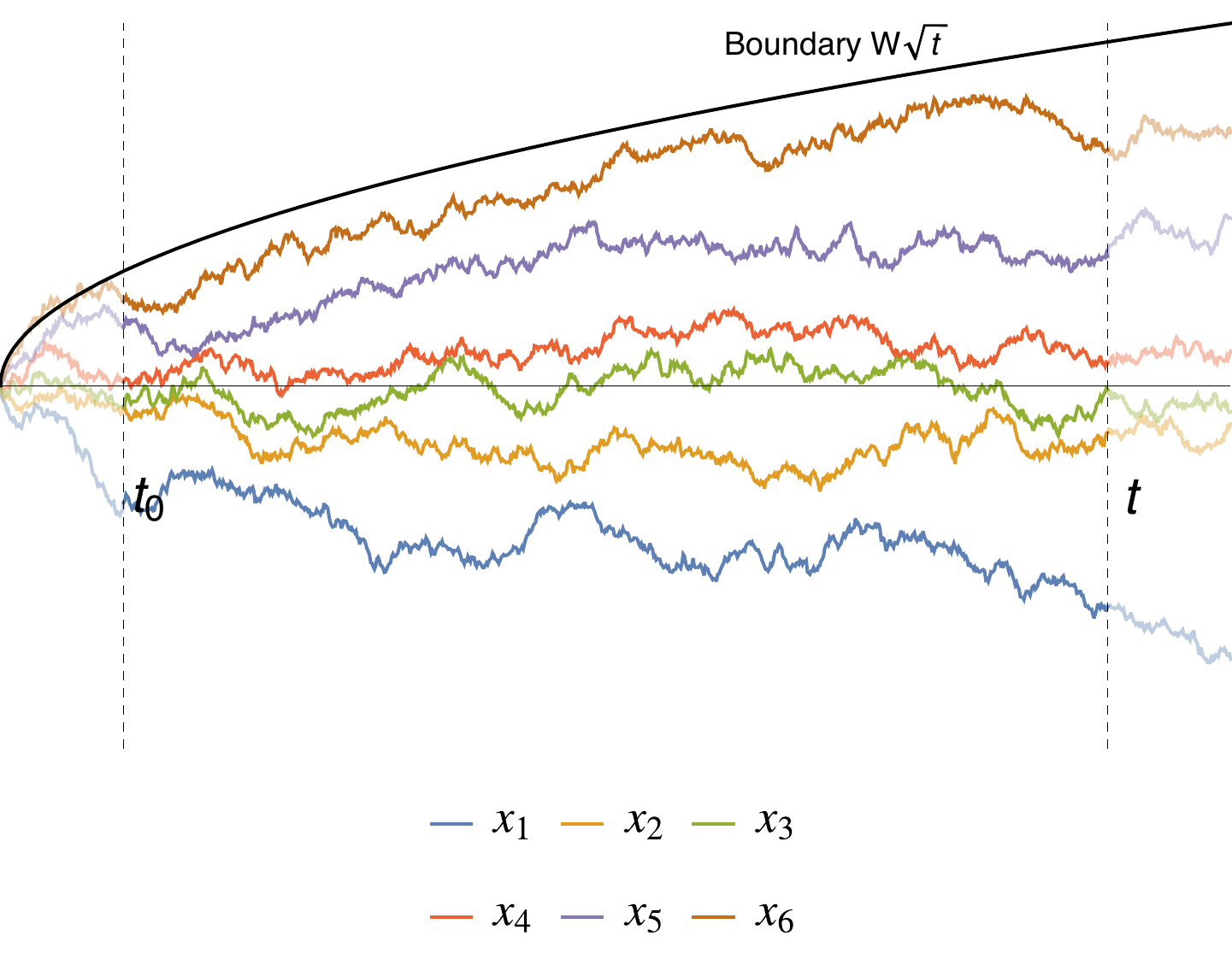}
\caption{A realization for $N=6$ brownian paths, which do not intersect and stay under the square-root boundary on $[t_0, t ]$. The diffusion coefficient was taken to be $D=1$, with a boundary prefactor $W=5$ with $t_0=1$ and $t=10$. In our work, the positions $\vec{x}_0$ at $t_0$ are given.}
\label{fig:brownien} 
\end{figure}

In this vicious walkers problem 
one studies $N$ independent Brownian motions and computes the
probability that they do not cross each other up to time $t$.
\footnote{A more general observable, the distribution of coincidences, was studied in \cite{KrajLacroix}.}
This probability decays as $t^{-\frac{N(N-1)}{4}}$ at late times, where the exponent $N(N-1)/4$ is known as the Fisher exponent~\cite{Fish1984, KratGutVien2000, HF84,BraWin2004}.
It is thus natural to consider these vicious walkers in the presence of an absorbing 
moving boundary $g(\tau)$. One defines the following survival probability
\be  \label{survival0} 
S(t) = {\rm Pr} ( x_1(\tau) < x_2(\tau) < \cdots < x_N(\tau) < g(\tau) , \ \forall \tau \in [t_0, t] ) \;,
\ee
i.e. the joint probability that the $N$ Brownian particles have not crossed {\it and} have remained under the barrier $g(\tau)$ between a fixed initial time $t_0$ and time $t$. It is illustrated in Fig. \ref{fig:brownien}. 
In the special case of a fixed barrier
$g(\tau)=0$, it was shown that at large time $t$ \cite{KratGutVien2000,BraWin2004,Katori2002}
\be
S(t) \sim t^{- \frac{N^2}{2}}
\ee
Note that in \cite{KratGutVien2000} the authors derived these results by establishing connections with the theory of Young tableaux. Further studies have focused on the extreme properties of non-intersecting Brownian paths and bridges \cite{KIK08,SMCR08,KIK08b,RS11,FMS11,Lie12,S12,SMCF13,Rem2017a,Rem2017b,Borodin2009}, relating them to the statistics of the largest eigenvalue of random matrices from Gaussian and Laguerre Orthogonal ensembles \cite{Rem2017a,Rem2017b}.

In this paper we consider $N$ vicious walkers in the presence of 
the critical square root $g(\tau)= W \sqrt{\tau}$ boundary. The physical motivation behind this problem is as follows. Studying the behavior of an interacting particle system in the presence of a time-dependent external potential is a central generic problem in non-equilibrium statistical physics, both for classical and quantum systems. Exact results are difficult to obtain in such systems. In this paper, we provide a solvable case of $N$ vicious walkers in the presence of a time-dependent hard wall potential. For this critical square-root boundary, we show that 
the survival probability defined in \eqref{survival0} decays at late times as 
\be 
S(t) \propto_{t \to \infty} t^{-\beta(N,W)} \;,
\ee
where the exponent $\beta(N,W)$ can be computed analytically. 
We show that this exponent can be written as 
\be \label{betaR} 
\beta(N,W) = \sum_{k=0}^{N-1} \epsilon_k(W) \;,
\ee
where the $\epsilon_k(W)$ are the eigenvalues of a Schr\"odinger operator
[see Eq. \eqref{eq:diffEqEig}]
describing a single quantum particle in a quadratic potential with an infinite
repulsive wall at $x=W$. 
In the limit $W=+\infty$, we recover the Fisher exponent $\beta(N, W \to \infty) = \frac{N(N-1)}{4}$
and for $W=0$ we recover the above result $\beta(N, W =0) = \frac{N^2}{2}$. 
We obtain explicit expressions in various limits (i) for any fixed $N$ and
in an expansion in $W$ near $W=0$, see Eqs. \eqref{pert1}-\eqref{a1}
(ii) for any fixed $N$ and $W$ large and negative, see Eq. \eqref{form2}, 
(iii) for any fixed $W<0$ and large $N$ see Eq. \eqref{asymp} 
and finally (iv) for $W$ and $N$ both large but the ratio $W/\sqrt{N}$ fixed.
This latter case is particularly interesting and allows for explicit 
results using the semi-classical analysis of the Schr\"odinger operator.
One finds
\be \label{scalb} 
\beta(N,W) \simeq \frac{N^2}{4} \, {\sf b}\left(\frac{W}{\sqrt{4 N}}\right)
\ee 
where the scaling function ${\sf b}(y)$ can be calculated exactly, see Eqs. \eqref{eq1} and 
\eqref{eq2} as well as Fig. \ref{fig:Plotb}.
In addition to the exponent $\beta(N,W)$, we also obtain the joint distribution of the
position of the surviving particles, see \eqref{general}. In the case
$W=0$ an explicit expression is given in Eq. \eqref{116} and is related
to the eigenvalues of the Laguerre Orthogonal Ensemble (LOE).

We then apply these results to the Dyson Browian motion (DBM) associated to the Gaussian Unitary Ensemble (GUE). The GUE-DBM, first introduced by Dyson in \cite{Dys62}, is the process of the $N$ real eigenvalues $x_i(t)$ of a $N \times N$ Hermitian matrix whose upper triangular entries (both real and imaginary parts independently) evolve according to independent Brownian motions in a fictitious time~$t$. Consequently the eigenvalues evolve via the Langevin equation 
\be 
\frac{d x_i(t)}{dt} =   \sum_{j\neq i} \frac{1}{x_i(t) - x_j(t)	} + \xi_i(t)
\ee    
where $\xi_i(t)$ are independent Gaussian white noises. By construction the
$x_i(t)$'s do not cross each other, i.e. the DBM trajectories are non-intersecting with probability one \cite{Tao2012}.
The question of interest is the probability $S_{DBM}(t)$ that the DBM stays below the
moving boundary $g(\tau)= W \sqrt{\tau}$. We show that it decays as 
\be
S_{DBM}(t) \sim t^{- \beta_c(N,W)} \quad , \quad \beta_c(N,W)= \beta(N,W) - \frac{N(N-1)}{4}  \;,
\ee
where $\beta(N,W)$ is given in Eq. \eqref{betaR}. In particular for $W=0$ one obtains the decay exponent
\be
\beta_c(N,W=0) = \frac{N(N+1)}{4} \;.
\ee 
We show that the propagator for the DBM under the barrier $g(\tau) = 0$ provides a realization of the
biorthogonal ensemble of random matrix theory \cite{Muttalib}. As a consequence, the
positions of the DBM particles in the presence of the absorbing wall at $W=0$ form a
determinantal point process with explicit expressions for the kernel. Using results
on biorthogonal ensembles \cite{Bor1998,BorotNadal2012,ClaeysBiorthogonal}
we obtain explicit expressions for the average density of the DBM particles in the
large $N$ limit, both near the boundary as well as in the bulk. 
In the bulk it takes the scaling form
\be \label{bulk2} 
\tilde \rho_N(x) = \frac{1}{\sqrt{2 N t}} \, \tilde r \left( \frac{x}{\sqrt{2 N t} } \right) \;,
\ee
where the scaling function $\tilde r(y)$ is given explicitly in \eqref{rtilde} and plotted in Fig. \ref{fig:plotbulk}. It diverges as $\tilde r(y) \sim y^{-1/3}$ near 
the boundary for $y \to 0$ which indicates that the particles accumulate near the barrier. The density in the edge region near the wall $x \sim \sqrt{t}/N$
is described by another scaling function as
$\tilde \rho_N(x) \simeq N \frac{|x|}{t} r_e\left(\frac{N^2 x^2}{2 t}\right)$
where the scaling function $r_e(\tilde z)$ has also been computed 
explicitly, see Eqs. \eqref{re1}, \eqref{re2} (see also Fig. \ref{fig:redge} for a plot). 

Finally, by exploiting a mapping between the non-crossing Brownian motions under a 
barrier $W \sqrt{\tau}$ and non-crossing Brownian bridges on the interval $[0,T]$ under 
a barrier $g_B(\tau)= W \sqrt{\tau(1- \frac{\tau}{T})}$, we obtain the
corresponding survival probability of the bridges, see definition \eqref{eq:DefS} and
result \eqref{SB}. In a second stage, we extend this result to non-crossing
Brownian motions (i.e. not bridges) 
under the same barrier $g(\tau)= W \sqrt{\tau(1- \frac{\tau}{T})}$,
see Eq. \eqref{BrB}.

Our results are obtained by using the Lamperti transform which maps the 
$N$ Brownian motions onto $N$ Ornstein-Uhlenbeck (OU) processes. 
The latter problem can be studied using a mapping to non-interacting
fermions in a harmonic potential (as in Ref. \cite{PLDMajSch2018,fermions_review}).
These transformations extend in the presence of a moving barrier
$g(\tau)=W \sqrt{\tau}$.
The corresponding fermion problem is the harmonic
oscillator in the presence of a fixed hard wall at position $W$.
At large time this system is dominated by its $N$ fermion ground state,
which leads to \eqref{betaR}.

This paper is organized as follows. In Section II, we detail the mappings that allow us to solve the problem of $N$ vicious Brownian motions under a square-root absorbing boundary. We map it to a system of $N$ non-interacting fermions in a quadratic potential, via a representation in terms of OU processes. In Section III, we use these mappings to compute the large time decay of the survival probability for $N$ vicious Brownian motions under a square-root absorbing boundary, as well as the exact form of the propagator of this process. 
In Section IV, we extend these results to the Dyson Brownian Motion under a square-root boundary. In section V, we extend similarly the results to Brownian Bridges. Conclusions and open problems are given in Section VI. 
Notations and standard results about Hermite polynomials and the harmonic oscillator wavefunctions are recalled in the appendices, together with some other details of the computations, such as the calculation of decay amplitudes, and the Coulomb gas calculation for the density of the constrained DBM.

\section{Mapping to fermions in a quadratic potential}
\label{sec:Mapping} 

In this section, we present the methods that allow to map the system of $N$ non-crossing Brownian motions under a square-root barrier to $N$ fermions in a quadratic potential with a hard wall. The results that can be obtained from this mapping will be detailed in the following sections.

\subsection{One particle}

We start with the demonstration of the mapping with one particle only. It relies on two successive transformations: first from the standard Brownian Motion to an Ornstein-Uhlenbeck (OU) process, and then from the OU process to a quantum particle in a harmonic potential.

\subsubsection{Lamperti mapping}

The first mapping is known as Lamperti mapping, since it is a special case of Lamperti's transformations for scale-invariant self-similar stochastic processes applied to the Brownian Motion \cite{Lam1972, BorAmbFla2005}. This mapping is also sometimes called Doob's transform \cite{Sal1988}, since it is given by the application of Doob's transformation theorem \cite{Doob1949} to the OU process. 
The mapping is as follows: 
\bigskip

Let $x(t)$, $t \in [0,+\infty)$, be a standard Brownian motion with $x(0)=0$ and $\eta$ a white noise such that 
\be
\frac{{d}x}{{d}t}  = \eta(t)  \;, \;\quad {\rm with}
\quad \quad \quad 
 \left\{
    \begin{array}{ll}
        \overline{\eta(t)} & = 0  \\
        \overline{\eta(t) \eta(t')}  & = \delta(t-t') \;.
    \end{array}
\right.
\quad \quad \quad 
\ee
In particular one has $\overline{x^2(t)}  = t $. Defining a new process $X(T)$ indexed by $T = \ln(t)$, $T \in (-\infty,+\infty)$, such that $X(T) = \frac{x(t)}{\sqrt{t}}$, we have 
\be
\frac{dX(T)}{dT} = \frac{d\left( x(e^T)e^{-\frac{T}{2}} \right)}{dT} = -\frac{1}{2} X(T) + e^{\frac{T}{2}} \eta(e^T) = -\frac{1}{2} X(T) + \xi(T) \;,
\ee
where $\xi(T)$ is a white noise of zero mean $\overline{\xi(T)}=0$ and delta correlations $\overline{\xi(T) \xi(T')} = \delta(T-T')$. From a Brownian motion $x(t)$, the Lamperti mapping thus gives an OU process $X(T)$ 
\be \label{Lamperti} 
 \quad \quad  \quad 
\frac{d x(t)}{dt}  =  \eta(t)   \quad  \xrightarrow[  t = e^T   ]{ X = \frac{x}{\sqrt{t}}    }  \qquad   \frac{{d}X(T)}{dT}  = -\frac{1}{2} X (T)+  \xi (T) \;.
\ee
With $T$-$t$ and $X$-$x$ pairs of variables related by the above mapping, this transformation gives for the single particle propagators 
\be
\label{eq:LinkPBrOU}
P^{ Br } ( x , t | x _ { 0 } , t _ { 0 } ) \ {d}x  = P^{ O U } ( X , T | X_ { 0 } , T _ { 0 }) \ {d}X \;,
\ee
where $P^{ Br } ( x , t | x _ { 0 } , t _ { 0 } ) \ {d}x$ is the probability that the Brownian
particle reaches $[x, x+ dx]$ at time $t$ starting from the initial position $x_0$ at time $t_0$
(and similarly for the OU process). Note that the initial conditions read
\be \label{initOU} 
P^{ Br } ( x , t_0 | x _ { 0 } , t _ { 0 } ) = \delta(x-x_0) \quad , \quad 
P^{ O U } ( X , T_0 | X_ { 0 } , T _ { 0 }) = \delta(X-X_0) \;.
\ee

The interest of this mapping will be to give a direct interpretation of the survival probabilities under a $W\sqrt{\tau}$ barrier as survival probabilities for OU processes under a \textit{fixed} barrier at $W$. This trick was used by Breiman in \cite{Brei1966} to study a similar problem for a single  particle, and in later works in the physics literature \cite{Maj1999,BrayMajSchehr2013,MajSirBraCor1996,DerHakZei1996, MS96, BraWin2004}. Indeed, let us define 
\be
P^{ Br }(x,t| x_{0}, t_{0}\, ; \ W\sqrt{t})  dx = {\rm Pr} ( x(t) \in [x,x+dx], x(\tau) < W \sqrt{\tau} , \ \forall \tau \in [t_0,t] | \mid x(t_0) = {x_0} )
\ee 
the corresponding propagator where the path is constrained to be below the barrier $W \sqrt{\tau}$ which
becomes the barrier $W$ for the OU process under the Lamperti mapping \eqref{Lamperti} 
\be
P^{ Br }(x,t| x_{0}, t_{0}\, ; \ W\sqrt{t}) \ {d}x  = P^{OU} ( X ,T|X_{0} , T_{0}; \, \ W) \ {d}X \;,
\ee
which we will now relate to a quantum problem.

\subsubsection{Mapping to a quantum problem}

{\it In the absence of a wall: free case}. The second mapping is between the propagator of the stochastic OU process to the imaginary time quantum propagator of a single particle in a harmonic potential. Indeed, 
following \cite{Risken}, let us define the single particle one dimensional quantum system described by the following Hamiltonian, with the specific choice of values of $\hbar$, $m$ and $\omega$ relevant to our problem:
\be
\label{eq:HamiltH}
\hat { H } = - \frac { \hbar ^ { 2 } } { 2 m } \frac { \partial ^ { 2 } } { \partial X ^ { 2 } } + \frac { 1 } { 2 } m \omega ^ { 2 } X ^ { 2 } - \frac { \hbar \omega } { 2 } \ \ = \  \  - \frac { 1 } { 2} \frac { \partial ^ { 2 } } { \partial X ^ { 2 } } + \frac { 1 } { 8 }  X ^ { 2 } - \frac { 1 } {4 }   \quad \;, \quad  \quad  \text{with} \, 
\begin{cases}
\hbar = m =1\\
\omega =\frac{1}{2}
\end{cases}  \;.
\ee
Given that $\hbar=1$ and $\omega=\frac{1}{2}$ the eigenvalues of $\hat H$ are
$\epsilon^{\rm free}_k=\frac{k}{2}$ for integer $k \geq 0$. Let us denote $\phi_k(X)$ the eigenfunction
associated with the eigenvalue $\epsilon^{\rm free}_k$. $\phi_k(X)$ is given by Hermite polynomials, and its explicit expression is recalled in Appendix \ref{app:HarmonicOsc}. The quantum propagator $G(X,T \mid X_0, T_0)$ in imaginary time, for this single particle system
can be expressed as \be \label{decompose} 
G(X,T \mid X_0, T_0)  = \bra{X} e^{-(T-T_0) \hat{H}} \ket{X_0}  =  \sum \limits_{k=0} ^{\infty} \phi_k(X) \phi_k^*(X_0) e^{- \frac{k}{2} (T-T_0)}  
\ee
whose explicit expression is recalled in Appendix \ref{app:HarmonicOsc} [see Eq. (\ref{propag_app})]. It satisfies the imaginary time Schr\"odinger equation 
\be
\label{eq:evolG}
\partial_T G = -  \hat { H } G \;,
\ee
together with the initial condition $G(X,T_0 \mid X_0, T_0)  = \delta(X-X_0)$.
The OU process propagator is then related to the quantum propagator as follows
\be
\label{eq:1partPropagators}
P^{OU} (X,T \mid X_0, T_0 ) = e^{ -  \frac{X^2}{4}  }  \  G(X,T \mid X_0, T_0)  \  e^{  \frac{X_0^2}{4}  }  \;.
\ee
Indeed one can verify that the above form satisfies the Fokker-Planck equation 
\bea 
\partial_T  P^{OU} = \frac{1}{2} \partial_X^2 P^{OU} + \frac{1}{2} \partial_X(X P^{OU}) \label{FP_OU}
\eea
together with the initial condition \eqref{initOU}.

\bigskip

{\it In the presence of a wall}. This link between a free OU process and a quantum harmonic oscillator still holds when a fixed absorbing barrier at $W$ is added to the OU process: this imposes the boundary conditions $P_{OU}(X=W,T \mid X_0,T_0)=0$ such that $G(X=W,T \mid X_0, T_0)=0 $, while verifying the same Fokker-Planck equation (\ref{FP_OU}) in the domain $X<W$. This boundary condition is implemented in the quantum problem by adding an infinite wall at $W$ 
\be 
\label{eq:Hw}
\hat{H}^W =  \begin{cases}
-\frac{1}{2} \frac{\partial^2}{\partial X^2}  + \frac{1}{8} X^2  - \frac{1}{4} &\text{for $X<W$}\\
 + \infty  &\text{for $X \geq W$} \;.
\end{cases}
\ee
We will denote by $\phi_k(X,W)$ and $\epsilon_k(W)$, labelled by $k \geq 0$, the eigenfunctions and
eigenvalues (in increasing order) of $\hat H^W$. The corresponding quantum propagator $  G(X,T \mid X_0, T_0 ; \ W)$ reads
\be \label{decomposeW} 
G(X,T \mid X_0, T_0;W)  = \bra{X} e^{-(T-T_0) \hat{H}^W} \ket{X_0}  =  \sum \limits_{k=0} ^{\infty} \phi_k(X,W) \phi_k^*(X_0,W) e^{-\epsilon_k(W) (T-T_0)}  \;.
\ee
We then obtain the mapping onto the OU propagator in the presence of a fixed absorbing boundary at $W$ 
\bea\label{identity_ou_fermion_single} 
P^{OU} (X,T \mid X_0, T_0 ; \ W ) = e^{ -  \frac{X^2}{4}  }  \  G(X,T \mid X_0, T_0 ; \ W )  \  e^{  \frac{X_0^2}{4}  } \;.
\eea
Therefore combining the Lamperti transformation and the mapping to the quantum propagator, we obtain the exact relation in the presence of an absorbing boundary
\be 
P^{ Br }(x,t| x_{0}, t_{0}; \ W\sqrt{t}) \ {d}x  =e^{ -  \frac{X^2}{4}  }  \  G(X,T \mid X_0, T_0 ; \ W )  \  e^{  \frac{X_0^2}{4}  }  \ {d}X  
\quad \quad
\left\{
    \begin{array}{ll}
        X = \frac{x}{\sqrt{t}} \ & ; \ X_0 = \frac{x_0}{\sqrt{t_0}}\\
        T = \ln t  \ & ; \  T_0 = \ln t_0 \;,
    \end{array}
\right.
\ee
which we now extend to $N$ particles. 

\subsection{$N$ particles}

\subsubsection{Constrained propagators and Lamperti mapping}

Consider now $N$ independent Brownian motions $x_i(t)$, $i=1,\dots,N$ whose joint positions are denoted by $\vec x(t)$. We can apply the Lamperti transformation to each of these $N$ particles independently.
This gives rise to $N$ independent OU processes whose joint positions are denoted by $\vec X(T)$. Consider what we call the {\it constrained propagator}, i.e. the probability 
$P_N^ { B r } ( \vec { x } , t | \vec { x _ { 0 } } , t _ { 0 }  )$ of the event where the $N$ Brownian particles starting from $\vec x_0$ at time $t_0$ arrive at $\vec x$ at time $t$ without crossing each other in the
time interval $[t_0,t]$
\be
\label{eq:DefP} 
P_N^{ B r } (\vec { x } , t | \vec { x _ { 0 } } , t _ { 0 }) d \vec x= {\rm Pr} \left[ x_i(t) \in [x_i,x_i+dx_i] , 1\leq i \leq N  \ ; \  x_1(\tau) < x_2 ( \tau) < \cdots < x_N(\tau), \ \forall \tau \in [t_0,t]  \mid x_i(t_0) = {x_0}_i  \right] \;.
\ee
Since the $N$ OU processes $\vec{X}$ are non-crossing on $[T_0,T]$ if and only if the original Brownian motions $\vec{x}$ are non-crossing on $[t_0,t]$, we have the relation 
\be
P_N^ { B r } ( \vec { x } , t | \vec { x _ { 0 } } , t _ { 0 }  ) \  {d} \vec { x } = P_N^{ O U } ( \vec { X } , T | \vec { X } _ { 0 } , T _ { 0 } ) \ {d} \vec { X } \;,
\ee
where $P_N^{ O U } ( \vec { X } , T | \vec { X } _ { 0 } , T _ { 0 } )$ is the OU constrained propagator, i.e. the probability 
of the event that $N$ OU processes starting from $\vec X_0$ at time $T_0$ arrive at $\vec X$ at time $T$ without crossing each other in the
time interval $[T_0,T]$. 

Similarly we define the {\it barrier-constrained propagator}, i.e the probability 
$P_N^ { B r } ( \vec { x } , t | \vec { x _ { 0 } } , t _ { 0 }  ; W \sqrt{t})$ of the event that the $N$ Brownian particles starting from $\vec x_0$ at time $t_0$ arrive at $\vec x$ at time $t$ without crossing each other 
{\it and} remaining under the barrier $W \sqrt{\tau}$ in the
time interval $[t_0,t]$
\bea
\label{eq:DefPw} 
&& P_N^ { B r } ( \vec { x } , t | \vec { x _ { 0 } } , t _ { 0 } \ ; \ W\sqrt{t} ) d \vec x \\
&& = {\rm Pr} \left[ x_i(t) \in [x_i,x_i+dx_i] , 1\leq i \leq N  \ ; \  x_1(\tau) < x_2 ( \tau) < \cdots < x_N(\tau) < W \sqrt{\tau}, \ \forall \tau \in [t_0,t]  \mid x_i(t_0) = {x_0}_i  \right] \;. \nonumber 
 \eea
A similar definition holds for the OU barrier constrained propagator.
The Lamperti mapping can also be applied in the presence of an absorbing boundary, leading to the relation
\be
\label{eq:LinkPBrOUNW}
P_N^ { B r } ( \vec { x } , t | \vec { x _ { 0 } } , t _ { 0 } ; \  W\sqrt{t} ) \  {d} \vec { x } = P_N^{ O U } ( \vec { X } , T | \vec { X } _ { 0 } , T _ { 0 } ; \ W ) \ {d} \vec { X } \;.
\ee
The barrier constrained OU process can now be studied using fermions. 

\subsubsection{Mapping to fermions}

{\it In the absence of a wall: free case}. This case corresponds to $W=+\infty$. To compute the probability that $N$ independent OU processes do not cross we use the Karlin-McGregor Theorem \cite{KarMcGreg1959} and obtain
\be
P_N^{OU} (\vec{X}, T \mid \vec{X_0}, T_0 )  = \det_{1 \leq i,j \leq N} P^{OU}(X_i, T \mid X_{0j}, T_0 )
\ee
where $P^{OU}(X_i, T \mid X_{0j}, T_0 )$ is the single particle propagator. 
Using Eq. (\ref{eq:1partPropagators}) we obtain
\be
\label{eq:preCauchyBinet}
P_N^{OU} (\vec{X}, T \mid \vec{X_0}, T_0  ) 
= e^{-\frac{1}{4} \sum\limits_{i=1}^{N} X_i^2   } \det_{1 \leq i,j \leq N}  G ( X_i , T | X _ { 0j } , T _ { 0 }  )  \, e^{\frac{1}{4} \sum\limits_{j=1}^{N} X_{0j}^2   }  \;.
\ee
Using the eigenstate decomposition of the single particle propagator given in Eq. \eqref{decompose},
together with the (reverse) Cauchy Binet formula, we can rewrite
\be
\label{eq:postCauchybinet}
\det_{1 \leq i,j \leq N}  G ( X_i , T | X _ { 0j } , T _ { 0 } ) = \frac{1}{N!} 
\sum _ {k _ { 1 },k _ { 2 }, \cdots, k _ { N } \geq 0 } ^ { } \det\limits_{1 \leq i,j \leq N} \phi _ { k _ { i } } \left( X _ { j } \right) \det\limits_{1 \leq i,j \leq N}\phi _ { k _ { i } } ^ { * } \left( X _ {0 j } \right) e ^ { - (T-T_0) 
\sum _ { \ell = 1 } ^ { N } \epsilon^{\rm free}_ { k _ { \ell } } } \;,
\ee
where we recall that $\epsilon^{\rm free}_ {k} = \frac{k}{2}$.

As we show now, the right hand side (rhs) of Eq. (\ref{eq:postCauchybinet}) is just the quantum propagator of $N$ noninteracting fermions in a quadratic potential defined in the previous subsection for the single particle system [see Eq. (\ref{eq:HamiltH})]. Consider the $N$-body Hamiltonian of this system:
\be
\mathcal{H}_N = \sum_{1 \leq i \leq N} \hat{H}_i
\ee
where $\hat H_i$ is defined in \eqref{eq:HamiltH} with the substitution $X \to X_i$.
Because of the Pauli exclusion principle for fermions, the set of many body energy levels $E_{\vec{k}}(N,+\infty)$
(which corresponds to $W=+\infty$), is obtained from the single particle energy levels $\epsilon^{\rm free} _ { k  }$ as 
\be 
\big\{   E_{\vec{k}}(N,+\infty)  = \sum_{1 \leq i \leq N }   \epsilon^{\rm free}_ { k _ { i } } , \ \vec{k} \in \Omega_N \big\} \ , \quad \quad \Omega_N = \{ \vec{k} \in \mathbb{N}^N \text{ such that } k_1 < k_2 < \cdots < k_N    \} \;.
\ee
For $\vec{k} \in \Omega_N$, the $N$-body eigenfunctions $\Psi_{\vec{k}} (\vec{X})$ associated to the energy $E_{\vec{k}}(N,+\infty)$ are obtained as Slater determinants built from the single particle eigenfunctions 
\be  \label{slater}
\Psi_{\vec{k}} (\vec{X}) = \frac{1}{\sqrt{N!}} \det_{1 \leq i,j \leq N} \phi_{k_i} (X_j) \;.
\ee
Therefore the $N$-particle quantum propagator is given by
\bea 
\label{eq:NpartQuantProp}
G_N(\vec{X}, T \mid \vec{X_0}, T_0 ) &=&
\sum_{\vec{k} \in \Omega_N} \Psi_{\vec{k}} (\vec{X})  {\Psi_{\vec{k}} } ^* (\vec{X_0}) e^{-(T-T_0)  E_{\vec{k}}(N,+\infty)  }  \\
&=& \frac{1}{(N!)^2} \sum_{k_1, k_2, \cdots, k_N\geq 0} \det\limits_{1 \leq i,j \leq N} \phi _ { k _ { i } } \left( X _ { j } \right) \det\limits_{1 \leq i,j \leq N}\phi _ { k _ { i } } ^ { * } \left( X _ {0 j } \right)  e^{-(T-T_0)  E_{\vec{k}}(N,+\infty)  }\;,
\eea
where, in the last equality we have used the expression of the many-body wave function (\ref{slater}) and replaced the constrained sum 
$\sum_{\vec{k} \in \Omega_N}$ by $1/(N!)  \sum_{k_1, k_2, \cdots, k_N\geq 0}$. By comparing Eqs. (\ref{eq:postCauchybinet}) and (\ref{eq:NpartQuantProp}) one thus finds 
\bea \label{identity_propag_gs}
G_N(\vec{X}, T \mid \vec{X_0}, T_0 ) = \frac{1}{N!} \det_{1 \leq i,j \leq N}  G ( X_i , T | X _ { 0j } , T _ { 0 } ) \;.
\eea
And therefore, by comparing Eq. (\ref{eq:preCauchyBinet}) and the latter identity (\ref{identity_propag_gs}), we establish a mapping between the probability for the absence of crossing of $N$ independent OU processes and the quantum propagator for a system of $N$ noninteracting fermions in a harmonic trap \cite{PLDMajSch2018,fermions_review}
\be
\label{eq:density}
 P^{OU}_N (\vec{X}, T \mid \vec{X_0}, T_0 ) = N! \ e^{-\frac{1}{4} \sum\limits_{i=1}^{N} X_i^2 }G_N(\vec{X}, T \mid \vec{X_0}, T_0 ) \ e^{\frac{1}{4} \sum\limits_{i=1}^{N} X_{0i}^2   } \;,
\ee
which generalises the identity in Eq. (\ref{identity_ou_fermion_single}) to the case of $N$ particles. 

{\it In the presence of a wall}. As in the single particle case, adding an absorbing barrier at $X=W$ for the OU processes is equivalent to adding an infinite wall at $X=W$ in the corresponding quantum system. In this case, the $N$-body Hamiltonian $\hat{H}_W$ is given by $\mathcal{H}_W  = \sum_{1 \leq i \leq N} \hat{H}^W_{i}$ where the single particle Hamiltonian 
$\hat{H}^W$ is given in Eq.~(\ref{eq:Hw}) with the substitution $X \to X_i$. Consequently Eq. \eqref{eq:density}, in the presence of a wall at $W$ reads
\be \label{prop3} 
 P_N^{OU} (\vec{X}, T \mid \vec{X_0}, T_0 ; \ W ) = N! \ e^{-\frac{1}{4} \sum\limits_{i=1}^{N} X_i^2 }G_N(\vec{X}, T \mid \vec{X_0}, T_0 ; \ W ) \ e^{\frac{1}{4} \sum\limits_{i=1}^{N} X_{0i}^2   } \;,
\ee
where the $N$-body quantum propagator is given by 
\bea \label{GNW}
G_N(\vec{X}, T \mid \vec{X_0}, T_0 ; \ W ) 
= \sum_{\vec{k} \in \Omega_N} \Psi_{\vec{k}} (\vec{X},W)  {\Psi_{\vec{k}} }^* (\vec{X_0},W) e^{-(T-T_0)  E_{\vec{k}}(N,W) } \;.
\eea
Here $E_{\vec{k}}(N,W) = \sum_{1 \leq i \leq N} \epsilon_{k_i}(W)$, $\vec k \in \Omega_N$, are the eigenvalues of $\mathcal{H}_W$ and $\Psi_{\vec{k}} (\vec{X},W)$ are given by Slater determinants as in Eq. (\ref{slater}) with the substitution $\phi_{k_i}(X_j) \to \phi_{k_i}(X_j,W)$.

In summary, the barrier constrained propagator $P_N^{ B r } ( \vec { x } , t | \vec { x _ { 0 } } , t _ { 0 } ; \  W\sqrt{t} )$ for $N$ Brownian motions defined in Eq. (\ref{eq:DefPw}) is related to the $N$-body fermionic quantum propagator $G_N(\vec{X}, T \mid \vec{X_0}, T_0 ; \ W )$ via 
\be 
\label{eq:MappingNpart}
P_N^{ B r } ( \vec { x } , t | \vec { x _ { 0 } } , t _ { 0 } ; \  W\sqrt{t} ) \  {d} \vec { x }  = N! \ e^{-\frac{1}{4} \sum\limits_{i=1}^{N} X_i^2 }G_N(\vec{X}, T \mid \vec{X_0}, T_0 ; \ W ) \ e^{\frac{1}{4} \sum\limits_{i=1}^{N} X_{0i}^2   }  \ {d}\vec{X}  
\quad \quad
\left\{
    \begin{array}{ll}
        \vec{X} = \frac{\vec{x} }{\sqrt{t}} \ & ; \ \vec{X_0} = \frac{ \vec{x_0} }{\sqrt{t_0}}\\
        T = \ln t  \ & ; \  T_0 = \ln t_0
    \end{array}
\right.
\ee
where $G_N$ is given in \eqref{GNW}. 

\section{Survival probability results for non-crossing Brownian motions}
\label{sec:Sprob} 

In this Section, we detail the results that can be obtained from the above mappings.
We consider $N$ independent Brownian motions $x_i(\tau)$ starting at the ordered
initial positions $x_i(t_0)=x_{0i}$, $x_{01} < x_{02} < \dots < x_{0N}$,
at time $t_0$. We are interested in the event such that the walkers do not cross each other
{\it and} stay below the moving boundary 
at $x= W \sqrt{\tau}$, for all $\tau \in [t_0,t]$. We call the probability of this
event the ``survival probability'' $S(t | \vec x_0,t_0;W)$ given by
\be 
\label{eq:DefS2}
S(t | \vec x_0,t_0;W) = {\rm Pr} \left[ x_1(\tau) < x_2 ( \tau) < \cdots < x_N(\tau) < W \sqrt{\tau} , \ \forall \tau \in [t_0,t] \ | \ x_{i}(t_{0})=x_{0 i} \right] \;.
\ee
Obviously it depends on $N$, but for the simplicity of notations we suppress
the explicit $N$ dependence.

It is also interesting to consider the conditional probability, {\it given that} they do not cross each other
in the time interval $[t_0,t]$, that they remain below the moving boundary 
$x= W \sqrt{\tau}$, for all $\tau \in [t_0,t]$
\be
\label{eq:DefSc2}
S_c(t | \vec x_0,t_0;W) = {\rm Pr} \left[ x_N(\tau) < W \sqrt{\tau}, \forall \tau \in [0,t] \ \mid  \ x_1(\tau)< x_2(\tau) < \cdots < x_N(\tau) , \forall \tau \in [t_0,t] \ , \ x_{i}(t_{0})=x_{0 i}  \right] \;.
\ee
This conditional probability can also be interpreted as the probability that $N$ non-crossing
Brownian paths (the so-called vicious walkers) remain below a moving boundary. 
These two probabilities $S$ and $S_c$ are related via
\be \label{ratio} 
S_c(t | \vec x_0,t_0;W) = \frac{ S(t | \vec x_0,t_0;W) }{\lim_{W \to + \infty} S(t | \vec x_0,t_0;W) } \;,
\ee 
since the denominator is precisely the probability that the walkers remain non-crossing up to
time $t$. We show below that both probabilities decay algebraically at late times $t \to +\infty$
\be
S(t | \vec x_0,t_0;W) \sim t^{- \beta(N,W)} \quad , \quad S_c(t | \vec x_0,t_0;W) \sim t^{- \beta_c(N,W)} \;,
\ee
where the decay exponents $\beta(N,W)$ and $\beta_c(N,W)$ are computed below.

\subsection{A general formula for the survival probability exponent $\beta(N,W)$} 

Using the mapping discussed in the previous section, the survival probability 
$S(t | \vec x_0,t_0;W)$ defined in \eqref{eq:DefS2} can be expressed in terms of the quantum propagator $G_N$ from (\ref{eq:MappingNpart}) where we  integrate over the final positions, i.e.
\bea
\label{eq:GtoS}
&& S(t | \vec x_0,t_0;W)  =  \int_{x_1<x_2<\dots<x_N<W \sqrt{t}} \, {d} \vec { x }
\, \, P_N^{ B r } ( \vec { x } , t | \vec { x _ { 0 } } , t _ { 0 } ; \  W\sqrt{t} )  \\
&& = 
N!  \left( \int_{X_1<X_2<\dots<X_N<W} 
{d} \vec{X}
\, e^{-\frac{1}{4} \sum\limits_{i=1}^{N} X_i^2}  G_N(\vec{X}, T \mid \vec{X_0}, T_0 ; \ W) \   \right)  e^{\frac{1}{4} \sum\limits_{i=1}^{N} X_{0,i}^2} \;.
\eea
Using Eq. (\ref{GNW}) we can express the quantum propagator as a sum over the many body
eigenstates of ${\cal H}^W$. Since large $t= e^T$ corresponds to large $T$, we see that this sum
is dominated by the contribution from the ground state of ${\cal H}^W$. This ground state
correspond to the index $\vec{k}_0 = (0, 1 , \cdots, N-1)$ which amounts to fill the $N$ lowest single
particle energy levels of $H^W_i$. Hence we obtain
\be 
G_N(\vec{X}, T \mid \vec{X_0}, T_0; W ) = \sum_{\vec{k} \in \Omega_N} \Psi_{\vec{k}} (\vec{X},W)  {\Psi_{\vec{k}} } ^* (\vec{X_0},W) e^{-(T-T_0)  E_{\vec{k}}(N,W)  } \ \simeq \ \Psi_{\vec{k_0}} (\vec{X},W)  {\Psi_{\vec{k_0}} } ^* (\vec{X_0},W) e^{-(T-T_0)  E_{\vec{k_0}}(N,W)  }
\ee
This is accurate for $T - T_0 \gg \Delta^{-1}$, where $\Delta$ is the gap between the many body ground state and the
first excited state which is labeled by $\vec k_1=(0,1,\dots,N-2,N)$. This gap
equals $\Delta=\epsilon_N(W) - \epsilon_{N-1}(W)$, where the $\epsilon_k(W)$ are the
single particle energy levels of $\hat H^W$. Recalling further that $T= \ln t$ and $T_0 = \ln t_0$ we
obtain for $T \gg T_0$ (or equivalently $t \gg t_0$)
\be
G_N(\vec{X}, T \mid \vec{X_0}, T_0 ; \ W) \simeq 
\Psi_{\vec{k}_0} (\vec{X},W)  {\Psi_{\vec{k}_0} } ^* (\vec{X_0},W)
\left( \frac{t_0}{t} \right) ^{E_{\vec{k}_0}(N,W) } \;.
\ee
The quantum mapping thus naturally shows that at large time the survival probability decays as a power-law
\be
\label{eq:Slargetime}
S(t | \vec x_0,t_0;W) \simeq A(\vec x_0,t_0;W) \, t^{- \beta(N,W) }  \;,
\ee
where the decay exponent is given by the ground state energy of the $N$-fermion system in the harmonic potential with an infinite wall at $W$, i.e. 
\be 
\label{eq:BetaE}
\beta(N,W) = E_{\vec{k_0}}(N,W)= \sum_{k=0}^{N-1} \epsilon_k(W) \;.
\ee
The amplitude in Eq. (\ref{eq:Slargetime}) is given by
\be \label{AA} 
A(\vec x_0,t_0;W)
 = N! \, t_0^{E_{\vec{k}_0}(N,W)} \left( \int_{X_1<X_2<\dots<X_N<W} 
  e^{-\frac{1}{4} \sum\limits_{i=1}^{N} X_i^2}  
  \Psi_{\vec{k_0}}(\vec{X},W) \ {d}\vec{X}   \right) {\Psi_{\vec{k}_0} }^*(\vec{X_0},W) e^{\frac{1}{4} \sum\limits_{i=1}^{N} X_{0i}^2}   \;,
\ee
where we recall that $X_{0i}= \frac{x_{0i}}{\sqrt{t_0}}$.

Obtaining explicitly the ground state energy of the harmonic oscillator in the presence of an infinite wall
at $X=W$, hence the exponent $\beta(N,W)$, is in general not easy. While there is a formal
expression for general $W$, discussed below, we can start by analyzing three simpler limiting cases,
$W \to +\infty$ and $W=0$ and $W \to -\infty$.

\begin{itemize}
\item In the limit $W\to +\infty$: $\hat{H}_W$ tends to $\hat{H}$, the simple harmonic oscillator hamiltonian without a wall, which has energy levels $\epsilon_k(W \to + \infty) \to \epsilon^{\rm free}_k = \frac{k}{2}$. Thus:
\be
\lim\limits_{W\to \infty} \beta(N,W) \quad = \quad \sum_{k=0}^{N-1} \frac{k}{2} \quad = \quad \frac{N(N-1)}{4} \;.
\ee
This yields back the Fisher exponent that characterizes the algebraic decay at late times of the
probability that $N$ independent Brownian walkers do not cross each other up to time $t$
\cite{Fish1984, KratGutVien2000, BraWin2004}. Note also that Eq. \eqref{ratio} implies that
\be \label{betabetac} 
\beta_c(N,W)= \beta(N,W) - \frac{N(N-1)}{4} \;,
\ee
which is valid for any $W$.

\item  $W=0$: In this case the hard wall imposes a zero of the wavefunction at $W=0$. The Hilbert space is composed of all odd wave-functions of the harmonic oscillator. The ground state energy is then obtained by populating the first $N$ odd levels of the free harmonic oscillator and therefore $\beta(N,W=0)$ in Eq. (\ref{eq:BetaE}) is given by
\be
\label{eq:BetaW0}
\beta(N,W=0) \quad = \quad \sum_{k=0}^{N-1} \frac{2k+1}{2} \quad = \quad \frac{N^2}{2} \;.
\ee
This gives the exponent for the temporal decay of the probability that $N$ independent
Brownian motions do not cross each other and remain below $W=0$ up to time $t$ and coincides with the
results previously obtained, using various different methods, in Ref. \cite{KratGutVien2000,KatTan2002,BraWin2004}. 
From \eqref{betabetac} we further find that for $W=0$
\be
\label{eq:BetacW0}
 \beta_c(N,W=0) = \frac{N(N+1)}{4}
\ee
which gives the probability that $N$ vicious walkers remain below $x=0$. 

\bigskip

\item In the limit $W \to -\infty$, the wall is very far to the left of the center at a potential energy 
$V(W)= \frac{W^2}{8}$. Expanding $X= W + \tilde X$ for small $\tilde X<0$ the potential energy is
$V(X)= \frac{W^2}{8} + \frac{W}{4} \tilde X + O(W^0)$. The energy levels are therefore
similar to the one of a particle in a linear potential with a slope $\frac{W}{4}$ and confined on the negative axis (i.e. with a hard-wall at $X=0$). 
The single particle energy levels are thus given by (see e.g. \cite{MajCom2005})
\be \label{form1}
\epsilon_k = \frac{W^2}{8}  + \frac{1}{2^{5/3}} \alpha_{k+1} |W|^{2/3} 
\ee 
where $- \alpha_m$ is the $m$-th zero of the Airy function. The eigenfunctions
are Airy functions centered at $X_k = \alpha_{k+1} (2/W)^{1/3} \ll W$ for fixed $k$ and $W \to - \infty$, 
hence in that limit we can indeed neglect the term $O(1)$ in the potential (which is $\propto X^2$)
and the linear potential approximation becomes exact. Hence we find the
asymptotics for fixed $N$ and $W \to -\infty$
\be \label{form2} 
\beta(N,W) \simeq \frac{N W^2}{8} +  A_N W^{2/3} \quad , \quad A_N = \frac{1}{2^{5/3}} \sum_{m=1}^N \alpha_m \;.
\ee
Later we will estimate this amplitude $A_N$ for large $N$ [see Eq. (\ref{largeW})].

\item General $W$: for arbitrary $W$, the single particle energy levels $\epsilon_k(W)$ are as follows.
The spectrum of the single particle Hamiltonian $\hat{H}_W$ can be obtained by solving
the eigenvalue equation with a Dirichlet boundary condition at $X=W$
\be  
\label{eq:diffEqEig}
\hat{H}_W \phi = \epsilon \phi   \quad \iff \quad \left\{
    \begin{array}{ll}
        \frac{\partial^2 \phi}{\partial X^2}  + (  2 \epsilon + \frac{1}{2} - \frac{1}{4} X^2   ) \phi = 0  \quad  \text{for $X<W$}   \\
        \\
        \phi (W) = 0  \quad , \quad  \phi (-\infty) = 0 \;.
    \end{array}
\right.
\ee
The solution of this differential equation which vanishes at $X \to - \infty $ is denoted
\be  \label{eigen} 
\phi ( X,W) = C_\epsilon \, D_{2 \epsilon}(-X) \;,
\ee
where $D_p(z)$ is the parabolic cylinder function of index $p$ and $C_\epsilon$ is
a normalization constant. Furthermore the boundary condition $\phi (W,W) = 0$ quantizes 
the eigenvalues $\epsilon=\epsilon_k(W)$ to be the $k+1$-th root in increasing order of the eigenvalue
equation 
\be
\label{eq:implicitEq}
D_{2 \epsilon} (-W) = 0 \;.
\ee
The series of energy levels $( \epsilon_k(W))_{ k \geq 0 } $ is thus the ordered sequence of solutions of this equation. 
The corresponding wavefunctions are
\be
\phi_k( X,W) = C_{\epsilon_k(W)} \, D_{2 \epsilon_k(W)}(-X) \;.
\ee
To guide the intuition, we show in Fig. \ref{fig:PlotD} a plot of the function 
$D_{2 \epsilon} (-W)$ versus $\epsilon$, for a given value of $W=6$, over the range $[0,7]$ where the first $13$ zeros can be found. The amplitude of the oscillations quickly increases.
We note that the first seven levels are very close to what would be expected for a harmonic oscillator $(0, \frac{1}{2}, 1, \frac{3}{2}, 2, \frac{5}{2}, 3 )$, but that the following levels start deviating from the $\frac{k}{2}$ levels, as the effect of the hard wall becomes more important for large $\epsilon$.

\begin{figure}[!h]
    \center
    \includegraphics[width=0.4\linewidth]{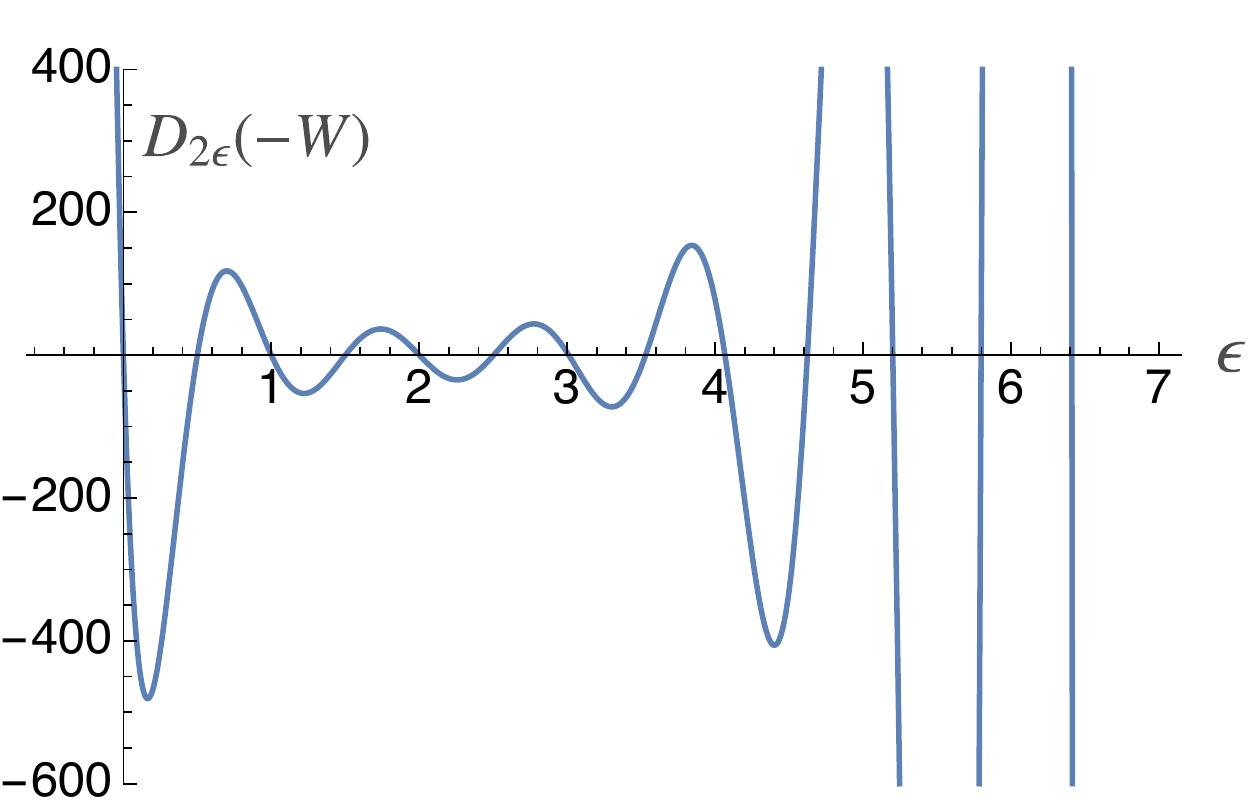}
    \caption{Plot of the function $D_{2 \epsilon} (-W)$ versus $\epsilon$ for $W=6$. }
	\label{fig:PlotD}
\end{figure}

In conclusion, for general $W$ the survival decay exponent $\beta(N,W)$
is given by the $\beta(N,W) = \sum_{0 \leq k \leq N-1}\epsilon_k(W) $, i.e. 
the sum of the $N$ smallest roots of the equation $D_{2 \epsilon} (-W) = 0$.

\end{itemize}

\subsection{Perturbative expansion of $\beta(N,W)$ around $W=0$}

\begin{figure}[t]
\includegraphics[width = \linewidth]{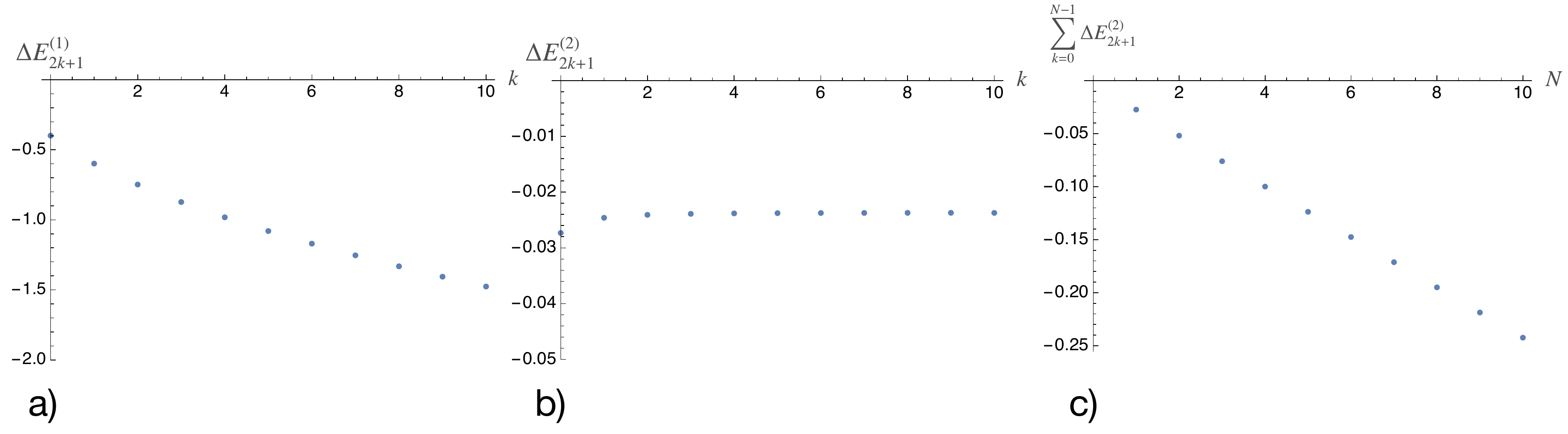}
\caption{Plots of the first values of $\Delta E_{2k+1}^{(1)}$,   $\Delta E_{2k+1}^{(2)}$  and of the second order coefficient in the perturbative expansion $\sum_{k=0}^{N-1} \Delta E_{2k+1}^{(2)}$, which is well approximated by a straight line.}\label{Fig_DeltaE}
\end{figure}

When the wall is close to the minimum of the harmonic trap, $|W| \ll 1$, we can calculate
$\epsilon_k(W)$ and the exponent $\beta(N,W)$ perturbatively for small $W$ and for any fixed $N$.
For $W=0$, the Hamiltonian reads [see Eq. (\ref{eq:Hw})]
\be
\hat H^{W=0} = -\frac{1}{2}\frac{\partial^2}{\partial X^2} + \frac{1}{8} X^2 - \frac{1}{4} \;, \; {\rm for} \; \; X < 0 \;,
\ee
while $\hat H^{W=0} = + \infty$ for $X>0$. Consequently, the eigenstates of this single particle Hamiltonian $\hat{H}^{W=0}$ are simply the odd eigenstates of the free harmonic oscillator with wavefunctions $\phi_{2k+1}(x)$, such that $\phi_{2k+1}(0)=0$, and 
energies $\epsilon_{k}(W=0) = \epsilon_{2 k+1}^{\rm free} = \frac{(2k+1)}{2}$. We recall that [see Eq. (\ref{eq:BetaW0})]   
\be 
\beta(N,W=0) =  \sum_{k=0}^{N-1} \frac{(2k+1)}{2} = \frac{N^2}{2} \;.
\ee
For $|W| \ll 1$, the wall is slightly offset from the minimum of the harmonic trap. 
Note that the range of $X$ is $X \in (-\infty,W]$ such that it is useful to make a change of variable,
$Y=X-W \in (-\infty,0]$, which brings back the wall at the origin, i.e. at $Y=0$. In the $Y$-coordinate the Hamiltonian becomes 
\bea \label{Prime} 
\hat{H}' = -\frac{1}{2}\frac{\partial^2}{\partial Y^2} + \frac{1}{8} (Y+W)^2 - \frac{1}{4} = \hat{H}^{W=0} + \frac{1}{4} Y W+\frac{1}{8}W^2 = \hat{H}^{W=0}  + \frac{1}{8}W^2 + W \Delta\hat{H} \;,
\eea
where we have defined $ \Delta\hat{H} = \frac{1}{4} Y$, which we will treat using perturbation theory since $W \Delta \hat{H}$ is small in (\ref{Prime}). The energy level $\epsilon_k(W)$ for small $W$ is thus obtained from the standard perturbative expansion of quantum mechanics and reads,
up to second order in $W$:
\be \label{perturb} 
\epsilon_{k}(W) - \epsilon_{k}(W=0) = \Delta E _{2k+1}
= \Delta E _{2k+1}^{(1)}  \ W+  \Delta E _{2k+1}^{(2)}  \ W^2 + \frac{1}{8}W^2 + \mathcal{O}(W^3)
\ee
with 
\be 
    \begin{cases}
       \Delta E _{2k+1}^{(1)} = \bra{2k+1} \Delta\hat{H} \ket{2k+1} \\
       \\
        \Delta E _{2k+1}^{(2)} = \sum\limits_{k' \geq 0, k'\neq k} \frac{\abs{ \bra{2k'+1}\Delta\hat{H} \ket{2k+1} }^2 }{E_{2k+1} -E_{2k'+1} }  \label{eq:DefDeltaE}
    \end{cases}
\ee
where $\ket{2k+1} $ denote the odd levels of the free harmonic oscillator, but normalized to unity over the half-space $(-\infty,0]$.
Computing the matrix elements on the half-space, see Appendix \ref{app:HarmonicOsc}, we obtain explicitly
\bea \label{deltaE_text}
\Delta E_{2k+1}^{(1)} = -\frac{1}{\sqrt{2\pi}2^{2k}} \frac{(2k+1)!}{k!^2} \;.
\eea
In Fig. \ref{Fig_DeltaE} a), we show a plot of $\Delta E_{2k+1}^{(1)}$ as a function of $k$. Similarly one obtains $\Delta E_{2k+1}^{(2)}$ in (\ref{eq:DefDeltaE}) under a slightly more complicated form (see Appendix \ref{app:HarmonicOsc} for details)
\bea \label{DeltaE2}
\Delta E_{2k+1}^{(2)} = \frac{(2k+1)!}{\pi \, 2^{2k+1}\, (k!)^2}\sum_{k'\geq 0, k' \neq k} \frac{1}{k-k'} \frac{(2k'+1)!}{2^{2k'} (k'!)^2} \frac{1}{(4(k-k')^2 - 1)^2} \;.
\eea
In Fig. \ref{Fig_DeltaE} b), we show a plot of $\Delta E_{2k+1}^{(2)}$ as a function of $k$ where one sees that  $\Delta E_{2k+1}^{(2)}$ quickly converges to a finite value for large $k$. In fact, one can show that (see Appendix \ref{app:HarmonicOsc})
\begin{eqnarray} \label{asympt_delta}
\lim_{k \to \infty} \Delta E_{2k+1}^{(2)} = \frac{1}{\pi^2} - \frac{1}{8} = -0.0236788 \ldots \;,
\end{eqnarray}
which is fully consistent with the plot shown in \ref{Fig_DeltaE} b). 

To obtain the decay exponent $\beta(N,W)=\sum_{k=0}^{N-1} \epsilon_{2 k +1}(W)$ we perform the summation 
over the first $N$ levels given in \eqref{perturb}, leading to 
\be \label{pert1} 
\beta(N,W) = \frac{N^2}{2} - \frac{ \sqrt{2}}{3 \sqrt{\pi}} \frac{1}{2^{2N} } \frac{N (2N+1)!}{(N!)^2} \  W + W^2 \left(\sum_{k=0}^{N-1} \Delta E_{2k+1}^{(2)}   + \frac{N}{8}\right) + \mathcal{O}(W^3) \;.
\ee
Specialising this formula (\ref{pert1}) to the case $N=1$, we obtain up to order $W^2$ 
\be  \label{pert2} 
\beta(N=1,W) = \frac{1}{2} - \frac{W}{\sqrt{2 \pi}} + \frac{1-\ln 2}{\pi}  W^2 +  \mathcal{O}(W^3)  \;,
\ee
which is in agreement with the first order result $\beta(1,W) =\frac{1}{2} - \frac{W}{\sqrt{2 \pi}} + {\cal O}(W^2)$ obtained in \cite{RedKrap1996}. Similarly, for the case $N=2$, we get from Eq. (\ref{pert1})
\be \label{pert3} 
\beta(N=2,W) = 2 - \frac{5 W}{2 \sqrt{2 \pi}} + \frac{23- 26 \ln 2}{8 \pi} W^2 +  \mathcal{O}(W^3)  \;.
\ee

Let us define the coefficients of the Taylor series expansion at small $W$
\be
\beta(N,W) = \frac{N^2}{2} + \sum_{p=1}^\infty a_p(N) W^p \;.
\ee
The first two coefficients $a_1(N)$ and $a_2(N)$ can be read off straightforwardly from Eq. (\ref{pert1}). It is interesting to study their large $N$ behaviors. The one of $a_1(N)$ can be obtained from Stirling's formula while the behavior of $a_2(N)$ can be obtained (at leading order for large $N$) from Eq. (\ref{DeltaE2}). This yields  
\bea  
&&a_1(N) = - \frac{2\sqrt{2}}{3 \pi} N^{\frac{3}{2}} -\frac{1}{2 \left(\sqrt{2} \pi \right)} \sqrt{N}+\frac{7 }{96 \sqrt{2} \pi } \frac{1}{\sqrt{N}} +  \mathcal{O}(N^{-3/2}) \label{a1} \\
&&a_2(N) = \frac{N}{\pi^2} + {\cal O}(\sqrt{N}) \;, \label{a2}
\eea
which we will compare below with the expansion for large $N$ and large $W \sim \sqrt{N}$. Before
that we first examine the expansion for large $N$ at fixed $W$.

\subsection{Large $N$ at fixed $W<0$}

In the case of fixed $W<0$, in the limit of large $k$, the energy level $\epsilon_k(W)$ obtained from 
\eqref{eq:implicitEq} can be approximated as follows. We first use the relation between the parabolic cylinder function with positive argument and the confluent hypergeometric function,
$U \left( - \frac { 1 } { 2 } \nu , \frac { 1 } { 2 } , \frac { x ^ { 2 } } { 2 } \right) = e ^ { x ^ { 2 } / 4 } 2 ^ { - \nu / 2 } D _ { \nu } ( x )$ for $x>0$. We apply this relation for $x=-W$ and $\nu = 2 \epsilon$.
Next we use the asymptotic estimate of the $n$-th root for $a$ of the equation  $U(a,b,z)=0$, with fixed $b$ and $z>0$, with $n$ counted from $1$, given in  \cite{DLMF} 
\be
a = - n - \frac { 2 } { \pi } \sqrt { z n } - \frac { 2 z } { \pi ^ { 2 } } + \frac { 1 } { 2 } b + \frac { 1 } { 4 } + \frac { z ^ { 2 } \left( \frac { 1 } { 3 } - 4 \pi ^ { - 2 } \right) + z - ( b - 1 ) ^ { 2 } + \frac { 1 } { 4 } } { 4 \pi \sqrt { z n } } +  \mathcal{O}\left( \frac{1}{n} \right) \;.
\ee
Substituting $a=- \epsilon_k(W)$, $z=W^2/2$, $n=k+1$ and $b=1/2$ we obtain the following estimate of the energy level $\epsilon_k(W)$ for large $k$
\be
\epsilon_k(W) =  k-\frac{\sqrt{2} \sqrt{k} W}{\pi }+\left(\frac{W^2}{\pi
   ^2}+\frac{1}{2}\right)+\frac{\sqrt{\frac{1}{k}}
   \left(\pi ^2 W \left(W^2-18\right)-12 W^3\right)}{24
   \sqrt{2} \pi
   ^3}+ \mathcal{O}\left(k^{-1}\right) \;.
\ee 
This gives the correct levels for $W=0$ where $\epsilon_k = \frac{2k+1}{2}$. Summing the series, and performing the expansion at large $N$, we obtain the asymptotic behavior of $\beta(N,W)$ for large $N$ and fixed $W<0$
\be  \label{asymp} 
\beta(N,W) = \sum_{k=0}^{N-1} \epsilon_k  = \frac{N^2}{2} - \frac{2 \sqrt{2} W}{3 \pi} N^\frac{3}{2} + \frac{W^2}{\pi^2} N + \frac{ \left(\pi ^2 W \left(W^2-6\right)-12 W^3\right)}{12 \sqrt{2} \pi^3}  N^{\frac{1}{2}} + o\left(N^{\frac{1}{2}}\right) \;.
\ee
Note that, in principle, it should be possible to obtain the large $N$ behavior of $\beta(N,W)$ for fixed $W>0$, but this analysis seems more complicated and is not presented here.

\subsection{The limit of large $N$ and large $W= {\cal O}(\sqrt{N})$: semiclassical analysis}

\begin{figure}[t]
\includegraphics[width =0.9 \linewidth]{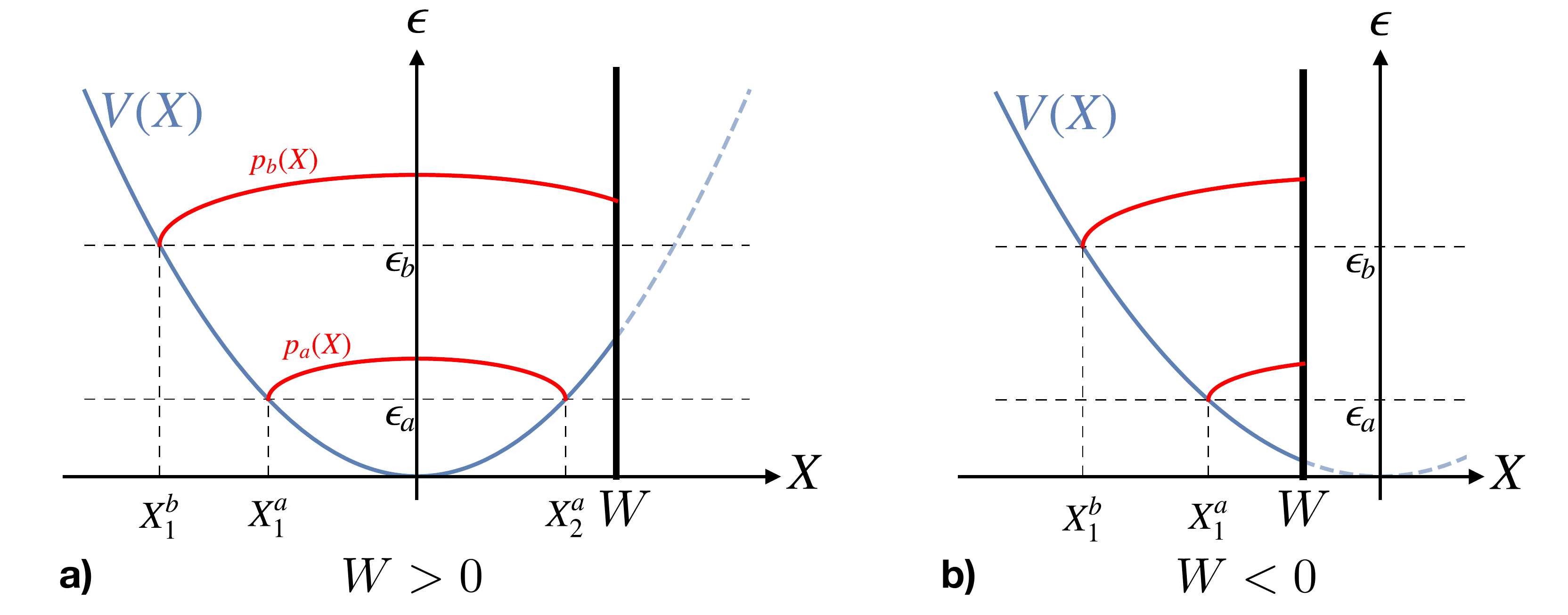}
 \caption{
    Quadratic potential $V(X)$ with a hard-wall in $W$ for a) $W>0$ and b) $W<0$. For a given energy $\epsilon$, the red line shows the semiclassical momentum $p(X)$. This function is supported on $[X_1, X_2]$ with $X_1 = - \sqrt{8 \epsilon} $ and $ X_2 = \min( \sqrt{8 \epsilon},W)$.}\label{Fig_semi_cl}
\end{figure}

Let us consider now the limit when $N$ and $|W|$ are simultaneously large (with $W = {\cal O}(\sqrt{N})$, see below) where the semi-classical approximation becomes exact. We consider the potential $V(X)=\frac{X^2}{8} - \frac{1}{4}$ as in \eqref{eq:Hw} with an infinite wall at $W$ (see Fig. \ref{Fig_semi_cl}). Let $\rho_W(\epsilon)$ denote the
density of single particle energy levels for this problem. Since we consider the ground state with $N$ particles, we
have the exact relation
\be \label{Nrho} 
N = \int_0^{\epsilon_F} {d} \epsilon \, \rho_W(\epsilon) \;,
\ee 
where $\epsilon_F$ denotes the Fermi energy, i.e. the energy of the highest occupied level. The survival
exponent $\beta(N,W)$ is given by Eq. \eqref{eq:BetaE}, i.e. it is equal to 
the ground state energy of the $N$ fermions. Thus one has
\be \label{betarho} 
\beta(N,W) = \int_0^{\epsilon_F} {d}\epsilon \, \epsilon \, \rho_W(\epsilon)  \;.
\ee
Hence $\beta(N,W)$ is obtained by eliminating $\epsilon_F$ from these two equations (\ref{Nrho}) and (\ref{betarho}).

Until now, these expressions (\ref{Nrho}) and (\ref{betarho}) are exact. In the limit of large $N$ and large $W$ we can compute $\beta(N,W)$
using the semi-classical approximation for the density of states. In this limit, the integrals in 
\eqref{Nrho} and \eqref{betarho} are then dominated by large values of $\epsilon \sim N$
as we show below. Under the semi-classical approximation
the energy level $\epsilon_k$ satisfies the Bohr-Sommerfeld quantization condition (setting $\hbar=1$) 
$\int_{X_1}^{X_2} p(X) {d}X = k \pi$, 
where the momentum $p(X) = \sqrt{2 (\epsilon_k - V(X))}$ and $X_1<X_2$ are the turning points of the
classical trajectories. These have a different form for $W>0$ and $W<0$ (see Fig. \ref{Fig_semi_cl}). 

From the figure one sees that one can write in all cases, $X_1=- \sqrt{8 \epsilon}$ and $X_2= \min( \sqrt{ 8 \epsilon} ,W)$. For the $k$th level, this leads to the condition
\be
\label{eq:semiclass}
\int _ {- \sqrt{ 8 \epsilon_k}}^ { \min( \sqrt{ 8 \epsilon_k} ,W) } \sqrt { 2 \left( \epsilon_k - \frac { X ^ { 2 } } { 8 } \right) } \  {d} X = k \pi  \;.
\ee
where we used the shortcut notation $\epsilon_k \equiv \epsilon_k(W)$. In this limit we can approximate $V(X) \simeq \frac{X^2}{8}$ since the integral is dominated
by large values of $X \sim \sqrt{N}$. By definition the density of states $\rho_W(\epsilon)$ is such that
\be
\rho_W(\epsilon_k) = \frac{dk}{d\epsilon_k} \;,
\ee 
Taking a derivative with respect to $\epsilon_k$ of \eqref{eq:semiclass} in the continuum
limit we obtain the semi-classical density of states
\bea\label{expr_sc}
\rho_W(\epsilon)  \simeq \rho^{\rm sc}_W(\epsilon)  = \frac{1}{\pi} \int _ {- \sqrt{ 8 \epsilon} }^ { \min( \sqrt{ 8 \epsilon} ,W) } \frac{{d}X}{ \sqrt { 2 \left( \epsilon - \frac { X ^ { 2 } } { 8 } \right) } } \;,
\eea 
where the superscript `sc' refers to `semi-classical'. The integral in (\ref{expr_sc}) can be performed explicitly and one obtains for $W>0$
\be \label{rhoW} 
\rho^{\rm sc}_{W>0}(\epsilon)  = \begin{cases} & 2 \;, \quad \hspace*{2.7cm} \epsilon < \frac{W^2}{8} \;, \\
& \\
& 1 + \frac{2}{\pi} \arcsin \left( \frac{W}{\sqrt{8 \epsilon}} \right) \;, \quad \epsilon > \frac{W^2}{8} \;,
\end{cases} 
\ee
which is plotted in Fig. \ref{fig:PlotRho}. Its asymptotic behaviors for large $\epsilon \gg W^2$ and near 
the point $\epsilon=W^2/8$ for $\epsilon > \frac{W^2}{8}$ are given respectively by
\be
\rho^{\rm sc}_{W>0}(\epsilon)  =  \begin{cases} & 1+\frac{W }{\pi \sqrt{2 \epsilon} 
   }+\mathcal{O}\left(\frac{1}{(\epsilon/W^2)^{3/2}} \right) \;, \quad  \hspace*{1.9 cm} \text{as } \epsilon \to \infty \;, \\
   &   \\
   & 2-\frac{4 \sqrt{2}}{\pi 
   W} \sqrt{\epsilon -\frac{W^2}{8}}+\mathcal{O}\left(\left(\epsilon
   -\frac{W^2}{8}\right)^{3/2}\right) \;, \quad \text{as } \epsilon \downarrow \frac{W^2}{8} \;.
   \end{cases} 
\ee 
Thus, for $W>0$, the density has a square root singularity at $\epsilon=W^2/8$. For $W<0$ one obtains
instead
\be \label{rhoW2} 
\rho^{\rm sc}_{W<0}(\epsilon)  = \begin{cases} & 0 \;, \quad  \hspace*{2.7cm} \epsilon < \frac{W^2}{8} \;, \\
& \\
& 1 + \frac{2}{\pi} \arcsin \left( \frac{W}{\sqrt{8 \epsilon}} \right) \;, \quad \epsilon > \frac{W^2}{8} \;,
\end{cases} 
\ee
which is also plotted in Fig. \ref{fig:PlotRho}.
Its asymptotics for large $\epsilon \gg W^2$ and near 
the point $\epsilon=W^2/8$ for $\epsilon > \frac{W^2}{8}$ are given respectively by
\be
\rho^{\rm sc}_{W<0}(\epsilon)  =  \begin{cases} & 1 - \frac{|W| }{\pi \sqrt{2 \epsilon} 
   }+\mathcal{O} \left(\frac{1}{(\epsilon/W^2)^{3/2}} \right)\;, \quad  \hspace*{1.45 cm} \text{as } \epsilon \to \infty \;, \\
   & \\
   & \frac{4 \sqrt{2}}{\pi 
   |W|} \sqrt{\epsilon -\frac{W^2}{8}}+\mathcal{O} \left(\left(\epsilon
   -\frac{W^2}{8}\right)^{3/2}\right) \;, \quad \text{as } \epsilon \downarrow \frac{W^2}{8} \;.
   \end{cases}  
\ee 
Inserting the expressions  \eqref{rhoW}, \eqref{rhoW2},
into \eqref{Nrho} and \eqref{betarho} we obtain:
\begin{figure}[t]
    \center
    \includegraphics[width=0.4 \linewidth]{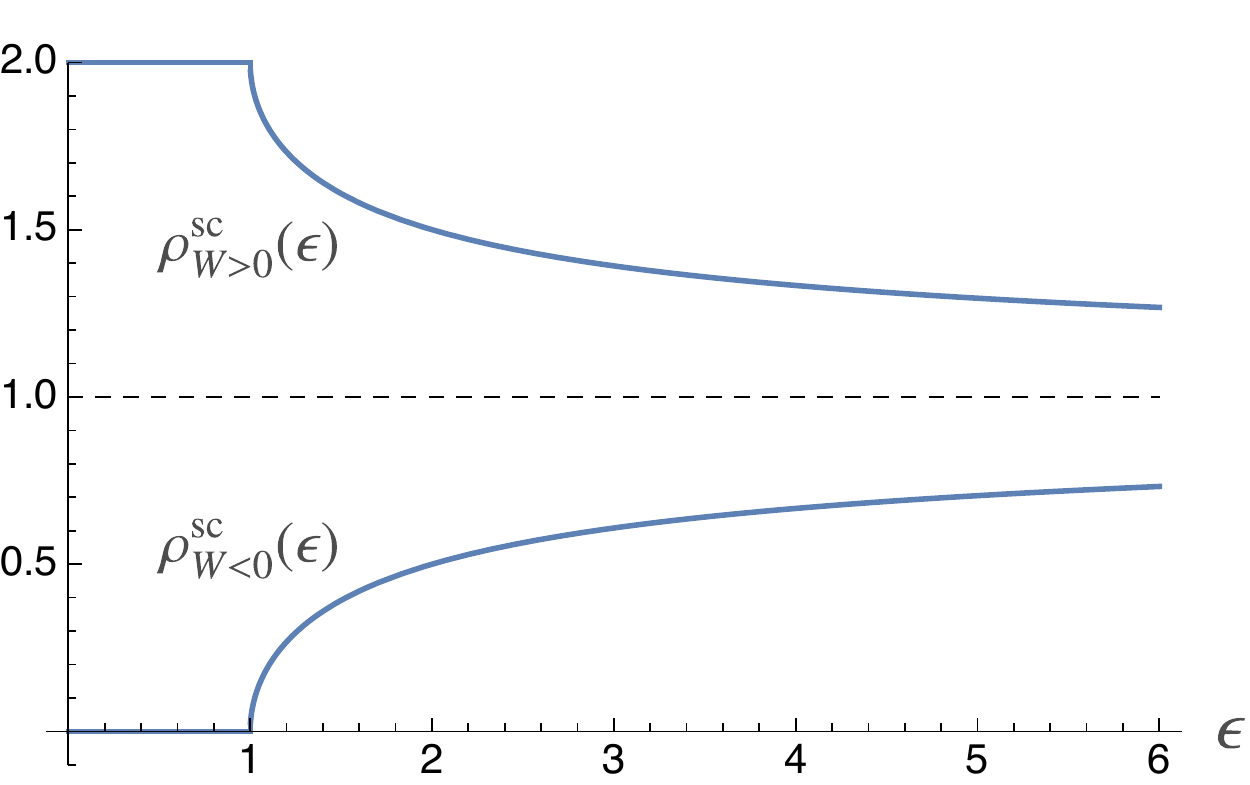}
    \caption{ 
    Plot of the semi-classical density of states $\rho^{\rm sc}_W(\epsilon)$ versus the energy $\epsilon$ in units of $\frac{W^2}{8}$. The top one corresponds to $W>0$ and the bottom one to $W<0$.
    Note that for $W<0$ the density vanishes for $\epsilon < W^2/8$. } 
    \label{fig:PlotRho}
\end{figure}
\begin{enumerate}

\item If $W \geq \sqrt{4 N}$, we obtain $\epsilon_F=N/2$ and
\bea \label{beta95}
\beta(N,W) = \epsilon_F^2 = \frac{N^2}{4} \quad , \quad \beta_c(N,W) = 0
\eea
In this case the system does not feel the wall.

\item If $0 \leq W \leq \sqrt{4 N}$, then $\epsilon_F > \frac{W^2}{8}$. Integrating (\ref{Nrho}) and (\ref{betarho}) using (\ref{rhoW}), we obtain $N$ and $\beta(N,W)$ as functions of the Fermi energy $\epsilon_F$ 

\bea 
&&N=\frac{W \sqrt{8 \epsilon_F -W^2}}{4 \pi }+\frac{2 \epsilon_F  \arcsin \left(\frac{W}{2 \sqrt{2 \epsilon_F
   }}\right)}{\pi }+\epsilon_F \;, \nonumber \\
&&\beta(N,W) = \frac{W}{96\pi} \sqrt{8 \epsilon_F-W^2} (4 \epsilon_F + W^2) + \frac{\epsilon_F^2}{\pi} \arcsin{\left( \frac{W}{\sqrt{8 \epsilon_F}}\right)}   + \frac{\epsilon_F^2}{2} \;.  \label{eq:two_eq}
\eea 
One can check that the two formulae (\ref{beta95}) and (\ref{eq:two_eq}) do coincide, as they should, for $W=\sqrt{4 N}$.

\item If $W \leq 0$, computing the integrals for $N$ and $\beta(N,W)$ with (\ref{rhoW2}) yields the same result as (\ref{eq:two_eq}), which thus holds for any $W \leq \sqrt{4N}$. The equation for $\beta(N,W)$ in (\ref{eq:two_eq}) can be slightly rearranged and written in the more compact form
\be
\label{eq:N-beta}
\beta(N, W)=\frac{N \epsilon_{F}}{2}-\frac{W}{96 \pi}\left(8 \epsilon_{F}-W^{2}\right)^{\frac{3}{2}} \;.
\ee
\end{enumerate}

\begin{figure}[t]
    \center
    \includegraphics[width=0.4\linewidth]{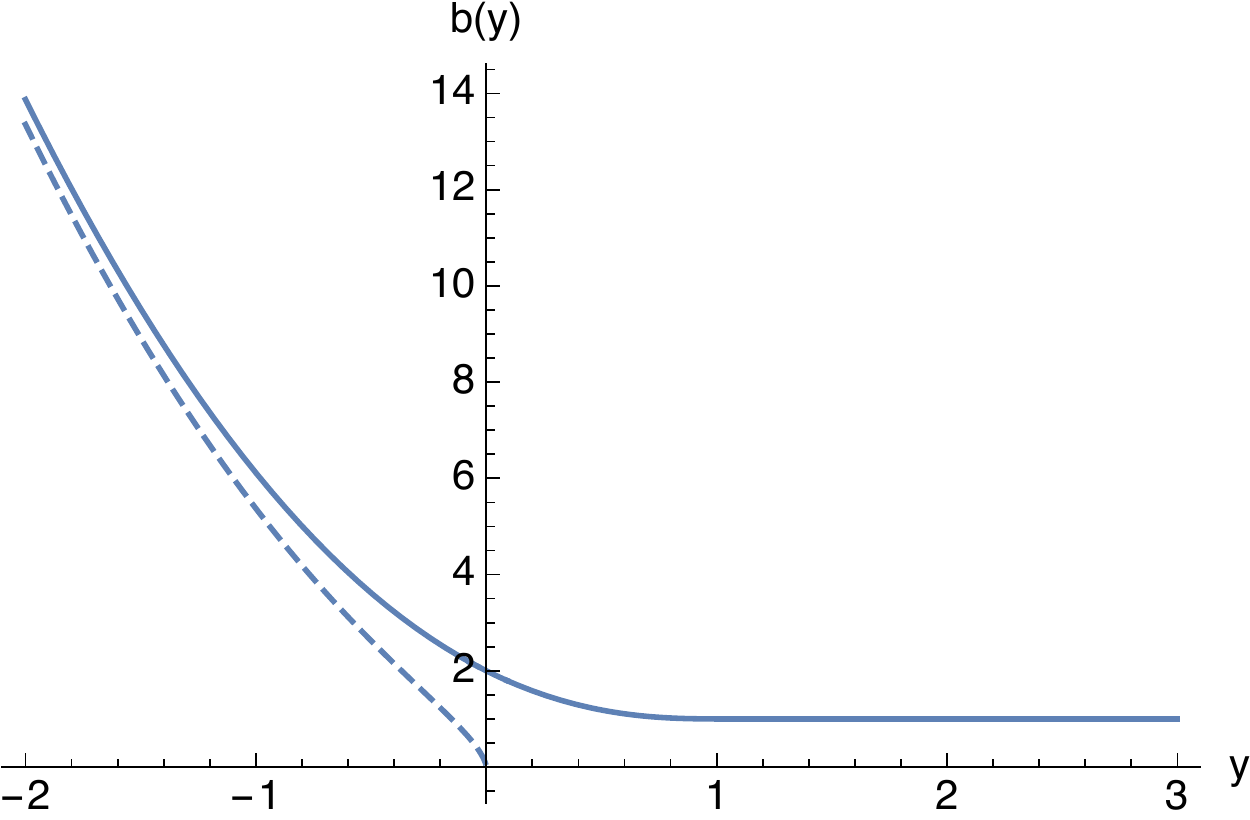}
    \caption{Plot of the scaling function ${\sf b}(y)$ versus $y$,
    which describes the scaling 
    form for the survival exponent
    $\beta(N,W) = \frac{N^2}{4} {\sf b}(y=\frac{W}{\sqrt{4 N}})$.  This function is obtained numerically by eliminating 
    the variable $u$ between Eqs. \eqref{eq1} and \eqref{eq2}.
    It is 
   compared to its asymptotic behaviour  $2 y^2 + \frac{3}{5} \sqrt[3]{2} (3 \pi )^{2/3} y^{2/3}$ for large negative
   argument. Note that for $y>1$ it is constant ${\sf b}(y)=1$, see Eq.~(\ref{bgeq1}), while close and to the left of the transition point 
   $y=1$ it behaves as ${\sf b}(y) -1 \propto (1-y)^{5/2}$, see Eq.~(\ref{5by2}).}
    \label{fig:Plotb}
\end{figure}

In fact, one can see from the two equations in (\ref{eq:two_eq}) that the survival exponent $\beta(N,W)$ takes the scaling form, in the limit 
$N,W \to +\infty$ keeping $y=\frac{W}{\sqrt{4 N}}$ fixed,
\be \label{scalb} 
\beta(N,W) \simeq \frac{N^2}{4} \, {\sf b}\left(\frac{W}{\sqrt{4 N}}\right)  \;.
\ee 
\begin{itemize}
\item
For  $y \geq 1$ the scaling function ${\sf b}(y)$ is given by
\bea \label{bgeq1}
{\sf b}(y)=1 \quad , \quad  y \geq 1 \;.
\eea

\begin{figure}[t]
    \center
    \includegraphics[width=0.4 \linewidth]{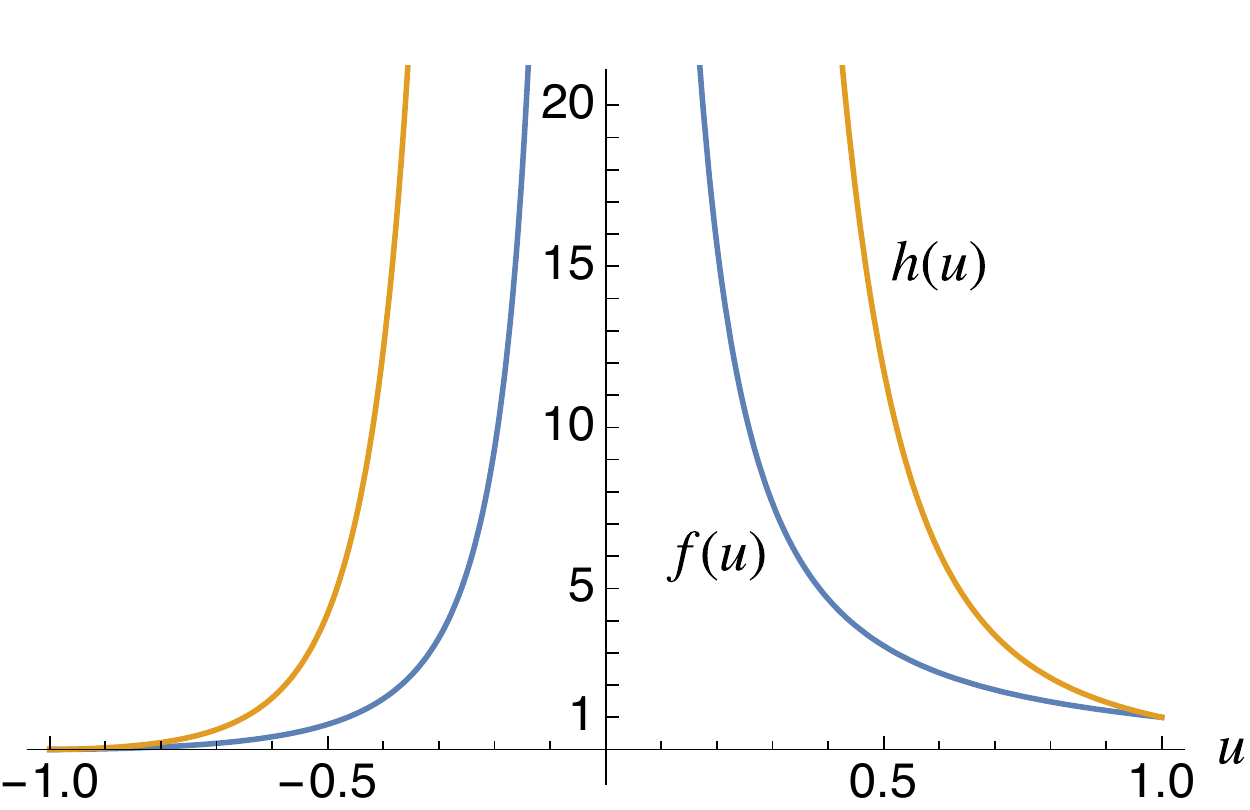}
    \caption{Plot of $f(u)$ and $h(u)$ defined respectively in Eq. (\ref{eq1}) and (\ref{eq2}) versus $u$ on $[-1,1]$.}
    \label{fig:Plotfh}
\end{figure}
\item
For $y \in (-\infty, 1]$ the function ${\sf b}(y)$ is obtained by expressing the two equations of (\ref{eq:two_eq}) in terms of $y$ and $u = \frac{W}{\sqrt{8 \epsilon_F}}$  ($-1 \leq u \leq 1$). In terms of these variables, the three equations of (\ref{eq:two_eq}) and (\ref{eq:N-beta}) give the following relations,
\bea \label{eq1} 
&&   \frac{1}{y^2} = \frac{1}{2u^2}+ \frac{\sqrt{1-u^2}}{\pi u} +  \frac{\arcsin(u)}{\pi u^2} = f(u) \;, \\
&& {\sf b}(y) = y^2 \left( \frac{1}{u^2} - \frac{2}{3 \pi} y^2  \frac{(1-u^2)^{3/2}}{u^3}  \right) = y^4 \left( \frac{1}{2 u^4} +  (1+2u^2) \frac{ \sqrt{1-u^2}}{3\pi u^3}+ \frac{ \arcsin(u)} {\pi u^4} \right) = y^4 h(u) \;, \label{eq2} 
\eea
\be 
\text{with} \quad
 \left\{
    \begin{array}{ll}
       y=\frac{W}{\sqrt{4 N}} \\ \\
       u = \frac{W}{\sqrt{8 \epsilon_F}} \;.
    \end{array}
\right.
\ee
Note that the first equation $f(u)=1/y^2$ for $y <1$ has two roots. For $0<y<1$ one
must choose the positive root denoted $u=u_+$. For $y<0$ one must instead choose 
the negative root $u=u_-$. The functions $f(u)$ and $h(u)$ are plotted in Fig. \ref{fig:Plotfh}. Eventually, 
by numerically inverting $f(u)$, we plot ${\sf b}(y)$ as a function of $y$ in Fig. \ref{fig:Plotb}.

\end{itemize}

To obtain the asymptotic behaviors of ${\sf b}(y)$ it is useful to study the ones of $f(u)$ and $h(u)$ in various limits. They are obtained straightforwardly from Eqs. (\ref{eq1}) and (\ref{eq2}) and they read
\begin{center}
\begin{tabular}{l|c|c}
  \hline
    & & \\[-1ex]
    & $f(u)$  & $h(u)$ \\ [2ex]
  \hline   
  & & \\[-1ex]
  $u \to 1^-$ & 	$1+ 2 (1- u) + \mathcal{O}((1- u)^{1/2})$ 	&		 $1+ 4 (1- u) + \mathcal{O}((1- u)^{1/2})$ \\[2ex]  
  $u \to 0$   & 		$\frac{1}{2 u^2}+\frac{2}{\pi  u}  -\frac{u}{3 \pi }+ \mathcal{O}(u^3)$		 & 	 $ \frac{1}{2u^4}  + \frac{4}{3 \pi u^3} + \frac{2}{3 \pi u}  + \mathcal{O}(u) 	$  \\ [2ex]
  $u \to -1^+$  & 	$	\frac{4 \sqrt{2} (u+1)^{3/2}}{3 \pi }+\frac{37 \sqrt{2} (u+1)^{5/2}}{15 \pi }+ \mathcal{O}((u+1)^{7/2}) 	$	&		 $	\frac{8 \sqrt{2} (u+1)^{3/2}}{3 \pi }+\frac{122 \sqrt{2} (u+1)^{5/2}}{15 \pi }+ \mathcal{O}((u+1)^{7/2}) 	$ \\[2ex]
  \hline
\end{tabular}
\end{center}

For $y \to 1^-$ we see that $u \to 1^-$, hence the formulas for ${\sf b}(y)$ with $y \leq 1$ and $y\geq 1$ match at
$y=1$. More precisely for $y \to 1$, $y<1$, ${\sf b}(y)$ behaves as
\bea \label{5by2}
{\sf b}(y)= 1+\frac{32 \sqrt{2} (1-y)^{5/2}}{15 \pi }+\frac{32  (1-y)^3}{9 \pi ^2}-\frac{8
   \left(\sqrt{2} \left(9 \pi ^2-28\right)\right)  (1-y)^{7/2}}{63 \pi
   ^3}+\mathcal{O}\left( (1-y)^4\right) \;.
\eea
All the successive derivatives of ${\sf b}(y)$ vanish for $y>1$ [see Eq. (\ref{bgeq1})]: therefore we see from (\ref{5by2}) that there is a $\frac{5}{2}$th order non-analyticity at $y=1$, where the wall starts to be felt by the system.

When $y \to 0$, $u \to 0$ as $u \simeq y/\sqrt{2}$ and ${\sf b}(y) \to 2$, which is the correct
result for $W=0$ [see Eq. (\ref{eq:BetaW0})].  One finds near $y=0$, for $y>0$ 
\bea 
{\sf b}(y) = 2-\frac{16 \sqrt{2} y}{3 \pi }+\frac{16 y^2}{\pi
   ^2}+\frac{4 \sqrt{2} \left(\pi ^2-12\right) y^3}{3
   \pi ^3}-\frac{8 \left(\pi ^2-8\right) y^4}{3 \pi
   ^4}-\frac{\left(240-40 \pi ^2+3 \pi ^4\right)
   y^5}{15 \left(\sqrt{2} \pi ^5\right)}-\frac{4
   y^6}{45 \pi ^2} + \mathcal{O}(y^7) \label{small_w}
\eea  
which leads for $W>0$, to
\be
\beta(N,W) \simeq \frac{N^2}{2}  -   \frac{2 \sqrt{2} }{3\pi} W N^{ \frac{3}{2}  } +\frac{1}{\pi^2}  W^2  N + \frac{(\pi^2 -12)  }{12 \sqrt{2} \pi^3}  W^3 \sqrt{N} + \mathcal{O}(W^4) \;.
\ee
At order $W^2 N$, this series agrees with the large $N$ expansion
of the Taylor coefficients $a_1(N)$ and $a_2(N)$ of the series in $W$ at fixed $N$ given in (\ref{a1}) and (\ref{a2}).

Finally, when $y \to - \infty$, $u \to - 1$ and one finds
\be 
{\sf b}(y) =  2 y^2 + \frac{6}{5} \left( \frac{3 \pi}{2}  \right) ^{ \frac{2}{3}} y^{\frac{2}{3}} + \frac{8}{35} \left(\frac{ 3 \pi^4}{2}\right)^{\frac{1}{3}}  \frac{1}{y^{\frac{2}{3}}}+ \frac{9 \pi^2}{700} \frac{1}{y^2} + \mathcal{O} (y^{-\frac{8}{3}}) \;.
\ee
Keeping the first two terms we thus find that, in the limit $W/\sqrt{4 N} \to - \infty$, the exponent
$\beta(N,W)$ in \eqref{scalb} is given by
\be \label{largeW}
\beta(N,W) \simeq \frac{N W^2}{8}  + \frac{3}{10} \left( \frac{3 \pi}{4}  \right) ^{ \frac{2}{3}}  N^{5/3} W^{2/3} + \mathcal{O} ( \frac{N^{7/3}}{W^{2/3}}  ) \;.
\ee 
It is interesting to compare this result with the exact asymptotics for $W \to - \infty$ at fixed $N$ obtained
in \eqref{form2}. Using the standard asymptotic results for the zeroes of the Airy function, i.e. 
$\alpha_k \simeq (3 \pi/2)^{2/3} k^{2/3}$
for $k \gg 1$, and performing the sum for large $N$, we obtain exactly the result given in \eqref{largeW}. 
The two expansions thus match smoothly for the first two leading terms. 

\bigskip

\subsection{Propagator at large time and survival amplitude}

\subsubsection{Propagator} 

Let us now give the form of the barrier constrained propagator in the large time limit.
Let us consider the expression given in Eq.~(\ref{prop3}) for the propagator for the $N$-particle OU process.
We consider the limit of large $t$, with $X_i = O(1)$ and $X_{0i}= O(1)$ (at most). Then
we can keep only the ground state contribution in \eqref{GNW} leading to
\be
G_N(\vec{X}, T \mid \vec{X_0}, T_0 ; \ W) \simeq  \Psi_{\vec{k}_0} (\vec{X},W)  {\Psi_{\vec{k}_0} } ^* (\vec{X_0},W) e^{- (T- T_0)  E_{\vec{k}_0} (N,W) }  \;.
\ee 
Inserting this form in (\ref{prop3}) and using the explicit form of the ground state wave function 
in term of Slater determinants of single particle wave functions discussed in 
\eqref{eigen} with single particle energies $(\epsilon_k(W))_{k\geq 0}$, we obtain
the large time behavior of the barrier constrained OU propagator for general $W$
\bea \label{general} 
\!\!\! P_N^{OU} (\vec{X}, T \mid \vec{X_0}, T_0 ; W ) \simeq  && C^2  \ e^{-\frac{1}{4} \sum\limits_{i=1}^{N}  X_i^2    }     \det_{1 \leq i,j \leq N}(D_{2 \epsilon_{i-1}(W) }(-X_j)) \nonumber \\
&&\times \det_{1 \leq i,j \leq N}(D_{2 \epsilon_{i-1}(W) }^*(-X_{0j}))  \ e^{\frac{1}{4} \sum\limits_{i=1}^{N} X_{0i}^2   }  \ \  e^{- (T- T_0)  E_{\vec{k}_0} (N,W) }   \;,
\eea
where $C=\prod_{i=1}^N C_{\epsilon_{i-1}(W)}$ (we recall that the normalisations $C_\epsilon$'s are defined in Eq. (\ref{eigen})). Using the Lamperti mapping we obtain the 
barrier constrained propagator for the $N$ Brownian motions under the $W\sqrt{\tau}$ absorbing boundary, through (\ref{eq:LinkPBrOUNW})

\be \label{general} 
P_N^{ B r } ( \vec { x } , t | \vec { x _ { 0 } } , t _ { 0 } ; \  W\sqrt{t} ) \simeq  \frac{C^2}{t^{N/2} }  \ e^{-\frac{1}{4 t} \sum\limits_{i=1}^{N}  x_i^2    }     \det_{1 \leq i,j \leq N}(D_{2\epsilon_{i-1}(W)}(-\frac{x_j}{\sqrt{t}}  )) \det_{1 \leq i,j \leq N}(D_{2 \epsilon_{i-1}(W)}^*(- \frac{x_{0j} }{\sqrt{t_0}}   )) \ e^{\frac{1}{4t_0} \sum\limits_{i=1}^{N} x_{0i}^2   }  \ \  \left(\frac{t_0}{t}\right)^{\beta(N,W) } 
\ee
where the additional factor of $t^{ - N/2}$ arises from the Jacobian of the transformation from $X_i$ to $x_i/\sqrt{t}$, and we have used \eqref{eq:BetaE}. We now analyze this expression in the two limiting cases $W \to +\infty$ and $W \to 0$. 

\begin{itemize}

\item $W \to +\infty$: In that limit, the eigenfunctions are those of the free harmonic oscillator. The 
Slater determinant can be expressed as
\be \label{107}
\det_{1 \leq i,j \leq N}(\phi_i(X_j)) =  \det_{1 \leq i,j \leq N} \left( c_{i-1} \ e^{-\frac{X_j^2}{4}} H_{i-1}(\frac{X_j}{\sqrt{2}}) \right)  =  \left( \prod_{0 \leq k \leq N-1} c_k \right) \  \ e^{-\frac{1}{4} \sum\limits_{i=1}^{N}  X_i^2    }  \det_{1 \leq i,j \leq N} \left(  H_{i-1}(\frac{X_j}{\sqrt{2}}) \right) \;,
\ee
where the normalization constant $c_k$ is given by (for details see the Appendix \ref{app:HarmonicOsc})
\be \label{ck}
\prod_{0 \leq k \leq N-1} c_k = \prod_{0 \leq k \leq N-1}  \frac{1}{ \sqrt{ \sqrt{2 \pi } 2^{k} k!}} \;.
\ee 
Recognizing that the determinant in Eq. (\ref{107}) is a Vandermonde determinant, as explained in Appendix \ref{app:Hermite}, one obtains
\be 
\det_{1 \leq i,j \leq N}(\phi_i(X_j)) =  \left( \prod_{0 \leq k \leq N-1} c_k  \right) \  \ e^{-\frac{1}{4} \sum\limits_{i=1}^{N}  X_i^2    }  2^{  \frac{N(N-1)}{4}}  \ \prod\limits_{1\leq i < j \leq N}  (X_j - X_i)  \;.
\ee
From the formula in (\ref{ck}), the product of the $c_k$'s can be written in terms of the Barnes' G function   
\be \label{barnes} 
\prod_{0 \leq k \leq N-1} c_k = \prod_{0 \leq k \leq N-1}  \frac{1}{ \sqrt{ \sqrt{2 \pi } 2^{k} k!}} = \frac{2^{-\frac{N^2}{4}} \pi^{-\frac{N}{4}}  }{ \sqrt{G(N+1) }} \quad , \quad G(N) = \prod_{i=0}^{N-2} \Gamma(i+1) \;.
\ee
Thus the constrained OU propagator with $W=+\infty$ reads at large $T$
\be
 P_N^{OU} (\vec{X}, T \mid \vec{X_0}, T_0 ; \ + \infty ) \simeq  \frac{1    }{  (2\pi) ^{\frac{N}{2}}  G(N+1)} \ e^{-\frac{1}{2} \sum\limits_{i=1}^{N}  X_i^2    } \  \prod\limits_{1\leq i < j \leq N}  (X_j - X_i) (X_{0j}-X_{0i})  \ \  e^{- (T- T_0) E_{\vec{k}_0} (N, \infty)    } \;.
\ee
Finally, the constrained propagator of the $N$ Brownian motions reads, using $E_{\vec{k}_0} (N,W \to+ \infty)  = \frac{N(N-1)}{4}$, 

\be
\label{eq:DensityWinf}
P_N^{ B r } ( \vec { x } , t | \vec { x _ { 0 } } , t _ { 0 } ; \ + \infty) \simeq \frac{1    }{  (2\pi) ^{\frac{N}{2}}  G(N+1)} \ e^{-\frac{1}{2t} \sum\limits_{i=1}^{N}  x_i^2    } \   \prod\limits_{1\leq i < j \leq N}  (x_j - x_i) (x_{0j}-x_{0i}) \quad t^{-\frac{N^2}{2}}  \;.   
\ee
This is simply the probability for $N$ vicious Brownian motions to end in $\vec{x}$ at time $t$, starting from $\vec{x}_0$ at time $t_0$. As shown in Appendix \ref{app:kmg}, this result can also be obtained straightforwardly from the Karlin-McGregor formula~\cite{KarMcGreg1959}. Note that this joint PDF (\ref{eq:DensityWinf}) of the $x_i$'s corresponds, up to a prefactor, to the joint PDF of the eigenvalues of the Gaussian Orthogonal Ensemble (GOE) \cite{mehta,forrester}.

\item $W=0$: In this limit, as already discussed above, the single particle 
eigenfunctions are the odd eigenfunctions of the free harmonic oscillator. Hence
we obtain the following expressions for the Slater determinants
\bea 
\det_{1 \leq i,j \leq N}(\phi_i(X_j,0)) &=&  \det_{1 \leq i,j \leq N} \left( \sqrt{2} c_{2i-1} \ e^{-\frac{X_j^2}{4}} H_{2i-1}(\frac{X_j}{\sqrt{2}}) \right) \;, \\
&=&  2^\frac{N}{2}  \left( \prod_{0 \leq k \leq N-1} c_{2k+1}  \right) \  \ e^{-\frac{1}{4} \sum\limits_{i=1}^{N}  X_i^2    }  \det_{1 \leq i,j \leq N} \left(  H_{2i-1}(\frac{X_j}{\sqrt{2}}) \right) \;.
\eea
The determinant can be evaluated and this leads to
\be 
\det_{1 \leq i,j \leq N}(\phi_i(X_j,0)) =   2^\frac{N}{2}  \left( \prod_{0 \leq k \leq N-1} c_{2k+1}  \right) \  \ e^{-\frac{1}{4} \sum\limits_{i=1}^{N}  X_i^2    }  \ 2^\frac{N^2}{2} \prod_{1 \leq i \leq N} X_i   \ \prod\limits_{1\leq i < j \leq N}  (X_j^2 - X_i^2) \;.
\ee
Thus the barrier constrained OU propagator in the case $W=0$ reads,  with $E_{{\vec k}_0}(N,W=0)
=\beta(N,W=0)= \frac{N^2}{2}$,
\be
 P_N^{OU} (\vec{X}, T \mid \vec{X_0}, T_0 ; \ 0 ) \simeq  
 D_N 
 \ e^{-\frac{1}{2} \sum\limits_{i=1}^{N}  X_i^2    } \ \prod_{1 \leq i \leq N} X_i X_{0i} \ \prod\limits_{1\leq i < j \leq N}  (X_j^2 - X_i^2) (X_{0j}^2-X_{0i}^2)  \ \  e^{- (T- T_0) \beta(N,0)  } \;. 
\ee
Thus the barrier constrained Brownian propagator in the case $W=0$ reads, 
\be \label{116} 
P_N^{ B r } ( \vec { x } , t | \vec { x _ { 0 } } , t _ { 0 } ;  0) \simeq  D_N
\ e^{-\frac{1}{2t} \sum\limits_{i=1}^{N}  x_i^2    } \  \prod_{1 \leq i \leq N} x_i x_{0i}  \prod\limits_{1\leq i < j \leq N}  (x_j^2 - x_i^2) (x_{0j}^2-x_{0i}^2)  \  \ t^{-N(N + \frac{1}{2})} \;.
\ee
Here the amplitude $D_N$ is defined as
\be 
D_N = 2^{N^2+N} \left( \prod_{0 \leq k \leq N-1} c_{2k+1}  \right)^2 = \frac{ \pi^\frac{1}{4} e^\frac{1}{8} 2^{ \frac{1}{24} -N(N-\frac{1}{2})} }{A^\frac{3}{2}  G(N+1) G(N+ \frac{3}{2})} \;,
\ee
where $c_k$ is given in Eq. \eqref{ck}, and the product can be expressed 
in terms of the Glaisher constant  $A=e^{\frac{1}{12} - \zeta'(-1)}$. Note that this joint PDF in (\ref{116}) is identical to the one
for the eigenvalues $\lambda_i$ of the Laguerre Orthogonal Ensemble 
(i.e. Wishart matrices with $\beta=1$) 
with the correspondence $\lambda_i = x_i^2$. As a consequence the 
density in the bulk is given by a half semi-circle. Note that similar results were obtained in
\cite{Katori2003}.

\end{itemize} 

\subsubsection{Survival amplitude}

In this section we calculate the survival amplitude  $A(\vec x_0,t_0;W)$ defined in \eqref{eq:Slargetime}
and given in \eqref{AA}. In general it can be expressed as a Pfaffian of a matrix, which in the case $W\to+\infty$ and $W=0$ takes an explicit form which we compute here. From Eq. \eqref{AA}, using the Slater determinant form one obtains 
\be
\label{eq:SFullExp}
A(\vec x_0,t_0;W) =  t_0^{ E^N_{\vec{k_0}}  }   \left( \int\limits_{\substack{X_1<\cdots<X_N <W }}     \det_{1 \leq i,j \leq N} \left( \phi_{i-1} (X_j,W) e^{-\frac{X_j^2}{4} }  \right)     \ {d}\vec{X}   \right) \det_{1 \leq i,j \leq N}  \left( \phi_{i-1} ^* ({X_0}_j,W)  e^{\frac{{X_0}_j^2}{4}  } \right) \;.
\ee
Such a multiple integral involving a single determinant was computed by de Bruijn in \cite{Bruijn}, in terms of the Pfaffian of an $N\times N$ matrix. Assuming $N$ even (if $N$ is odd, this can also be written as a Pfaffian, with the subtlety that the last column and row are different from the general term), de Bruijn's formula gives 
\bea 
\label{eq:PfaffianTerm}
&&\int\limits_{X_1<\cdots<X_N<W }     \det_{1 \leq i,j \leq N} \left( \phi_{i-1} (X_j,W) e^{-\frac{X_j^2}{4} }  \right)     \ {d}\vec{X}   \nonumber \\
&&=   \underset{1 \leq i,j \leq N}{\mathrm{Pf}} \left( \int\limits_{\substack{X<W \\ \tilde{X}<W }}  \phi_{i-1} (X,W) \phi_{j-1} (\tilde{X},W) e^{-\frac{X^2+ \tilde{X}^2}{4} } \mathrm{sgn}(X-\tilde{X})  {d}X {d}\tilde{X}         \right) \;.
\eea 
This can be computed explicitly for the special limiting cases $W \to + \infty$ and $W = 0$ (see Appendix \ref{app:PrefactorPfaffian} for details):
\begin{itemize}
\item In the case $W\to \infty$ (for $N$ even) one finds
\bea \label{ampli_Winf}
A(\vec x_0,t_0;W \to \infty) =  t_0^{ \frac{N(N-1)}{4} }   \underset{   \substack{0 \leq i,j \leq N-1 \\ i+j \notin 2\mathbb{N}} }{\mathrm{Pf}} \left(    \frac{(-1)^{i + \frac{i+j-3}{2}}  }{2^{i+j-3/2}}   \frac{ (i+j-1)!   }{ \sqrt{i! j! } (\frac{i+j-1}{2})!  }           \right)    \det_{1 \leq i,j \leq N}  \left( \phi_{i-1} ^* ({X_0}_j)  e^{\frac{{X_0}_j^2}{4}  } \right) \;,
\eea
where the Pfaffian is taken over a matrix $\left( A_{i,j}^\infty \right)_{i,j}$ with $A_{i,j}^\infty =0$ if $i+j$ is even.\\
The Pfaffian term can be evaluated explicitly for the first few values of $N$, reading
\begin{center}
\begin{tabular}{c|c|c|c|c|c}
  \hline
   	& & & & &   \\[-1ex]
    $N$ & 2 & 4 & 6  & 8 & 10    \\ [2ex]
  \hline   
  & & & & &   \\[-1ex]
    ${ \mathrm{Pf}} \left( A_{i,j}^\infty \right)$ & $\sqrt{2}$  & $ \frac{1}{2 \sqrt{3}}$  & $\frac{1}{16 \sqrt{30}}$ & $\frac{1}{1024 \sqrt{105}}$ &  $\frac{1}{393216 \sqrt{210}}$  \\[2ex]
  \hline
\end{tabular}
\end{center}
This term is plotted logarithmically in Figure \ref{fig:AmpNum} for even values of $N$ up to 34.
\item In the case $W\to 0$, one obtains
\bea 
A(\vec x_0,t_0;W\to 0) = t_0^{ \frac{N^2}{2} }  \underset{ 0 \leq i,j \leq N-1 }{ \mathrm{Pf}} \left( A_{i,j}^0 \right)    \det_{1 \leq i,j \leq N}  \left( \phi_{i-1} ^* ({X_0}_j)  e^{\frac{{X_0}_j^2}{4}  } \right)
\label{Ampli_W0}
\eea
where (for $N$ even), without conditions on the parity of $i+j$
\begin{eqnarray}\label{explicit_A0}
&  A_{i,j}^0 = \frac{(-1)^{i+j+1}}{\sqrt{ \frac{\pi}{2} (2i+1)! (2j+1)!  }} \bigg[    
(\frac{1}{4})^{i+j} \frac{(2i+2j+1)!}{(i+j)!} {}_2F_1 \big(  \frac{1}{2},-i-j;\frac{3}{2}, -1  \big)   - (2i-1)!!(2j-1)!!  \nonumber \\
& - (2)^{i+j+1} \bigg(  \frac{2 \Gamma(\frac{3}{2}+i+j ) }{\sqrt{\pi}}  {}_2F_1 \big( 1,\frac{3}{2}+i+j;\frac{3}{2}, -1 \big)   
 + (-1)^{1+i} \frac{\Gamma(\frac{3}{2}+j)  }{\Gamma(\frac{3}{2}-i) }   {}_2F_1 \big( 1,\frac{3}{2}+j;\frac{3}{2}-i , -1 \big) \bigg)  \bigg] \;,
\end{eqnarray} 
where ${}_2F_1(a,b; c, z)$ is the standard hypergeometric function. This Pfaffian in (\ref{Ampli_W0}) can be evaluated explicitly for the first few values of $N$. In the case $N=2$, this yields
\be 
(A_{i,j}^0) = 
\begin{pmatrix} 
0 & \frac{1}{\sqrt{3 \pi}} \\
 - \frac{1}{\sqrt{3 \pi}}  & 0 
\end{pmatrix}
\quad 
\quad
\quad \;, \;\quad \quad
{ \mathrm{Pf}} \left( A_{i,j}^0 \right)    = \frac{1}{\sqrt{3 \pi}}  \;,
\ee
while for $N=4$ it reads 
\be 
(A_{i,j}^0) = 
\begin{pmatrix} 
0 & \frac{1}{\sqrt{3 \pi}} &- \frac{2}{\sqrt{15 \pi}} & \frac{\sqrt{7}}{2 \sqrt{10 \pi}} \\
 - \frac{1}{\sqrt{3 \pi}}  & 0  & \frac{1}{2\sqrt{10 \pi }} & -\frac{2}{\sqrt{105 \pi}}  \\
\frac{2}{\sqrt{15 \pi}}  & -\frac{1}{2\sqrt{10 \pi }}  & 0 & \frac{\sqrt{3}}{ 8 \sqrt{7 \pi }} \\
 -\frac{\sqrt{7}}{2 \sqrt{10 \pi}} & \frac{2}{\sqrt{105 \pi}}  &  -  \frac{\sqrt{3}}{ 8 \sqrt{7 \pi }} & 0
\end{pmatrix}
\quad 
\quad
\quad \;, \;\quad \quad
{ \mathrm{Pf}} \left( A_{i,j}^0 \right)    =
\frac{1}{30 \sqrt{7} \pi } \;.
\ee
The following (even) values of $N$ give
\begin{center}
\begin{tabular}{c|c|c|c|c|c}
  \hline
   	& & & & &   \\[-1ex]
    $N$ & 2 & 4 & 6  & 8 & 10    \\ [2ex]
  \hline   
  & & & & &   \\[-1ex]
  ${ \mathrm{Pf}} \left( A_{i,j}^0 \right)  $ & $\frac{1}{\sqrt{3 \pi}}$  & $\frac{1}{30 \sqrt{7} \pi }$ & 	$\frac{1}{3780 \sqrt{385} \pi ^{3/2}}$ 	&	$\frac{1}{227026800 \sqrt{33} \pi ^2}$   & $\frac{1}{16557064524000 \sqrt{627} \pi ^{5/2}}$     \\[2ex]
  \hline
\end{tabular}
\end{center}
This term is plotted logarithmically in Figure \ref{fig:AmpNum} for even values of $N$ up to 34.

\begin{figure}[t]
    \center
    \includegraphics[width=0.85\linewidth]{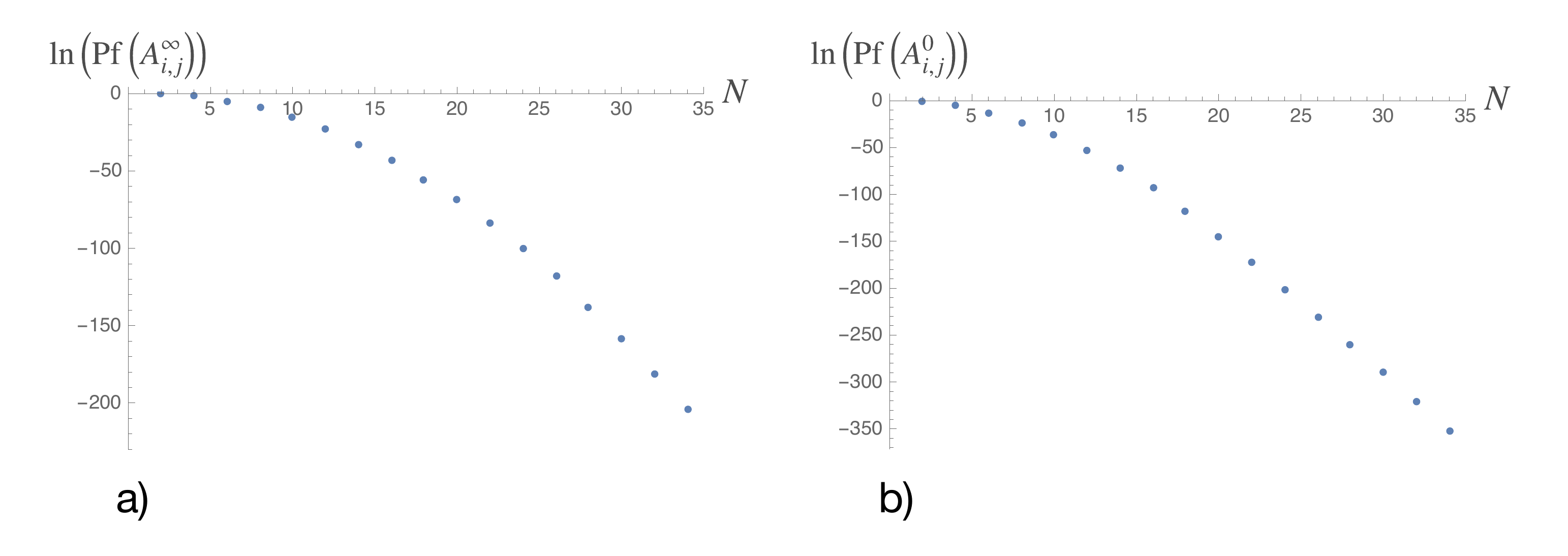}
    \caption{Logarithmic plots of the Pfaffian term are presented for even values of $N$, in the limits of a) $W\to \infty$ and b) $W \to 0$}
    \label{fig:AmpNum}
\end{figure}

\end{itemize} 
Of course, evaluating these Pfaffians explicitly for arbitrary value of $N$ remains a challenging task.

\section{Dyson Brownian Motion} 
\label{sec:DBM} 

\subsection{Dyson Brownian motion with a moving boundary} 

We now apply the above results to the Dyson Brownian motion (DBM). We investigate the
probability that the DBM remains below the barrier $W \sqrt{\tau}$. The
DBM \cite{Dys62}, with Dyson index $\beta$, is the process describing the evolution of the $N$ eigenvalues $(x_1(t), \cdots, x_N(t))$ of a matrix whose entries evolve as Brownian motions with the appropriate symmetry constraints: $N\times N$ real symmetric for $\beta =1$, $N\times N$ complex hermitian for $\beta = 2$, $2N\times 2N$ complex self-dual for $\beta = 4 $. The DBM evolves according to the Langevin equations~\cite{Tao2012}
\bea  \label{Langevin}
\frac{d x_i(t)}{dt} = \frac{\beta}{2}  \sum_{j\neq i} \frac{1}{x_i(t) - x_j(t)	} + \xi_i(t) \;, \; \;\; \mathrm{for} \;\;\;\; 1 \leq i \leq N  \;,
\eea    
where $\xi_i(t)$ are independent Gaussian white noises with zero mean and correlators
$\overline{\xi_i(t) \xi_j(t')} =  \delta_{ij}\,\delta(t-t')$. 
For $\beta =2$ one can show \cite{Tao2012} that 
 this process is equivalent to $N$ independent Brownian motions conditioned not to cross each other for all 
 times $t \in [0, \infty)$. Since we are interested in non-crossing Brownian 
 motions for finite time windows, 
 we will use instead a relation between the propagator ${ P }_N^{ \mathrm {DBM } }$ of the DBM for $\beta = 2$, and 
 the constrained propagator ${ P }_N^{Br }$ for the $N$ Brownian motions,
 for initial conditions $x_{0i}$ at time $t_0$. This relation reads \cite{SchehrRambeau} 
 \be
  { P }_N^{ {DBM } } ( \vec { x } , t | \vec { x }_0 , t_0 ) = \frac { \prod _ { i < j } \left( x _ { j } - x_ { i } \right) } { \prod _ { i < j } \left( x_ { 0j } - x_ { 0i } \right) } \   { P }^{Br}_N ( \vec { x } , t | \vec {x }_0 , t_0 ) \;.
\ee
One can actually show that this relation can be extended to the barrier constrained propagators 
see Appendix \ref{app:DBM-Noncrossing}
\be \label{relGreg2} 
P_N^{ DBM } \left( \vec { x } , t | \vec { x } _ { 0 } , t _ { 0 } ;  W\sqrt{t} \right)  
=   \frac { \prod _ { i < j } \left( x _ { j } - x _ { i } \right) } { \prod _ { i < j } \left( x _ {0 j } - x _ { 0i } \right) }     P_N^{ Br } \left( \vec { x } , t | \vec { x } _ { 0 } , t _ { 0 } ;  W\sqrt{t} \right) \;.
\ee

We can now use the result for the previous Section, in Eq. (\ref{general}), for the large time limit of 
$P_N^{ Br }$ to obtain, for $t \gg 1$,
\bea \label{largetDBM} 
&& P_N^ { DBM } \left( \vec { x } , t | \vec { x } _ { 0 } , t _ { 0 } ;  W\sqrt{t} \right)  \\
&& \simeq   \frac { \prod _ { i < j } \left( x _ { j } - x _ { i } \right) } { \prod _ { i < j } \left( x _ {0 j } - x _ { 0i } \right) }  
\ e^{-\frac{1}{4 t} \sum\limits_{i=1}^{N}  x_i^2    }     \det_{1 \leq i,j \leq N}(D_{2 \epsilon_{i-1}(W)}(-\frac{x_j}{\sqrt{t}}  )) \det_{1 \leq i,j \leq N}(D_{2 \epsilon_{i-1}(W)}^*(- \frac{x_{0j} }{\sqrt{t_0}}   )) \ e^{\frac{1}{4t_0} \sum\limits_{i=1}^{N} x_{0i}^2   }  \ \  (\frac{t_0}{t})^{\beta(N,W)} \ t^{ - N/2} \;. \nonumber
\eea\\
From this formula we can obtain the survival probability, $S_{DBM}(t | \vec x_0,t_0;W)$, i.e. the probability that the DBM remains below the barrier $W \sqrt{t}$ up to time $t$
\bea
\label{eq:GtoS}
&& S_{DBM}(t | \vec x_0,t_0;W)  =  \int_{x_1<x_2<\dots<x_N<W \sqrt{t}} \, {d} \vec { x }
\, \, P_N^{DBM} ( \vec { x } , t | \vec { x _ { 0 } } , t _ { 0 } ; \  W\sqrt{t} )  \;.
\eea
One finds that its decay at large time is given by the exponent $\beta_c(N,W)$, i.e.,  
\be
S_{DBM}(t | \vec x_0,t_0;W) \sim t^{- \beta_c(N,W)} \;,
\ee 
where we recall from \eqref{betabetac}  that 
$
\beta_c(N,W)= \beta(N,W) - \frac{N(N-1)}{4} \;.
$

We first study the limit $W \to +\infty$ where the DBM's are unconstrained, i.e. without a barrier.
Inserting the large time limit
\eqref{eq:DensityWinf} of the constrained Brownian propagator directly into \eqref{relGreg2} 
we obtain
\be
P_N^{ DBM } \left( \vec { x } , t | \vec { x } _ { 0 } , t _ { 0 }  \right)  \simeq 
\frac{1    }{  (2\pi) ^{\frac{N}{2}}  G(N+1)} \ e^{-\frac{1}{2t} \sum\limits_{i=1}^{N}  x_i^2    } \   \prod\limits_{1\leq i < j \leq N}  (x_j - x_i)^2 \quad t^{-\frac{N^2}{2}} \;,     
\ee
which is independent of the initial condition. Here $G(N)$ is the Barnes function defined
in \eqref{barnes}. This is the classical result for the large time
limit of the DBM in the absence of a wall, i.e. the PDF of the eigenvalues of the
Gaussian Unitary Ensemble (GUE) -- note that this is at variance with the case of non-intersecting Brownian motions which correspond to GOE (\ref{eq:DensityWinf}). Note also the difference in normalization
since $P_N^{ DBM } \left( \vec { x } , t | \vec { x } _ { 0 } , t _ { 0 }  \right)$ is normalized
to unity over the sector $x_1<\dots<x_N$. 

\subsection{Dyson Brownian motion with a boundary at $W=0$} 

\subsubsection{Propagator} 

We now discuss the result for the large time limit \eqref{largetDBM} of the constrained propagator of the DBM 
for a fixed barrier at $W=0$, i.e. the DBM's in the half-space. For $W=0$ the same mapping \eqref{relGreg2}, using the result \eqref{116} for the $N$ Brownian motions leads to the following result in the large $t$ limit (for  $\vec{x} \in \left(- \infty , 0\right]^N  $)
\be
\label{eq:JointPDBMzero}
P_N^ { DBM } \left( \vec { x } , t | \vec { x } _ { 0 } , t _ { 0 } ; 0 \right) \simeq D_N  \ e^{-\frac{1}{2t} \sum\limits_{i=1}^{N}  x_i^2    } \  \prod_{1 \leq i \leq N} \abs{x_i }   \prod _ { i < j } \left( x _ { j } - x _ { i } \right) (x_j^2 - x_i^2)   \prod_{1 \leq i \leq N}  \abs{x_{0i} }     \prod\limits_{1\leq i < j \leq N} (x_{0j} + x_{0i})  \  \ t^{-N(N + \frac{1}{2})} \;.
\ee
Note that here the quantity on the left hand side (lhs) corresponds to the probability that 
$N$ DBM remain below $W=0$ up to time $t$. It satisfies the Fokker-Planck
equation corresponding to the following Langevin process
\be
\frac{dx_i(t)}{dt}= \frac{1}{2 x_i(t)} + \sum_{j \neq i} \left( \frac{1}{x_i(t)-x_j(t)} + \frac{1}{2} \frac{1}{(x_i(t)+x_j(t))} \right) 
+ \xi_i(t) \;, \; \;\; \mathrm{for} \;\;\;\; 1 \leq i \leq N  \;,
\ee 
where $\xi_i(t)$ are again independent Gaussian white noises with zero mean and correlator
$\overline{\xi_i(t) \xi_j(t')} = \delta_{ij} \delta(t-t')$. This process is known as the
DBM of type $C$ symmetry class \cite{OConnell2005,Borodin2009,Katori2003}. Note the slight difference
in the numerical factors in Eq. (1.2) of Ref. \cite{Borodin2009} (a factor `$1$' on each interaction
as well as in the $1/x$ term) which describes Brownian
particles conditioned to never collide with each other or the wall (up to infinite time).
Here by contrast, $P_N^ { DBM } \left( \vec { x } , t | \vec { x } _ { 0 } , t _ { 0 } ; 0 \right)$
corresponds to the probability that $N$ Brownian
particles never collide with each other or the wall up to time $t$. 
This results in slightly different joint distributions.

\subsubsection{Kernel and density of constrained DBM} 

To study the density of DBM walkers which have survived until time $t$, we first 
change variables to $z_i = x_i^2 / 2 t$. The propagator reads
\be \label{PNN} 
P_N^ { DBM } \left( \vec { z } , t | \vec { x } _ { 0 } , t _ { 0 } ; 0 \right) \simeq D(t, \vec{x}_0, t_0)  \ e^{- \sum\limits_{i=1}^{N}  z_i    }   \prod _ { i < j } \left( z_ { j } ^{\frac{1}{2}} - z_ { i }^{\frac{1}{2}} \right) (z_j - z_i)  \;,
\ee
with 
\be 
 D(t, \vec{x}_0, t_0) = D_N \prod_{1 \leq i \leq N}  \abs{x_{0i} }     \prod\limits_{1\leq i < j \leq N} (x_{0j} + x_{0i})  \   
 2^{\frac{3}{4} N(N-1)}  t^{ - \beta_c(N,W=0)} \;,
\ee
where we recall that $\beta_c(N,W=0)=\frac{N(N+1)}{4}$ is the survival exponent of the DBM
for $W=0$.

\begin{figure}[t]
    \includegraphics[width=0.6\linewidth]{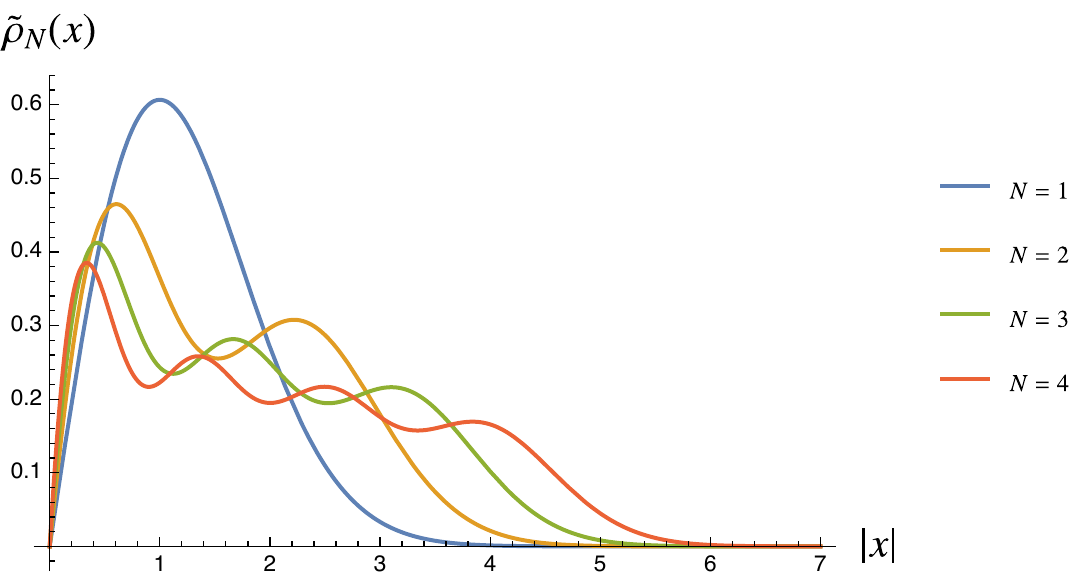}
    \caption{Density $\tilde \rho_N(x)$ of the surviving DBM particles $x_i$ for various values of $N$ and at $t=1$. }
    \label{fig:rhotilde}
\end{figure}

Joint distributions of the type \eqref{PNN} are known to belong to
the so-called biorthogonal ensembles
\cite{Muttalib,Bor1998,Tierz2007}. 
With $\vec{z} \in [ 0 , \infty )^N  $, the probability distribution \eqref{PNN} for the $z_i$ variables
coincides with the one of the biorthogonal Laguerre ensemble defined in Ref. \cite{Bor1998}. Using the 
same notations as in Ref. \cite{Bor1998}, the weight 
function is here $\omega(z) = e^{-z}$ and the parameters are $\alpha = 0$ 
and $\theta=\frac{1}{2}$. From this work \cite{Bor1998} we thus know 
that the $z_i$'s form a determinantal point process with a kernel given by [see formula (4.4) in \cite{Bor1998}]
\be \label{KN} 
K_{N}(z, z')= \frac{1}{2} \sum_{k, i=0}^{N-1} \sum_{r=i}^{N-1} \frac{\Gamma\left(N+2(i+1)\right)}{\Gamma(\frac{k}{2}+1) \Gamma\left(2 (i+1)\right)} \,  \frac{(-1)^{i+k}}{k !(N-k-1) ! i !(r-i) !} \frac{z^{\frac{k}{2}} (z')^{r}}{\frac{k}{2}+i+1} e^{-\frac{z+z'}{2}} \;.
\ee
This implies that the $n$-point correlation function is given by $\det_{1 \leq i,j \leq n} K^{}(z_i,z_j)$. 
In particular the mean density is $\rho_N(z) = \frac{1}{N} K^{}_N(z,z)$. For the first values of $N$ one finds explicitly 
\bea
&& \rho_1(z) = e^{-z} \quad \ \ ,\  \quad \ \rho_2(z) = \frac{e^{-z} \left(\sqrt{\pi } (3-2 z)+4 (z-1) \sqrt{z}\right)}{2 \sqrt{\pi }} \;,
\\
&&  \rho_3(z) = 
\frac{e^{-z} \left(\sqrt{\pi } (z (z (2 z-5)-6)+6)-8 \sqrt{z} ((z-4) z+2)\right)}{3
   \sqrt{\pi }}
\eea 
with the normalization $\int_0^{+\infty} dz \rho_N(z) =1$. From the change
of variable $\rho_N(z) dz = \tilde \rho_N(x) dx$, one also obtains the density of the $x_i$'s
\bea
\tilde \rho_N(x) = \frac{|x|}{t} \rho_N\left(\frac{x^2}{2 t}\right) \;.
\eea 
The density $\tilde \rho_N(x)$ of the surviving DBM particles is plotted in Fig. \ref{fig:rhotilde} for various values of $N = 1, 2, 3, 4$
and $t=1$. It vanishes at the origin as
\be
 \tilde \rho_N(x) = \frac{N+1}{2}  \frac{|x|}{t} + {\cal O}(x^2) \;.
\ee 

\begin{figure}[t]
\includegraphics[width = 0.4\textwidth]{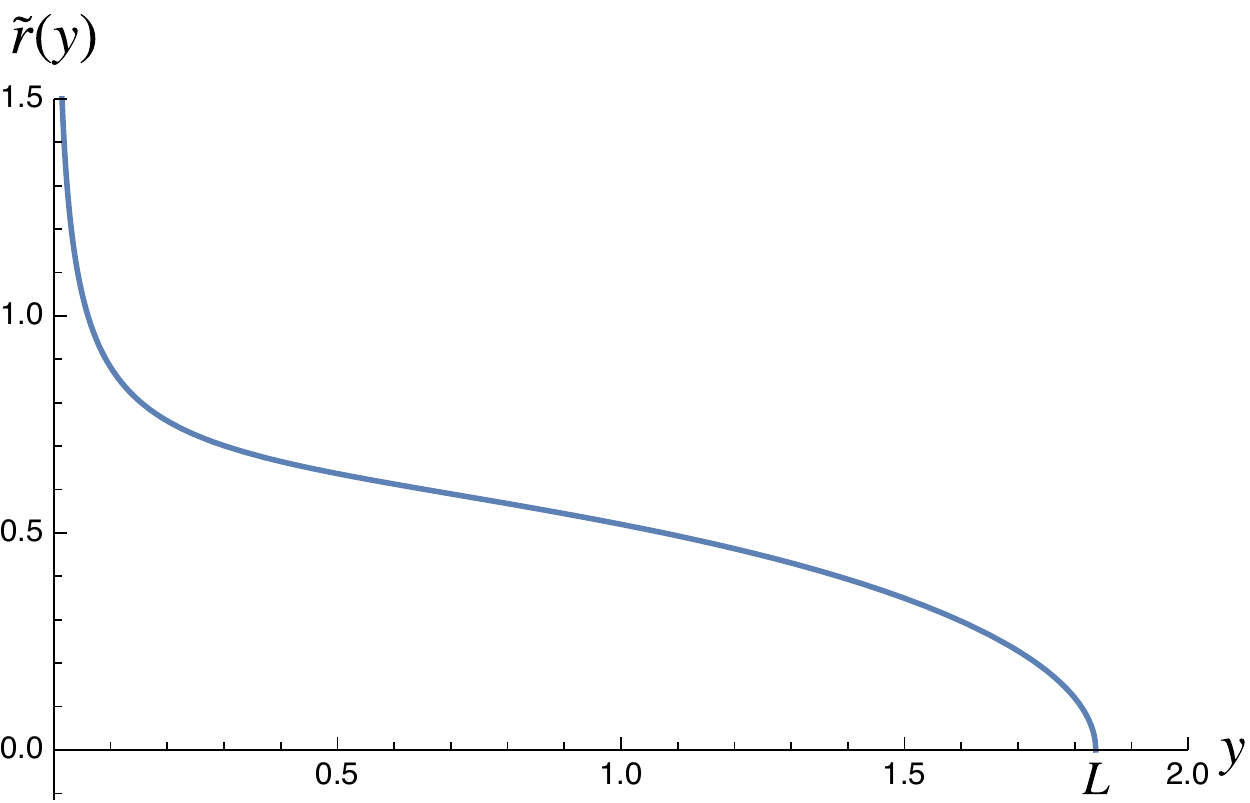}
\caption{Density of the surviving DBM particles in the bulk, in the large $N$ limit, as given in Eq. (\ref{rtilde}).}
\label{fig:plotbulk}
\end{figure}

In the limit of large $N$ we show below that the density converges to the following limiting forms.
In the bulk it takes the form
\be \label{bulk1} 
\rho_N(z) \simeq \frac{1}{N} r_{\rm bulk}\left(\frac{z}{N}\right) \quad , \quad \tilde \rho_N(x) \simeq \frac{1}{N }
\frac{|x|}{t}  \, r_{\rm bulk}\left(\frac{x^2}{2 N t}\right) \;,
\ee 
with $\int_0^{+\infty} dy \, r_{\rm bulk}(y)=1$.
There are two edge regimes. The first is a soft edge at $x_e \sim\sqrt{2 N t}$, i.e. $z_e \sim N$,
located at the upper end of the support
of the density, see Fig. \ref{fig:plotbulk}. There is another hard edge at $x=0$.
Near this hard edge, i.e. in a layer of width $z \sim \frac{1}{N^2}$, i.e. $\abs{x} \sim \frac{\sqrt{t}}{N}$ the density takes another scaling form
\be\label{edge1} 
\rho_N(z) \simeq N r_e(N^2 z) \quad , \quad \tilde \rho_N(x) \simeq N \frac{|x|}{t} r_e\left(\frac{N^2 x^2}{2 t}\right) \;,
\ee 
with $r_e(0)= \frac{1}{2}$. Below we will calculate these two scaling functions exactly.
To match the two scaling forms \eqref{bulk1} and \eqref{edge1}
(assuming that there are no additional intermediate regime) we can 
insert a power law behavior in both scaling functions
\be
r_e(\tilde z) \sim \frac{1}{\tilde z^\nu} 
\quad , \quad \tilde z \to +\infty \quad , \quad r_{\rm bulk}(y) \sim \frac{1}{y^{\nu}} \quad , \quad y \to 0
\ee 
Inserting these forms into Eqs. \eqref{bulk1} and \eqref{edge1} scaling form and imposing that the powers of $N$ match leads to the prediction $\nu = 2/3$. As we show below this is confirmed by an exact 
calculation in both regimes.

\begin{figure}[t!]
    \includegraphics[width=0.4\linewidth]{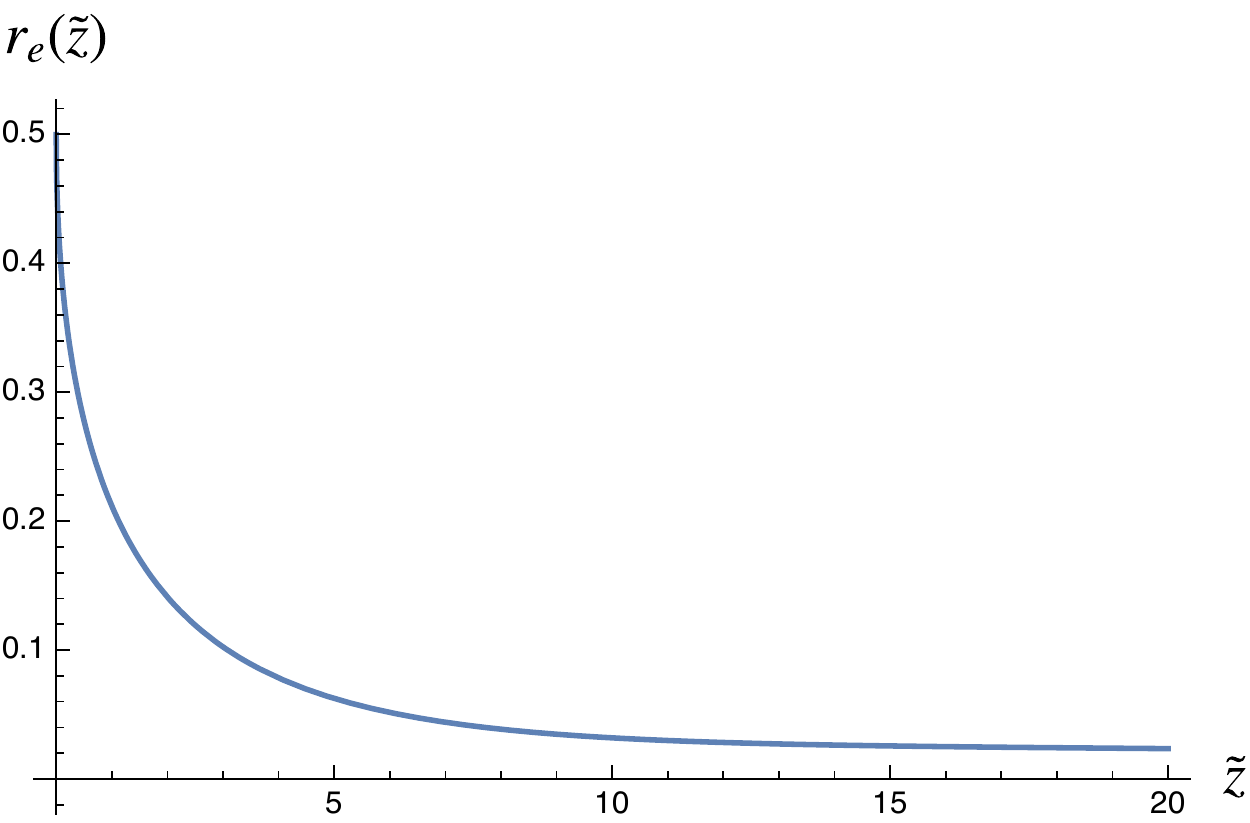}
    \includegraphics[width=0.4\linewidth]{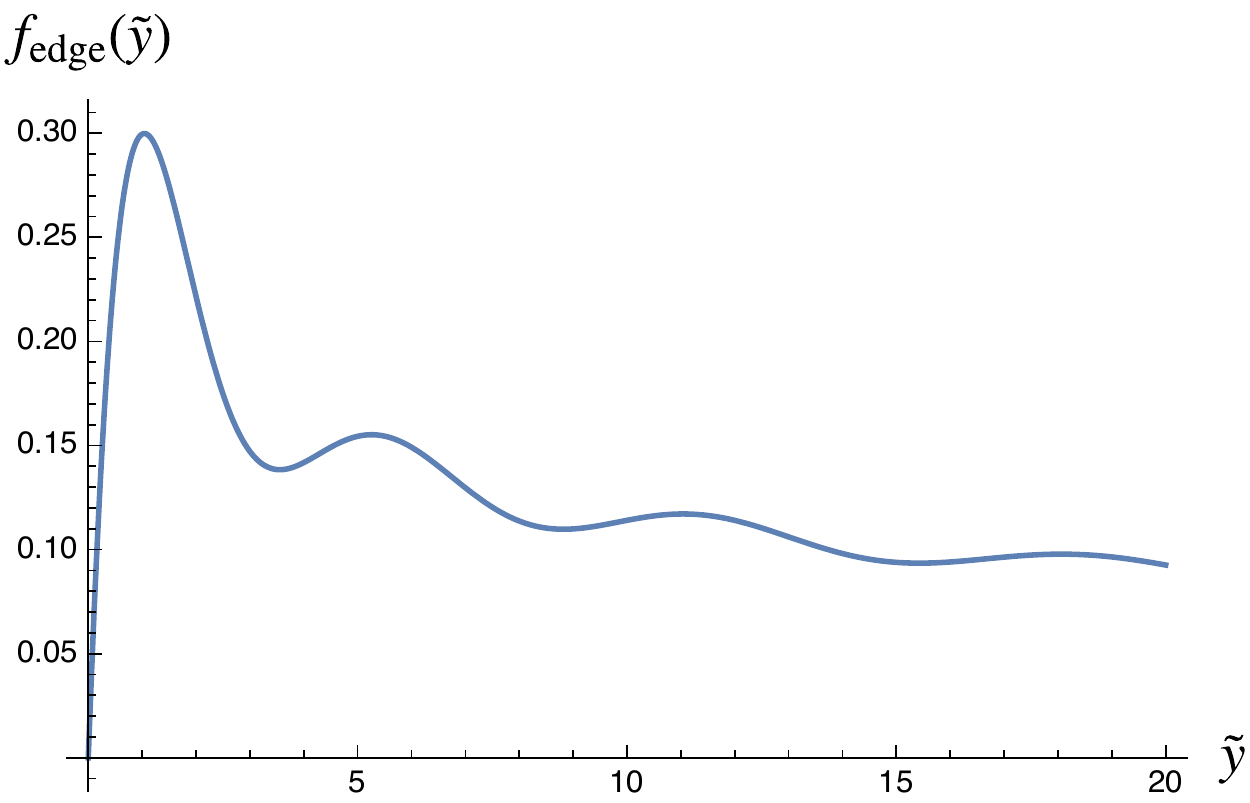}
    \caption{{\bf Left:} Plot of the scaling function $r_e (\tilde z)$ [given in Eqs. (\ref{re2})-(\ref{re1})]. {\bf Right:} Plot of the scaling function $f_{\rm edge}(\tilde y)= \sqrt{2} \tilde y \, r_e(\tilde y^2)$ which describes the hard edge scaling limit of $\tilde \rho_N(x)$, for $|x| \sim \sqrt{2t}/N$ [see the second line of Eq. (\ref{bulk_scaling})] -- at this scale, individual particles appear as oscillations in $f_{\rm edge}$, which cannot be seen in $r_e$ due to the overall decay. See also Fig. \ref{fig:rhotilde} for a plot of $\tilde \rho_N(x)$.}
    \label{fig:redge}
\end{figure}

Thus, to summarize, the density $\tilde \rho_N(x)$ of the DBM particles,
as a function of $x$ (for fixed large $N$ and $t$), has two different
behaviors (bulk and edge) depending on the scale of $x$. We find
\begin{eqnarray}
\tilde \rho_N(x)\simeq
\begin{cases}
& \frac{1}{\sqrt{2Nt}}\, {\tilde r} \left(\frac{|x|}{\sqrt{2Nt}}\right)\, ,
\quad\quad {\rm for} \quad \quad |x|\sim  \sqrt{2Nt} \, , \label{bulk_scaling}
\\
\\
& \frac{1}{\sqrt{t}}\, f_{\rm edge}\left(\frac{|x|N}{\sqrt{2t}}\right)\, ,
\quad\quad {\rm for} \quad\quad |x|\sim  \sqrt{2t}/N\, \;,
\end{cases}
\label{edge_scaling}
\end{eqnarray}
where the bulk scaling function ${\tilde r}(y)$ is related to $r_{\rm bulk}(y)$ defined in (\ref{bulk1}) by the simple relation
${\tilde r}(y)= 2 y \, r_{\rm bulk}(y^2)$.  An explicit expression of ${\tilde r}(y)$ is given in Eq. (\ref{rtilde}) and is plotted in Fig. \ref{fig:plotbulk}.
The edge scaling function $f_{\rm edge}(\tilde y)$ can be conveniently expressed as $f_{\rm edge}(\tilde y)= \sqrt{2} \tilde y \, r_e(\tilde y^2)$, where
the function $r_e(\tilde z)$ is computed explicitly in Eqs. (\ref{re2})-(\ref{re1}) and is plotted in the left panel of Fig. \ref{fig:redge} -- while the function  $f_{\rm edge}(\tilde y)$ itself
is plotted in the right panel of Fig. \ref{fig:redge}.

\medskip

{\bf Density in the bulk}

\medskip

In principle one can calculate the exact bulk density by an asymptotic large $N$ analysis of
the kernel $K_N(z,z)$ given in Eq. \eqref{KN}. However it turns out to be more convenient
to extract the bulk density using a Coulomb gas method developed for these
biorthogonal ensembles and the $O(n)$ matrix models
\cite{Sommers2003,Sommers2004,Kostov,Kostov2,Kostov3,BorotNadal2012,TheseNadal2012,ClaeysBiorthogonal}.
We find that the average density in the bulk takes the scaling form at large $N$
\be \label{bulk2} 
\tilde \rho_N(x) = \frac{1}{\sqrt{2 N t}} \, \tilde r \left( \frac{|x|}{\sqrt{2 N t} } \right)
\ee
where the scaling function $\tilde r(y)$ has a finite support $[0,L]$ with $L = \left( \frac{3}{2} \right) ^{3/2}$
and reads (see Appendix \ref{App:Bueckner} for details)

\bea \label{rtilde} 
&& \tilde r(y) =
\frac{1}{2 \pi \sqrt{2} }   \left(  g_-(y) - g_+(y) + 3 \sqrt{1 - \frac{y^2}{L^2} } \, \bigg( g_-(y) + g_+(y) \bigg) \right) \;, \\
&& g_\pm(y) = \left( \frac{L}{y} \pm \sqrt{\frac{L^2}{y^2} -1 }\right) ^{1 /3}  \nonumber \;.
\eea
This result coincides with Proposition 2.5 of \cite{KuijlaarsMolag}, where the authors also derive the bulk density for this $\theta = \frac{1}{2}$ biorthogonal ensemble, through a calculation based on a vector equilibrium problem (see Appendix \ref{App:Bueckner} for the exact mapping between the two formulas). In Fig. \ref{fig:plotbulk} we show a plot of the bulk density $\tilde r(y)$. 
Its asymptotic behaviors near the hard and soft edges read respectively 
\be 
\tilde r(y) \simeq \begin{cases} & \frac{\sqrt{3}}{2^{2/3} \pi } y^{-1/3} \quad , \quad y \to 0  \;, \label{y0} \\
& \\
& \frac{8}{3 \pi } \left( \frac{2}{3} \right)^{3/4}  \sqrt{L - y } \quad , \quad y \to L^- \;,
\end{cases}
\ee
which are consistent with the general asymptotic results obtained in \cite{ClaeysBiorthogonal} for
general biorthogonal ensembles.

\medskip

{\bf Density at the (hard) edge}

\medskip

Near the hard edge we can use the limiting kernel obtained in Ref. \cite{Bor1998}.
The determinantal point process $z_i$ 
in the variable $\tilde z = N^2 z$ is described by the following kernel for large $N$
 \be  \label{re2} 
\mathcal{K}^{\rm edge} (\tilde z, \tilde z') =
\sum_{j,k=0}^{\infty} \frac{(-1)^{k} \tilde z^{k}}{k ! \Gamma\left(2(1+k)\right)} \frac{(-1)^{j} (\tilde z')^{\frac{j}{2}}}{j ! \Gamma(1+\frac{j}{2})} \frac{1}{2(1+k)+ j} \;,
 \ee
given in Eq. (3.6) of \cite{Bor1998}. The density at the hard edge is described by the
scaling function
\be \label{re1} 
r_e(\tilde z) = \mathcal{K} ^{\rm edge}(\tilde z, \tilde z) \;,
\ee 
with $r_e(0)=\frac{1}{2}$ and is plotted in Fig. \ref{fig:redge}.

We can now show that the large $\tilde z$ limit matches the one in the bulk. To perform this asymptotic analysis, it is useful to 
write the kernel as \cite{Bor1998} 
\be
\mathcal{K} (\tilde z,\tilde z) = \frac{1}{2} \int_0^1 {d}t J_{2,2}(\tilde z t) J_{1, \frac{1}{2}} (\sqrt{\tilde z t}) \;,
\ee 
in terms of the so called Wright's generalized Bessel functions (see e.g. \cite{Wright})
\be
J_{a,b}(x) = \sum_{k=0}^{+\infty} \frac{(-x)^k}{k! \Gamma(a +b k)} \;.
\ee 
In particular, for large $x>0$, one has \cite{Wright} 
\be
J_{a,b}(x) \simeq 2 {\rm Re} \frac{1}{\sqrt{2 \pi (1+b)}} [ ( b x)^{\frac{1}{1 + b}} e^{i \frac{\pi}{1+b}} ]^{\frac{1}{2} -a} 
e^{ (1 + \frac{1}{b}) ( b x)^{\frac{1}{1 + b}} e^{i \frac{\pi}{1+b}} } \;,
\ee
which yields for the specific values of $a$ and $b$ of interest here
\be
J_{2,2}(x) \simeq 2 {\rm Re} \frac{1}{\sqrt{6 \pi}} [ ( 2 x)^{\frac{1}{3}} e^{i \frac{\pi}{3}} ]^{- \frac{3}{2}} 
e^{ \frac{3}{2} ( 2 x)^{\frac{1}{3}} e^{i \frac{\pi}{3}} } \quad , \quad 
J_{1,\frac{1}{2}}(x) \simeq 2 {\rm Re} \frac{1}{\sqrt{3 \pi}} 
[ ( \frac{1}{2} x)^{\frac{2}{3}} e^{i \frac{2 \pi}{3}} ]^{- \frac{1}{2}} 
e^{ 3 ( \frac{1}{2} x)^{\frac{2}{3}} e^{i \frac{2 \pi}{3}} } \;.
\ee
From these asymptotics we find that 
\be 
\mathcal{K} (\tilde z,\tilde z) \simeq \frac{2^\frac{1}{3}}{3 \pi \tilde z } \int_0^{\tilde z} {d}t \ t^{-\frac{2}{3}} \sin({\cal A}t^\frac{1}{3}) \cos({\cal A}t^\frac{1}{3} - \frac{\pi}{3} ) \;,
\ee
with $ {\cal A} = \frac{3\sqrt{3}}{2^\frac{5}{3}  }$, which, neglecting subdominant oscillating terms,
leads to the following decay of the edge density
\be
\mathcal{K} (\tilde z, \tilde z) \underset{\tilde z \to +\infty}{\simeq} \frac{\sqrt{3} }{ 2^\frac{5}{3} \pi } \ \frac{1}{\tilde z^{2/3}}  + \mathcal{O}\left(\frac{1}{\tilde z}\right) \;.
\ee
Using that 
$\tilde z = \frac{N^2 x^2}{2 t}$ we find that the density at the edge decays for $x \gg 1/N$ as
\be
\tilde \rho_N(x) \simeq N \frac{|x|}{t}  \frac{\sqrt{3} }{ 2^\frac{5}{3} \pi } (\frac{2 t}{N^2 x^2})^{2/3} = 
\frac{1}{(N t)^{1/3}} \frac{\sqrt{3} }{2 \pi} x^{-1/3} \;,
\ee 
which coincides with the behavior at small argument $x \ll \sqrt{N}$ of the bulk density (using 
\eqref{bulk2} and \eqref{y0}).

\section{Brownian Bridges}

Let us now study the problem where the $N$ particles are Brownian Bridges. We will use the following mapping to translate our results for Brownian motions in this setting. The Brownian bridge $Y(\tau)$ is a Brownian motion $B(\tau)$ conditioned to hit zero at time $T$: $ Y( \tau): = \left( B(\tau) | B (T) = 0 \right) , \tau \in [ 0 , T ] $.
A Brownian bridge on $[0,T]$ can be obtained from a Brownian motion as \cite{ManYor2008}  
\be 
\label{eq:BridgeMapping}
Y(\tau) = \frac{T-\tau}{T} B\left(\frac{T\tau}{T-\tau}\right) \;.
\ee
Conversely, a Brownian motion $B(t)$, $t \in [0,+\infty)$, can be obtained from a Bridge $Y(\tau)$ as 
\be
B(t) = \frac{T+t}{T} Y\left(\frac{Tt}{T+t}\right) \;. 
\ee
Since these processes are Gaussian, this mapping can be checked by computing the two time covariance. A simple computation yields
\be
{\rm Cov}[ Y(\tau), Y(\tau') ] = \frac{T-\tau}{T}  \frac{T-\tau'}{T}  
{\rm Cov}[ B\left(\frac{T\tau}{T-\tau}\right) , B\left(\frac{T\tau'}{T-\tau'}\right) ] 
= \frac{T-\tau}{T}  \frac{T-\tau'}{T}  \frac{T \min(\tau,\tau')}{T- \min(\tau,\tau')}
= \min(\tau,\tau') - \frac{\tau \tau'}{T} \;,
\ee 
where we used that ${\rm Cov}[ B(t) , B(t') ] = \min(t,t')$. This indeed recovers the
standard covariance function of the Brownian bridge.

\begin{figure}[t]
    \includegraphics[width=0.5\linewidth]{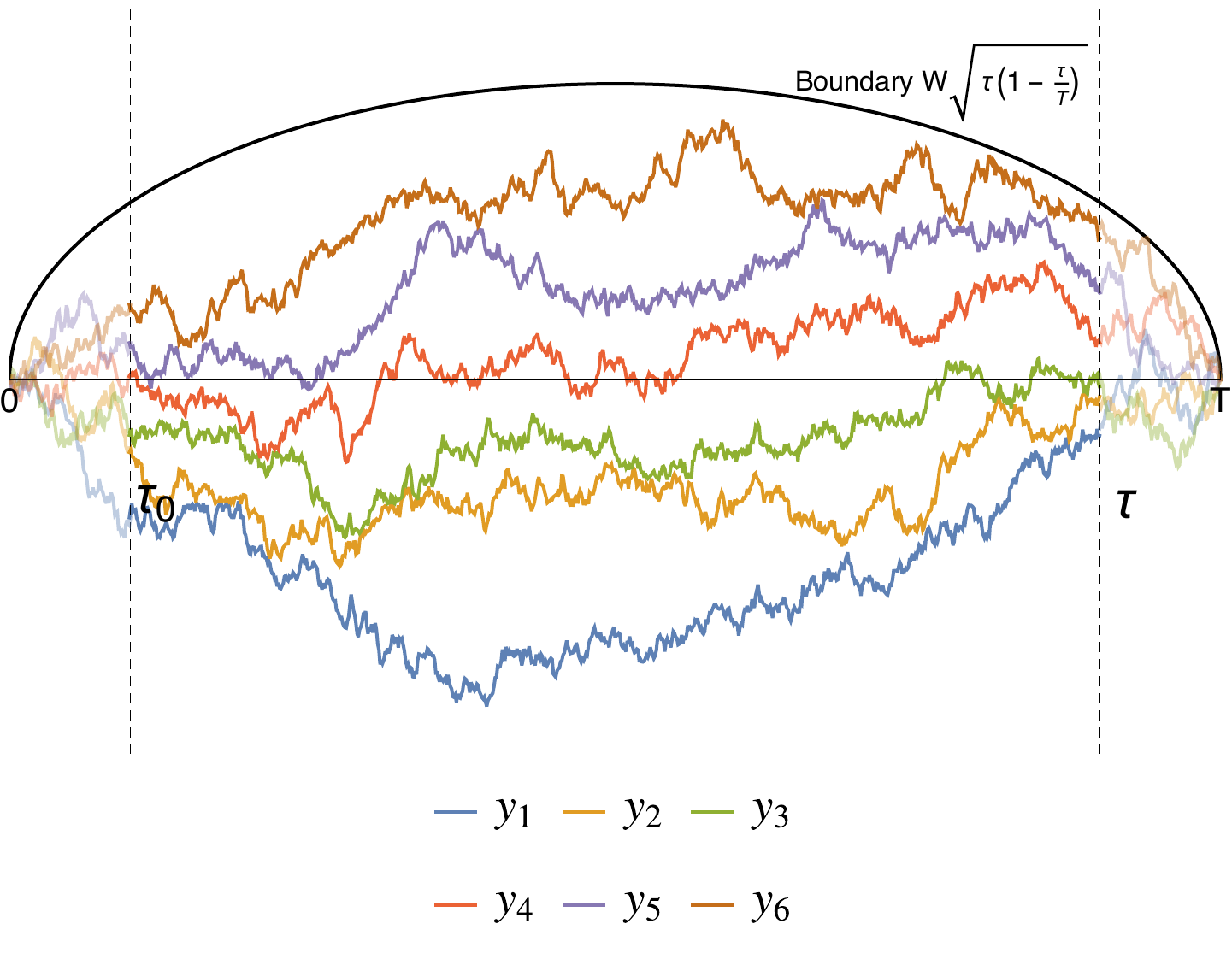}
    \caption{ Plot of $N=6$ Brownian bridge trajectories conditioned not to cross and to remain below the barrier in the time interval $[\tau_0,\tau]$ as discussed in the text. Note that in the time interval $[0,\tau_0]$ and $[\tau,T]$ they are independent bridges. }
    \label{fig:bridge}
\end{figure}

As a consequence of this mapping, the results obtained previously for $N$ non-crossing Brownian motions $x_i(t)$ under a moving barrier $g(t) = W\sqrt{t}$, $t \in [0,+\infty)$, can be translated to results for $N$ non-crossing Brownian bridges $y_i(\tau) $, $\tau \in [0,T]$, under a moving barrier $g_B(\tau) = W \sqrt{\frac{\tau (T-\tau) }{T}}$:
\be  \label{124}
\left\{
    \begin{array}{ll}
        t \\
        x_i(t) \\
        g(t) = W \sqrt{t}
    \end{array}
\right.
\quad
\xrightarrow{ \quad t =\frac{T \tau}{T- \tau} \quad }
\quad
\quad
\left\{
    \begin{array}{ll}
        \tau = \frac{T t }{T+t} \\
        y_i(\tau)  = \frac{T- \tau}{T} x_i(\frac{T\tau}{T-\tau}) \\
        g_B(\tau) = W \sqrt{\frac{\tau}{T}  (T - \tau ) }
    \end{array}
\right. \;.
\ee
The asymptotic results obtained for $N$ Brownian motions for $t \to \infty$ can now be applied to $N$ Brownian bridges $y_i(\tau)$ on $[0,T]$, for~$\tau \to T$.

\subsection{Survival probability}

Let us define $S_B(\tau|\vec y_0,\tau_0;W)$ as the probability that $N$ Brownian bridges stay under the absorbing boundary $g_B(\tau)$ and do not cross each other in the time interval $[\tau_0,\tau]$ with $\tau_0>0$ and $\tau <T$, given they are
at positions $\vec{y}_0$ at $\tau_0$:
\be 
\label{eq:DefS}
S_B(\tau|\vec y_0,\tau_0;W) = {\rm Pr} \left( y_1(\tilde{\tau}) < y_2 ( \tilde{\tau}) < \cdots < y_N(\tilde{\tau}) < W \sqrt{\frac{\tilde{\tau}}{T}  (T - \tilde{\tau} ) }, \ \forall \tilde{\tau} \in [\tau_0,\tau]  \ | \ y_{i}\left(\tau_{0}\right)=y_{0 i}    \right)
\ee
The main result is that under the mapping given in \eqref{124} one has
\be
\label{eq:MappingSBridge}
S_B(\tau|\vec y_0,\tau_0;W) = S(t|\vec x_0,t_0;W)
\ee
where $S(t|\vec x_0,t_0;W)$ is the same object for the Brownian and defined in equation
\eqref{eq:DefS2}. From the results on the power-law decay of $S(t)\propto t^{-\beta(N,W)}$ for large $t$ in (\ref{eq:Slargetime}), we obtain that for $\tau \to T$ the survival probability of the bridges vanishes as a power law
\be \label{SB} 
S_B(\tau) \propto \left(T - \tau  \right)^{\beta(N,W)}
\ee
where $\beta(N,W)$ has been computed in the previous sections. Similarly as in the 
Brownian case one can define and compute the associated probability
$S_{B_c}(\tau)$ conditioned on non-crossing trajectories, which decays as 
\mbox{$S_{B_c}(\tau) \propto \left(T - \tau \right)^{\beta_c(N,W)}$}.

\subsection{Distribution under the boundary}

By the pathwise mapping, we have the following relationship for the barrier constrained
propagators for the $N$ Brownian bridges and the $N$ Brownian motions (at times respectively $\tau$ and $t$)
\be 
P^{ B }_N \left( \vec { y } , \tau | \vec { y } _ { 0 } , \tau _ { 0 } ; \ W \sqrt{\frac{\tau}{T}  (T-\tau ) } \right) \ {d} \vec { y } = P_N^{ B r } \left( \vec { x } , t | \vec { x _ { 0 } } , t _ { 0 } ; \  W\sqrt{t} \right) \  {d} \vec { x } 
\ee
where $P_N^{Br}$ was defined in \eqref{eq:DefPw} and $P_N^B$ has a similar definition but for bridges. More explicitly one has
\be
P^{ B }_N \left( \vec { y } , \tau | \vec { y } _ { 0 } , \tau _ { 0 } ; \ W \sqrt{\frac{\tau}{T}  (T-\tau ) } \right) = \left(\frac{T}{T - \tau}\right)^N \  P^{ B r }_N \left( \frac{T}{T- \tau} \vec { y } , \frac{T\tau}{T- \tau }| \frac{T}{T- \tau_0}  \vec { y _ { 0 } } , \frac{T\tau_0}{T- \tau_0 } ; \  W\sqrt{t} \right) \;.
\ee
Using now the formula \eqref{general} for the Brownian, we obtain the 
barrier constrained propagator for the $N$ Brownian bridges $y_i$, under the absorbing boundary $g_B(\tau)= W \sqrt{ \tau (T-\tau)/T}$ in the limit $\tau \to T$
\begin{eqnarray}
\label{eq:PropBridge}
P_N^{ B } \left( \vec { y } , \tau | \vec { y } _ { 0 } , \tau _ { 0 } ; \ g_B(\tau) \right) \simeq C^2 \ e^{-\frac{1}{4 H^2(\tau)} \sum\limits_{i=1}^{N}  y_i^2    }     \det_{1 \leq i,j \leq N}(D_{2 \epsilon_{i-1}(W)}(-\frac{y_j}{ H(\tau)  }  ))
&& \det_{1 \leq i,j \leq N}(D^*_{2 \epsilon_{i-1}(W)}(- \frac{y_{0j} }{H(\tau_0)  }   )) \ e^{\frac{1}{4 H^2(\tau_0)} \sum\limits_{i=1}^{N} y_{0i}^2   } \nonumber \\  && \
 \left(\frac{\tau_0 (T-\tau)  }{\tau (T-\tau_0)    } \right) ^{\beta(N,W)} \ \left( \frac{T}{\tau(T - \tau)} \right)^{N/2} ,
\end{eqnarray}
where we denote $H(\tau)= \sqrt{ \tau (T-\tau)/T}$. Note that if one integrates over the $\vec y$ one recovers the survival probability, which coincides with \eqref{eq:SFullExp} by the change of variables $(y,\tau) \to (X= \frac{y}{H(\tau)}, t = \frac{T \tau}{T - \tau})$, and one can thus check \eqref{eq:MappingSBridge}.

\subsection{Application to Brownian motions under $g_B(\tau)$}

In the previous subsections, we have expressed the survival probability and constrained propagators of Brownian bridges under a barrier 
$g_B(\tau) = W \sqrt{\frac{\tau (T-\tau) }{T}}$. Recalling that Brownian bridges are Brownian paths conditioned to hit zero at time $T$, we can now obtain results about standard Brownian motions $x(\tau)$ under the same barrier $g_B(\tau)$, between times $\tau_0$ and $\tau<T$.

Expressing the conditioning explicitly, we can write the probability that bridges go from $(\vec{y}_0,\tau_0) $ to $(\vec{y},\tau)$ while remaining under the barrier and not crossing each other for $\tilde \tau \in [\tau_0,\tau]$,
as the probability of the same event for standard Brownian motions conditioned on returning to $\vec{0}$ at time $T$,
see Fig. \ref{fig:bridge}, i.e.
\be
P_N^{ B } \left( \vec { y } , \tau | \vec { y } _ { 0 } , \tau _ { 0 } ; \ g_B(\tau) \right) = 
P_N^{ Br } \left( \vec { y } , \tau | \vec{y}(T) = \vec{0}  ; \ \vec { y } _ { 0 } , \tau _ { 0 } ; \ g_B(\tau) \right) \;.
\ee
This can be written explicitly, in terms of the probabilities $P_N^{Br,0}$ that $N$ independent Brownian motions go from $(\vec{y},\tau)$ to $(\vec{0},T)$, and from $(\vec{y}_0,\tau_0)$ to $(\vec{0},T)$ in the numerator and denominator respectively \cite{MO15}
\be
P_N^{ B } \left( \vec { y } , \tau | \vec { y } _ { 0 } , \tau _ { 0 } ; \ g_B(\tau) \right) = 
\frac{
P_N^{ Br } \left( \vec { y } , \tau | \vec { y } _ { 0 } , \tau _ { 0 } ; \ g_B(\tau) \right) 
P_N^{Br,0} \left( \vec{0}, T  |   \vec { y } , \tau \right)
}{
P_N^{Br,0} \left( \vec{0}, T  |   \vec { y }_0 , \tau_0 \right)
} \;,
\ee
which reads
\be
P_N^{ B } \left( \vec { y } , \tau | \vec { y } _ { 0 } , \tau _ { 0 } ; \ g_B(\tau) \right) = \frac{
P_N^{ Br } \left( \vec { y } , \tau | \vec { y } _ { 0 } , \tau _ { 0 } ; \ g_B(\tau) \right) 
\left(  \frac{1}{\sqrt{2 \pi (T-\tau)}^N} e^{-\sum_i \frac{y_i^2}{2(T-\tau)}}  \right)
}{
\left(  \frac{1}{\sqrt{2 \pi (T-\tau_0)}^N} e^{-\sum_i \frac{y_{0i}^2}{2(T-\tau_0)}}  \right) 
}\;.
\ee
Finally, from the expression of $P_N^{ B } \left( \vec { y } , \tau | \vec { y } _ { 0 } , \tau _ { 0 } ; \ g_B(\tau) \right)$ from \eqref{eq:PropBridge}, we obtain the barrier constrained probability for $N$ Brownian motions with the
barrier $g_B(\tau) = W \sqrt{\frac{\tau (T-\tau) }{T}}$
\begin{eqnarray}
&P_N^{ Br } \left( \vec { y } , \tau | \vec { y } _ { 0 } , \tau _ { 0 } ; \ g_B(\tau) \right) 
\simeq
e^{\frac{1 - \frac{T}{2 \tau}}{2 (T-\tau)}  \sum_{i=1}^N y_i^2}
e^{- \frac{1 - \frac{T}{2  \tau_0}}{2 (T-\tau_0)}  \sum_{i=1}^N y_{0i}^2} \nonumber \\
&\times \det\limits_{1 \leq i,j \leq N}(D_{2 \epsilon_{i-1}(W)}(-\frac{y_j}{ h(\tau)  }  )) \det\limits_{1 \leq i,j \leq N}
(D_{2 \epsilon_{i-1}(W)}^*(- \frac{y_{0j} }{h(\tau_0)  }   )) \
\left( \frac{T}{\tau (T-\tau_0)} \right)^\frac{N}{2} \left( \frac{\tau_0(T-\tau)}{\tau (T- \tau_0)} \right) ^{\beta(N,W)} \;.
\end{eqnarray}
Note that this result is valid only in the $\tau \to T$ limit from the large-time approximations on the quantum propagator.

\begin{figure}[t]
    \includegraphics[width=0.5\linewidth]{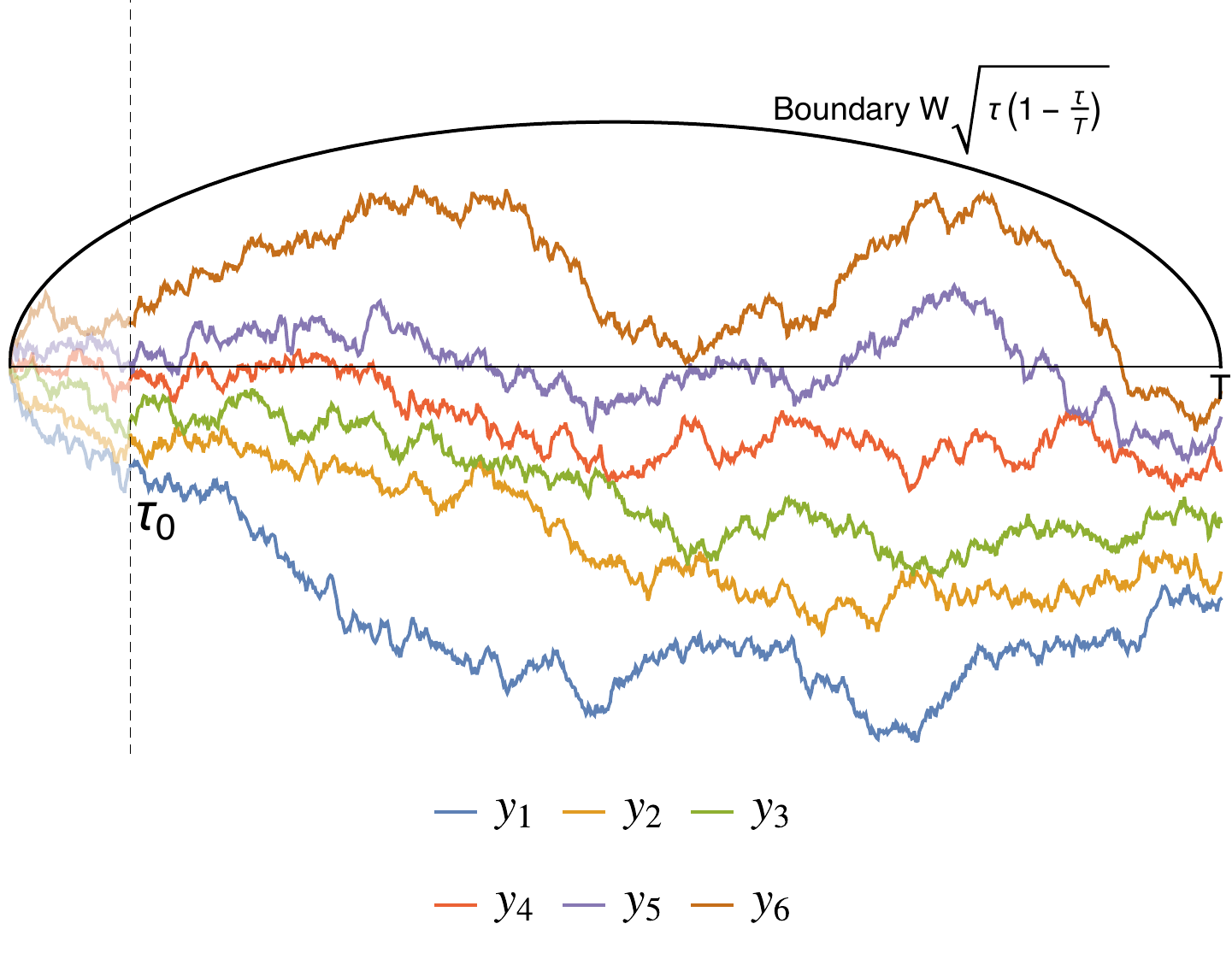}
    \caption{ Plot of $N=6$ Brownian trajectories conditioned not to cross and to remain below the $g_B(\tau)$ barrier in the time interval $[\tau_0,T]$. }
    \label{fig:brownians-gb}
\end{figure}

This formula has a finite limit as $\tau \to T$. It represents the probability that the $N$ Brownian motions $y_i(\tau)$
starting from $y_{0i}$ do not cross and stay below the barrier $g_B(\tau)$ in the
time interval $[\tau_0,T]$ and arrive at $y_i(T)=y_i < 0$. This is represented in
the Fig. \ref{fig:brownians-gb}. Using the asymptotics 
$D_{\alpha}(x) = e^{-\frac{x^2}{4}} (x^\alpha + \mathcal{O}(x^{\alpha-2}))$ for $x>0$,
one finds
\be \label{BrB} 
P_N^{ Br } \left( \vec { y } , \tau | \vec { y } _ { 0 } , \tau _ { 0 } ; \ g_B(\tau) \right) 
= B(\vec y_0, \tau_0, T,W) \, e^{ - \sum_i \frac{y_i^2}{2 T} } 
\det\limits_{1 \leq i,j \leq N}(|y_j|^{2 \epsilon_{i-1}(W)} )
\ee
with
\bea
B(\vec y_0, \tau_0, T,W) = 
C^2 e^{- \frac{1 - \frac{T}{2  \tau_0}}{2 (T-\tau_0)}  \sum_{i=1}^N y_{0i}^2}
\det\limits_{1 \leq i,j \leq N}
(D_{2 \epsilon_{i-1}(W)}^*(- \frac{y_{0j} }{h(\tau_0)  }   )) \
\left( \frac{1}{ (T-\tau_0)} \right)^\frac{N}{2} \left( \frac{\tau_0}{T (T- \tau_0)} \right) ^{\beta(N,W)}   \;.
\eea
The analysis of this joint distribution (\ref{BrB}) in the large $N$ limit seems rather challenging and is left for future studies.

\section{Conclusion} \label{sec:conclusion}

In this paper, we have studied a system of $N$ interacting Brownian motions in one-dimension, in the presence
of a moving boundary $g(\tau) = W \sqrt{\tau}$. Our first result is to compute the probability $S(t)$ of the event that
$N$ independent Brownian motions do not cross each other and stay below the boundary $g(\tau)$ up to time $t$. In this case, the
walkers are not directly interacting with each other but the event counts the probability that they remain non-intersecting as well
as below $g(\tau)$ up to time $t$. For the case $g(\tau) = W \sqrt{\tau}$, we showed that $S(t)$ decays as a power law at late times
$S(t) \sim t^{-\beta(N,W)}$ where the exponent $\beta(N,W)$ is a non-trivial function of both $N$ and $W$. We showed that
$\beta(N,W)$ is exactly the ground-state energy of $N$ spinless non-interacting fermions in a harmonic potential cut-off by an
infinite hard wall at $x= W$. We have provided analytical estimates of $\beta(N,W)$ in various limits of $N$ and $W$. In particular, in the
asymptotic scaling limit where both $N$ and $W$ are large with $W/\sqrt{4N}$ fixed, we showed that $\beta(N,W) \approx (N^2/4) {\sf b}(W/\sqrt{4N})$
where the scaling function ${\sf b}(y)$ can be computed analytically. We then extended our results to another type of interacting Brownian
motions, namely the DBM corresponding to the Gaussian Unitary Ensemble. In this case, the $N$ walkers repel each other
pairwise with a force that is inversely proportional to the distance between them. In this case, we again calculated the probability 
$S_{DBM}(t)$ that these walkers stay below the moving boundary $g(\tau) = W\sqrt{\tau}$. We showed that $S_{DBM}(t) \sim t^{-\beta_c(N,W)}$ where
$\beta_c(N,W) = \beta(N,W) - N(N-1)/4$. Furthermore, in this case, we also computed the joint distribution of the positions at (large) time $t$ below the boundary. For $W=0$, we showed that this joint distribution of the positions of the DBM are in one-to-one correspondence with the eigenvalues 
of the Laguerre biorthogonal ensemble of random matrix theory, as well those
of the $O(n)$ matrix model with $n=-1$. For this special case, using this connection to random matrix theory, we could
compute explicitly the density of the DBM near the barrier, as well as in the bulk away from the barrier. 
Interestingly we found that the mean density of surviving DBM walkers diverges
with a power $-1/3$ near the boundary. Finally, we also extended our results to
the case of $N$ Brownian bridges over the time interval $[0,T]$ under a moving  boundary $g_B(\tau) = W \sqrt{\tau(1-\tau/T)}$.  

We obtained these results by using a variety of analytical tools. We first generalized the so-called Lamperti transformation.
This transformation maps a single Brownian motion to an OU process, which in turn can be mapped to a quantum harmonic oscillator, using a path integral method. For $N$ Brownian motions, that are not allowed to cross each other, this Lamperti generalisation leads to a $N$-body
quantum fermion problem in a harmonic well. We have shown that this mapping still works for a moving boundary $g(\tau) = W \sqrt{\tau}$ and the corresponding
quantum problem corresponds to $N$ non-interacting spinless fermions in a harmonic potential, but having in addition an infinite hard-wall at the position $W$. The ground state energy of this $N$ fermion problem (that corresponds to the exponent $\beta(N,W)$) can then be estimated by several
methods well known in quantum mechanics, including the semi-classical method valid for large $N$ and large $W$.

We expect that the methods presented here can be extended in several directions. For instance, it will be interesting to compute the
survival probability of $N$ Brownian motions in the presence of two moving boundaries that enclose them. For example $g_1(\tau) = W_1 \sqrt{\tau}$ and
$g_2(\tau) = - W_2 \sqrt{\tau}$, with $W_1, W_2  \geq 0$. The other natural extension would be to consider the survival probability of $N$ Brownian motions 
in the presence of a constant linear drift, and one or two moving boundaries of the type mentioned above.
A few other extensions would be of interest. In the limit of large $N,W$ with $W/\sqrt{N}$ fixed
there is a transition at $W/\sqrt{4 N}=1$ where $\beta_c(W,N)$ vanishes. It would be interesting
to explore this transition on a finer scale, i.e. in the critical regime, where we expect that the
methods developed for fermions in \cite{UsHardBox, UsHardBoxLong} 
will be useful. Another classical problem is the one of a single Brownian
walker conditioned
to remain below a circle \cite{FerrariSpohn,Nechaev,Smith}. Our calculation for
$N$ Brownian walkers under the barrier $g(\tau) = W \sqrt{\tau(1-\tau/T)}$
is thus a generalization of that problem. Although here we have only studied
the distribution of the endpoints, it would be interesting to connect to these
works to compute the probability density at intermediate times.

\acknowledgements
We thank A. Krajenbrink for numerous interactions during the preparation of this manuscript, and B. R\'egaldo-Saint Blancard for technical help during the review process.
We are grateful to P. Krapivsky for useful discussions.
We thank A. Borodin, N. O'Connell and  L. Turban for pointing out useful references.
We acknowledge support from ANR grant ANR-17-CE30-0027-01 RaMaTraF.

\begin{appendix}

\section{Hermite polynomials}
\label{app:Hermite}

In this Appendix, we recall some basic properties of the Hermite polynomials $H_n$ defined as 
\bea
H_n(x) = (-1)^n e^{x^2} \frac{{d}^n}{{d}x^n}e^{-x^2} \;. \label{def_H}
\eea
The first ones read  
\be
\begin{aligned} H_{0}(x) &=1 \\ H_{1}(x) &=2 x \\ H_{2}(x) &=4 x^{2}-2 \\ H_{3}(x) &=8 x^{3}-12 x \;.\end{aligned}
\ee
They solve the following differential equation
\be 
\label{eq:DiffEq}
\frac{d}{d x}\left(e^{-x^{2}} \frac{{d} H_{n}(x)}{{d} x}\right)+2 n  \ e^{-x^{2}}  H_{n}(x)=0 \;.
\ee
They also satisfy the recurrence relation 
\be
\label{eq:RecRel}
H_{n+1}(x)=2 x H_{n}(x)-2 n H_{n-1}(x) 
\ee
as well as the following relation for the derivative of the $n$-th Hermite polynomial 
\be
\label{eq:DerivHermite}
H_{n}^{\prime}(x)=2 n H_{n-1}(x) \;.
\ee
For our purpose, it is also interesting to write $H_n(\frac{X}{\sqrt{2}})$, from the definition (\ref{def_H}), as 
\be 
H_n\left(\frac{X}{\sqrt{2}}\right)  = (-1)^n 2^\frac{n}{2} e^\frac{X^2}{2} \frac{{d}^n}{{d}X^n}e^{-\frac{X^2}{2}} \;.
\ee
Furthermore, introducing the double factorial $(n-1) ! ! = (n-1)(n-3)\cdots1$, the Hermite numbers are given by
\be
\label{eq:HermiteNumbers}
H_{n}= H_n(0)=\left\{\begin{array}{ll}{0 \;,} & {\text { if } n \text { is odd }} \\ {(-1)^{n / 2} 2^{n / 2}(n-1) ! ! \;.} & {\text { if } n \text { is even }}\end{array}\right.
\ee
We will also use the explicit formula for the generating function of Hermite polynomials
\be
e^{2 x t-t^{2}}=\sum_{n=0}^{\infty} H_{n}(x) \frac{t^{n}}{n !}\;.
\label{eq:GeneratingFunction}
\ee
In addition to these standard formulae for Hermite polynomials, we will need the following result 
\be
\label{eq:HalfSpaceScalarProd}
\int_{-\infty}^{0} H_{m}(x) H_{n}(x) e^{-x^{2}} d x=\frac{1}{2(m-n)}\left(H_{m}(0) H_{n}^{\prime}(0)-H_{n}(0) H_{m}^{\prime}(0)\right) \;.
\ee
This identity can be proved by
multiplying (\ref{eq:DiffEq}) by $H_m(x)$ and integrating over $x$ to obtain 
\bee
\int_{-\infty}^{0} H_{m}(x) \frac{{d}}{{d} x}\left(e^{-x^{2}} \frac{{d} H_{n}(x)}{{d} x}\right) {d}x=-2 n \int_{-\infty}^{0} e^{-x^{2}} H_{m}(x) H_{n}(x) {d} x \;.
\eee
Then, integrating by parts yields
\bea \label{rel_app1}
 - \int_{-\infty}^{0} \frac{{d}  H_{m}(x) }{{d} x}   e^{-x^{2}} \frac{{d} H_{n}(x)}{{d} x}{d}x + H_m(0) H_n'(0) =-2 n \int_{-\infty}^{0} e^{-x^{2}} H_{m}(x) H_{n}(x) {d} x \;.
\eea
By permuting $m$ and $n$ and substracting to (\ref{rel_app1}) one obtains the relation in (\ref{eq:HalfSpaceScalarProd}).

Another useful property of Hermite polynomials is that the determinant of the matrix $\left(  H_{i-1}(x_j) \right)_{1 \leq i,j \leq N} $ is a Vandermonde determinant. Indeed, as the Hermite polynomials form a sequence of orthogonal polynomials with successive degrees, the polynomials in the determinant simplify to their leading monomials and the determinant reads
\be 
 \det_{1 \leq i,j \leq N} \left(  H_{i-1}(x_j) \right)=  \det_{1 \leq i,j \leq N} \left( 2^{i-1} x_j^{i-1} \right) = 2^{\frac{N(N-1)}{2}} \det_{1 \leq i,j \leq N} ( x_j^{i-1} ) = 2^{\frac{N(N-1)}{2}}  \prod_{1\leq i < j \leq N} (x_j - x_i) \;.
\ee

\section{Harmonic oscillator wavefunctions}
\label{app:HarmonicOsc}

The harmonic oscillator without a wall (\ref{eq:HamiltH}) with $\hbar = m =1, \omega =\frac{1}{2}$ (and zero ground-state energy) is described by the Hamiltonian 
\be
\hat { H } = \  \  - \frac { 1 } { 2} \frac { \partial ^ { 2 } } { \partial X ^ { 2 } } + \frac { 1 } { 8 }  X ^ { 2 } - \frac { 1 } {4 }   
\ee
The eigenfunctions are expressed in terms of Hermite polynomials (see e.g. \cite{DalBasd2002})
\be 
\phi_k(X) =  c_k \ H_k\left(\frac{X}{\sqrt{2}}\right) \ e^{-\frac{X^2}{4}}
\ee
with the normalization constant $c_k$ and energy level $E_k$ given by
\bea\label{ck_app}
\begin{cases}
c_k =(\sqrt{2 \pi} 2^{k} k !)^{-\frac{1}{2}} \;,\\
E_k = \frac{k}{2} \;.
\end{cases}
\eea
The quantum propagator in imaginary time is then defined as 
\be
G(X,T \mid X_0, T_0)  = \bra{X} e^{-(T-T_0) \hat{H}} \ket{X_0}  =  \sum \limits_{k=0} ^{\infty} \phi_k(X) \phi_k^*(X_0) e^{- \frac{k}{2} (T-T_0)}  
\ee
and it can be computed explicitly (using Mehler's formula), leading to
\bea 
G(X,T \mid X_0, T_0)  =\frac{1}{\sqrt{2\pi\left(1-e^{-\left(T-T_{0}\right)}\right)}} \exp \left(- \frac{1}{2} \frac{\left(X-X_{0} e^{-\frac{1}{2} \left(T-T_{0}\right)}\right)^{2}}{1-e^{-\left(T-T_{0}\right)}}+ \frac{X^2 - X_0^2}{4}\right) \;.\label{propag_app}
\eea

In the presence of a hard wall at position $W=0$, the wall imposes a zero of the wavefunction at $X=0$. Thus, the wavefunctions of this system are the odd wavefunctions of the harmonic oscillator, with an extra $\sqrt{2}$ factor due to the normalization 
\bea \label{def_PhiW0}
\phi_k(X,W=0) = \sqrt{2} \ \phi_{2k+1}(X) = \sqrt{2} \  c_{2k+1} \ H_{2k+1}\left(\frac{X}{\sqrt{2}} \right) \ e^{-\frac{X^2}{4}} \;.
\eea
Let us compute the half-space integrals for the system with a hard wall in $W=0$, which are needed for the perturbative expansion around $W=0$ in Eq. (\ref{eq:DefDeltaE}) in the text, namely
\bea \label{DeltaE2}
    \begin{cases}
       &\Delta E _{2k+1}^{(1)} = \bra{2k+1} \Delta\hat{H} \ket{2k+1} \\
       & \\
       & \Delta E _{2k+1}^{(2)} = \sum\limits_{k'\neq k} \frac{\abs{ \langle 2k'+1 |\Delta\hat{H}| {2k+1}\rangle }^2 }{E_{2k+1} -E_{2k'+1} } 
    \end{cases} \;,
\eea
where $\Delta \hat H = X/4$. 
\begin{itemize}
\item The first matrix element reads
\bea \label{B8}
\bra{2k+1} \Delta\hat{H} \ket{2k+1}  = \int_{-\infty}^{0} 2 \;\phi_{2k+1}(X)^{2} \frac{X}{4} d X = \frac{1}{2} c_{2k+1}^2 \int_{-\infty}^{0} X \ H_{2 k+1}^{2}\left(\frac{X}{\sqrt{2}}\right) e^{-\frac{X^{2}}{2}} {d}X \;,
\eea  
where we have used the explicit expression of $\phi_k(X,W=0)$ given in Eq. (\ref{def_PhiW0}). Performing the change of variable $X \to \sqrt{2} X$ and using the recurrence relation for Hermite polynomials (\ref{eq:RecRel}) one gets
\bea
\int_{-\infty}^{0} X \ H_{2 k+1}^{2}\left(\frac{X}{\sqrt{2}}\right) e^{-\frac{X^{2}}{2}} {d}X =  
\int_{-\infty}^{0} 2 X \ H_{2 k+1}^{2} ( X ) e^{- X^{2}  } {d}X  \\
= \int_{-\infty}^{0} ( H_{2k+2}(X) + 2(2k+1) H_{2k}(X) ) H_{2k+1}(X)  e^{- X^{2}  }  {d}X \;. \label{B10}
\eea
These integrals can then be evaluated using the identities in (\ref{eq:DerivHermite}), (\ref{eq:HermiteNumbers}) and (\ref{eq:HalfSpaceScalarProd}), namely   
\be
\begin{cases}
&\int_{-\infty}^{0} H_{2 k+1}(X) H_{2 k}(X) e^{-X^{2}} d X =-(2 k+1) \dfrac{(2 k) !^{2}}{k !^{2}}\;, \\
&\\
&\int_{-\infty}^{0} H_{2 k+2}(X) H_{2 k+1}(X) e^{-X^{2}} d X= -\dfrac{(2 k+2) !(2 k+1) !}{k !(k+1) !} \;.
\end{cases}
\ee
Finally injecting these results in Eqs. (\ref{B8}) and (\ref{B10}) and using the explicit expression of the coefficients $c_k$ (\ref{ck}), one obtains
\bea \label{deltaE_app}
  \Delta E _{2k+1}^{(1)} = \bra{2k+1} \Delta\hat{H} \ket{2k+1} = -\frac{1}{\sqrt{2 \pi} 2^{2 k}} \frac{(2 k+1) !}{k !^{2}} \;,
\eea
which is the result given in Eq. (\ref{deltaE_text}) in the text. 

\item The more general matrix element needed for the computation of $\Delta E _{2k+1}^{(2)}$ in Eq.~(\ref{DeltaE2}) can be computed similarly and it yields
\bea
\langle{2k'+1}|\Delta\hat{H} |{2k+1}\rangle  &=& \frac{1}{2} c_{2k+1} c_{2k'+1} \int_{-\infty}^{0} X \ H_{2 k+1}\left(\frac{X}{\sqrt{2}}\right) H_{2 k'+1}\left(\frac{X}{\sqrt{2}}\right) e^{-\frac{X^{2}}{2}} {d}X \\
&=& \frac{1}{2} c_{2k+1} c_{2k'+1}  \int_{-\infty}^{0} H_{2 k' +1 }(X)\left(\frac{1}{2} H_{2 k+2}(X)+(2 k+1) H_{2 k}(X)\right) e^{-X^{2} } {d}X \\
 &=& \frac{(-1)^{k+k'}}{\sqrt{2 \pi} \, 2^{k+k'}\left(4(k-k')^{2}-1\right)} \frac{\sqrt{(2 k+1) !(2 k'+1) !}}{k ! k' !} \;,
\eea 
where, again, we have used the recurrence relation (\ref{eq:RecRel}) together with the 
identities in (\ref{eq:DerivHermite}), (\ref{eq:HermiteNumbers}) and (\ref{eq:HalfSpaceScalarProd}) as well as the explicit expression of $c_k$ in (\ref{ck_app}). Finally, one obtains $\Delta E _{2k+1}^{(2)}$ by injecting this expression in the second line of Eq. (\ref{DeltaE2}), which yields the expression (\ref{DeltaE2}) given in the text. With these expressions, $\beta(N,W)$ can be evaluated up to order $W^2$, as given in the text in Eq. (\ref{pert1}). 

\end{itemize}

We close this section by presenting the asymptotic analysis of $\Delta E _{2k+1}^{(2)}$, yielding the result (\ref{asympt_delta}) given in the text. We start with the formula (\ref{DeltaE2}) given in the text
\bea \label{DeltaE2_app1}
\Delta E_{2k+1}^{(2)} = \frac{(2k+1)!}{\pi \, 2^{2k+1}\, (k!)^2}\sum_{k'\geq 0, k' \neq k} \frac{1}{k-k'} \frac{(2k'+1)!}{2^{2k'} (k'!)^2} \frac{1}{(4(k-k')^2 - 1)^2} \;.
\eea
In the sum, we perform the change of variable $q=k'-k$, which yields
\bea \label{DeltaE2_app2}
\Delta E_{2k+1}^{(2)} = -\frac{(2k+1)!}{ \pi\, 2^{4k+1}\, (k!)^2}\sum_{q\geq -k, q \neq 0} \frac{1}{q} \frac{(2k+2q+1)!}{2^{2q} ((k+q)!)^2} \frac{1}{(4q^2 - 1)^2} \;.
\eea
We now use the large $k$ expansions, obtained from Stirling's formula
\bea \label{DeltaE2_app3}
\frac{(2k+2q+1)!}{((k+q)!)^2} = \frac{2^{2k} 2^{2q+1}}{\sqrt{\pi}} \sqrt{k} + \frac{2^{2k} 2^{2q-2}(3+4q)}{\sqrt{\pi}} \frac{1}{\sqrt{k}} + {\cal O}(k^{-3/2}) \;.
\eea
Inserting this expansion (\ref{DeltaE2_app3}) into Eq. (\ref{DeltaE2_app2}) and using the large $k$ expansion (obtained again from Stirling's formula) 
\bea \label{DeltaE2_app4}
\frac{(2k+1)!}{2^{2k} (k!)^2} = \frac{2}{\sqrt{\pi}} \sqrt{k} + {\cal O}(1) \;,
\eea
one obtains that 
\bea \label{DeltaE2_app5}
\Delta E_{2k+1}^{(2)} = - \frac{1}{4\pi^2} \left( 8 k \sum_{q\geq-k, q \neq 0}    \frac{1}{q(4q^2-1)^2}  + \sum_{q\geq-k, q \neq 0}    \frac{3+4q}{q(4q^2-1)^2} + {\cal O}(k^{-1})\right) \;.
\eea
In the first sum, one notices that the summand $1/(q(4q^2-1)^2)$ is an odd function of $q$ and, therefore, for large $k$, it is easy to see that this first term is actually of order ${\cal O}(k^{-3})$. The leading term, for large $k$, is thus the second sum in Eq. (\ref{DeltaE2_app5}), which yields
\bea
\lim_{k \to \infty} \Delta E_{2k+1}^{(2)} = -\frac{1}{4\pi^2} \sum_{q \in \mathbb{Z}^*} \frac{3+4q}{q(4q^2-1)^2} = \frac{1}{\pi^2} - \frac{1}{8} \;,
\eea
as announced in Eq. (\ref{asympt_delta}). 

\section{Derivation of the constrained propagator via the Karlin-McGregor formula} 
\label{app:kmg} 

In this section, we give another derivation of the probability for $N$ vicious Brownian motions to have survived and be at $\vec{x}$ at time $t$, starting from $\vec{x}_0$ at time $t_0$ which was obtained in \eqref{eq:DensityWinf}. From the Karlin McGregor formula, we can write this propagator for $N$ non-crossing particles as an $N \times N$ determinant involving only the single particle propagators 
\bea \label{C1}
P_N^{ B r } ( \vec { x } , t | \vec { x _ { 0 } } , t _ { 0 } ; + \infty) = \det_{ 1 \leq i,j \leq N} \left( P^{ B r }( x_j , t | x _ { 0i }, t _ { 0 } )      \right)
\eea
where $P^{ B r }( x_j , t | x _ { 0i }, t _ { 0 } )$ is the propagator of the free Brownian motion
\bea \label{C2}
P^{ B r }( x_j , t | x _ { 0i }, t _ { 0 } ) = \frac{1}{\sqrt{2 \pi (t-t_0) }} e^{ - \frac{ (x_j - x_{0i})^2 }{2(t-t_0)} } \;.
\eea
therefore, by injecting (\ref{C2}) in (\ref{C1}) and factoring out the common factors of the determinants, one finds 
\be
P_N^{ B r }( \vec { x } , t | \vec { x _ { 0 } } , t _ { 0 } ; + \infty) = \frac{1}{\sqrt{2 \pi (t-t_0)}^N}  e^{ - \frac{1}{2(t-t_0)} \sum_{i=1}^N (x_i^2 + x_{0i}^2  )  }      \det_{ 1 \leq i,j \leq N} \left(   e^{\frac{x_j x_{0i} }{t-t_0}}    \right)
\ee 
As in the main text, we analyse this propagator in the limit $t \to \infty$, $x \to \infty$ keeping $x/ \sqrt{t}$ fixed, while $x_0$ and $t_0$ are fixed and of order ${\cal O}(1)$. In this limit, we truncate the exponential factor, such that the determinant, to leading order for large $t$, can be computed by using the identity 
\bea
\bigg\vert \sum_{k=0}^{N-1} x_j^k  \frac{(x_{0i}/t)^k }{k! }   \bigg\vert_{1\leq i,j \leq N}  \
&& =  
\ \begin{vmatrix} 
1 & x_1  &  \cdots & x_1^{N-1} \\
1 & \cdots  &  \cdots & \cdots \\
1 & x_N  &  \cdots & x_N^{N-1} 
\end{vmatrix}
\times
\begin{vmatrix} 
1 & 1 &  1 \\
(x_{01}/t) & \cdots  & (x_{0N}/t)  \\
\cdots & \cdots  & \cdots   \\
\frac{(x_{01}/t)^{N-1}}{(N-1)!} & \cdots  & \frac{(x_{0N}/t)^{N-1}}{(N-1)!}
\end{vmatrix}
\\
&&= 
\ \Delta(\vec{x}) \ \frac{1}{t^{N(N-1)/2} \prod_{k=0}^{N-1} k!   }  \ \Delta(\vec{x}_0) \;,
\eea
where $\Delta(\vec{x}) = \prod_{1\leq i < j \leq N} ( x_j - x_i ) $ is the Vandermonde determinant. Finally, the large time limit of the Karlin-McGregor formula for $N$ non-crossing Brownian motions yields
\be 
P_N^{ B r } ( \vec { x } , t | \vec { x _ { 0 } } , t _ { 0 } ; + \infty) \simeq \frac{1    }{  (2\pi) ^{\frac{N}{2}}  G(N+1)} \ e^{-\frac{1}{2t} \sum\limits_{i=1}^{N} x_i^2     } \   \prod\limits_{1\leq i < j \leq N}  (x_j - x_i) (x_{0j}-x_{0i}) \quad t^{-\frac{N^2}{2}}     
\ee
in agreement with (\ref{eq:DensityWinf}) given in the text.

\section{Computation of the survival amplitude in terms of Pfaffians}
\label{app:PrefactorPfaffian}

We detail here the computations of the survival probability prefactors in the special cases $W \to \infty$ and $ W = 0$ (see Eq. (\ref{eq:PfaffianTerm}) and below in the main text).

\subsection{$W\to \infty$ }

Equation (\ref{eq:PfaffianTerm}) can be computed in the case where there is no wall. In this case, the eigenfunctions are simply those of the harmonic oscillator. The $(i,j)$ term in the Pfaffian is then:

\bea
A^\infty_{i,j}  &=& \int\limits_{\mathbb{R}^2}   c_i c_j \ e^{-\frac{X^2 + \tilde{X}^2}{2}}   \ H_i(\frac{X}{\sqrt{2}}) H_j(\frac{ \tilde{X} }{\sqrt{2}})  \ \mathrm{sgn}(X-\tilde{X})  \ {d}X {d}\tilde{X}   \label{eq:D1} \\
 &=& (-1)^{i+j} \ c_i c_j \ 2^\frac{i+j}{2}  \int\limits_{\mathbb{R}^2}  \   (\frac{{d}^i}{{d}X^i} e^{-\frac{X^2}{2}} ) (\frac{{d}^j}{{d}\tilde{X}^j} e^{-\frac{\tilde{X}^2}{2}} ) \  \mathrm{sgn}(X-\tilde{X})  \ {d}X {d} \tilde{X}  
\eea
where we have used the definition of Hermite polynomials given in \eqref{def_H}. Integrating by parts with respect to the $X$ variable:
\bea
A^\infty_{i,j}   &=&   (-1)^{i+j+1} \ c_i c_j 2^\frac{i+j}{2} \int\limits_{\mathbb{R}^2} \  (\frac{d^{i-1}}{{d} X^{i-1}} e^{-\frac{X^2}{2}} ) (\frac{d^j}{d\tilde{X}^j} e^{-\frac{\tilde{X}^2}{2}} ) \  2 \delta(X-\tilde{X}) \ {d}X {d} \tilde{X} \\
&=&   (-1)^{i+j+1} \ c_i c_j 2^\frac{i+j+2}{2}  \int\limits_{\mathbb{R}} \  (\frac{d^{i-1}}{{d} X^{i-1}} e^{-\frac{X^2}{2}} ) (\frac{d^j}{ {d} X^j} e^{-\frac{X^2}{2}} )   \ {d} X   
\eea 
Integrating by parts again $i-1$ times:
\bea
A^\infty_{i,j} &=&   (-1)^{j} \ c_i c_j 2^\frac{i+j+2}{2}  \int\limits_{\mathbb{R}} \   e^{-\frac{X^2}{2}} (\frac{{d}^{i+j-1} }{{d}X^{i+j-1} } e^{-\frac{X^2}{2}} )   \ {d}X  \\
&=&   (-1)^{i-1} 2^{\frac{3}{2}} \ c_i c_j   \int\limits_{\mathbb{R}} \   e^{-X^2} H_{i+j-1}(\frac{X}{\sqrt{2}})   \ {d}X   \\
A^\infty_{i,j}  &=&     (-1)^{i-1} 4 \ c_i c_j  \int\limits_{\mathbb{R}} \   e^{-2 X^2} H_{i+j-1}(X)  {d}X \label{eq:d7}
\eea
The term is nonzero only if $i$ and $j$ have opposite parity, as can be read from (\ref{eq:d7}). This ensures the antisymmetry of $A^\infty_{i,j}$ upon exchanging $i$ and $j$, as expected from (\ref{eq:D1}). This integral can be evaluated starting from the identity for the generating function of Hemite polynomials (\ref{eq:GeneratingFunction}), as explained in Appendix  \ref{app:Integrals}, yielding 
\be 
\label{eq:GenFuncIntegral}
\int\limits_{\mathbb{R}} \  e^{-2X^2} H_{2m}(X)   \ {d} X  = \sqrt{\frac{\pi}{2}} \left(-\frac{1}{2} \right)^m \frac{(2m)!}{m!} \;.
\ee
The Pfaffian term of the prefactor in the survival probability is then 
\be 
\underset{  0 \leq i,j \leq N-1 }{\mathrm{Pf}}  \left(    A^\infty_{i,j}   \right) 
=
\underset{  0 \leq i,j \leq N-1 }{\mathrm{Pf}}  \left(    \frac{(-1)^{i + \frac{i+j-3}{2}}  }{2^{i+j-3/2}}   \frac{ (i+j-1)!   }{ \sqrt{i! j! } (\frac{i+j-1}{2})!  }  \   \mathbf{1}_{ i+j \notin 2\mathbb{N}}       \right)  \;.
\ee
And the survival amplitude is, in the case $W \to \infty$, given by
\be
A(\vec{x_0},t_0;W \to \infty) =   t_0^{ \frac{N(N-1)}{4} }   \underset{   \substack{0 \leq i,j \leq N-1 \\ i+j \notin 2\mathbb{N}} }{\mathrm{Pf}} \left(    \frac{(-1)^{i + \frac{i+j-3}{2}}  }{2^{i+j-3/2}}   \frac{ (i+j-1)!   }{ \sqrt{i! j! } (\frac{i+j-1}{2})!  }           \right)    \det_{1 \leq i,j \leq N}  \left( \phi_{i-1} ^* ({X_0}_j)  e^{\frac{{X_0}_j^2}{4}  } \right) \;,
\ee
as given in Eq. (\ref{ampli_Winf}) in the text. 

\subsection{$W = 0$ }
As explained in Appendix \ref{app:HarmonicOsc}, in the $W=0$ case, the $k$-th wavefunction is the $(2k+1)$-th wavefunction of the harmonic oscillator:
\be
\phi_k(X) = \sqrt{2} \ c_{2k+1}   \ H_{2k+1} (\frac{X}{\sqrt{2}} ) \ e^{-\frac{X^2}{4}}
\ee
The generic term $A_{i,j}^0$ in the Pfaffian in Eq. (\ref{Ampli_W0}) of the main text reads
\bea
A_{i,j}^0  &=& 2 \ c_{2i+1} c_{2j+1} \ \int\limits_{({\mathbb{R}^-})^2}   e^{-\frac{X^2 + \tilde{X}^2}{2}}   \ H_{2i+1}(\frac{X}{\sqrt{2}}) H_{2j+1}(\frac{ \tilde{X} }{\sqrt{2}})   \ \mathrm{sgn}(X-\tilde{X})  \ {d} X  {d} \tilde{X} \\
 &=& 2^{i+j+2}  \ c_{2i+1} c_{2j+1}  \int\limits_{({\mathbb{R}^-})^2}  (\frac{d^{2i+1}}{{d}X^{2i+1}} e^{-\frac{X^2}{2}} ) (\frac{d^{2j+1}}{d\tilde{X}^{2j+1}  } e^{-\frac{\tilde{X}^2}{2}} )  \ \mathrm{sgn}(X-\tilde{X})  \ {d}X {d}\tilde{X} \\
&=& 2^{i+j+2}  \ c_{2i+1} c_{2j+1} \left(- 2  \int\limits_{-\infty}^0  (\frac{d^{2i}}{{d}X^{2i}} e^{-\frac{X^2}{2}} ) (\frac{d^{2j+1}}{{d}X^{2j+1}} e^{-\frac{X^2}{2}} )  {d} X + \frac{H_{2i}(0) H_{2j}(0)}{2^{i+j}} \right)
\eea
where we have used \eqref{def_H} and integrated by parts with respect to $X$. Integrating by parts $2i$ times, we compute the integral as:
\bea
 \int\limits_{-\infty}^0  (\frac{{d}^{2i}}{{d}^{2i}X} e^{-\frac{X^2}{2}} ) & & (\frac{{d}^{2j+1}}{{d}^{2j+1}X}   e^{-\frac{X^2}{2}} )  {d} X  =   \int\limits_{-\infty}^0  e^{-\frac{X^2}{2}}  (\frac{d^{2i+2j+1}}{{d}X^{2i+ 2j+1}} e^{-\frac{X^2}{2}} )  {d} X  - \sum_{m=1}^{i} \frac{H_{2i-2m}(0) H_{2j+2m}(0)    }{ 2^{i+j} } \\
& &=  -\frac{1}{2^{i+j}\sqrt{2}  }    \int\limits_{-\infty}^0  e^{-X^2}  H_{2i+2j+1}( \frac{X}{\sqrt{2}}  ) {d} X  - (-1)^{i+j} \sum_{m=1}^{i} (2i-2m-1)!! (2j+2m-1)!!  \\
& & =  -\frac{1}{2^{i+j} } \int\limits_{-\infty}^0  e^{-2 X^2}  H_{2i+2j+1}(X) {d} X  - (-1)^{i+j} \sum_{m=1}^{i} (2i-2m-1)!! (2j+2m-1)!!  
\eea
The remaining integral and the discrete sum can be evaluated explicitly (see Eq. (\ref{eval_2F1}) below for the computation of the integral) 
\be 
\begin{cases}
&\int\limits_{-\infty}^0  e^{-2 X^2}  H_{2n + 1}(X) {d} X   \quad  = \quad  \frac{(-1)^{n+1} (2n+1)! }{2^{n+1} n!  } \ {}_2F_1(\frac{1}{2}, -n , \frac{3}{2}, -1) \\
&\overset{i}{\underset{m=1}{\sum}} (2i-2m-1)!! (2j+2m-1)!!  \\
&=  2^{i+j} \bigg( \frac{2 \Gamma(\frac{3}{2}+i+j ) }{\sqrt{\pi}}  {}_2F_1\left(1,\frac{3}{2}+i+j;\frac{3}{2}, -1\right)   -(-1)^{i} \frac{\Gamma(\frac{3}{2}+j)  }{\Gamma(\frac{3}{2}-i) }   {}_2F_1\left(1,\frac{3}{2}+j;\frac{3}{2}-i , -1\right)   \bigg) \;,
\end{cases}
 \label{eq:GenFuncIntegral2}
\ee
where ${}_2F_1(a,b; c, z)$ is the standard hypergeometric function. Finally, the generic term $A_{i,j}^0$ in the Pfaffian in Eq.~(\ref{Ampli_W0}) reads
\begin{align}
  A_{i,j}^0 = \frac{(-1)^{i+j+1}}{\sqrt{ \frac{\pi}{2} (2i+1)! (2j+1)!  }} \bigg[    
(\frac{1}{4})^{i+j} \frac{(2i+2j+1)!}{(i+j)!} {}_2F_1 \big(  \frac{1}{2},-i-j;\frac{3}{2}, -1  \big) &  - (2i-1)!!(2j-1)!!  \nonumber \\
 - (2)^{i+j+1} \bigg(  \frac{2 \Gamma(\frac{3}{2}+i+j ) }{\sqrt{\pi}}  {}_2F_1 \big( 1,\frac{3}{2}+i+j;\frac{3}{2}, -1 \big)   
 + (-1)^{1+i} \frac{\Gamma(\frac{3}{2}+j)  }{\Gamma(\frac{3}{2}-i) } &  {}_2F_1 \big( 1,\frac{3}{2}+j;\frac{3}{2}-i , -1 \big) \bigg)  \bigg] \;,
\end{align}
as given in Eq. (\ref{explicit_A0}) in the main text.

\section{Computation of some integrals}
\label{app:Integrals}

The generating function of the Hermite polynomial (\ref{eq:GeneratingFunction}) enables us to compute some integrals which are useful for the computations presented in Appendix \ref{app:PrefactorPfaffian}. 

\begin{itemize}
\item By multiplying both sides of Eq. (\ref{eq:GeneratingFunction}) by $e^{-2x^2}$ and integrating over $x \in {\mathbb{R}}$ one obtains
\be 
\sum_{n=0}^{\infty}  \frac{t^n}{n!} \int\limits_{\mathbb{R}} \  e^{-2x^2} H_{n}(x)   dx   =
\int\limits_{\mathbb{R}} \  e^{-2x^2} e^{2 x t-t^{2}} {d}x  = \sqrt{\frac{\pi}{2}} e^{-\frac{t^2}{2} } = \sum_{m=0}^{\infty} \sqrt{\frac{\pi}{2}}  \left(-\frac{1}{2}\right)^m \frac{t^{2m}}{m!} \;.
\ee
By identifying the powers of $t$, one obtains the identity given in Eq. (\ref{eq:GenFuncIntegral}).

\item By evaluating the same integral on $\mathbb{R}^-$ one gets
\be 
\sum_{n=0}^{\infty} \frac{t^n}{n!} \int_{-\infty}^{0} e^{-2x^2}H_n(x){d}x  
= e^{-\frac{t^2}{2}} \int_{-\infty}^{-t/2} e^{-2x^2} {d}x = \frac{\sqrt{\pi}}{2 \sqrt{2}} e^{-\frac{t^2}{2}} \text{erfc}\left(\frac{t}{\sqrt{2}}\right) \;.
\ee
Using the series expansion $ \text{erfc}(x) = 1 -  \text{erf}(x) = 1 -\frac{2}{\sqrt{\pi}}  \sum_n \frac{(-1)^n x^{2n+1}  }{ n! (2n+1)    } $: 
\be 
\sum_{n=0}^{\infty} \frac{t^n}{n!} \int_{-\infty}^{0} e^{-2x^2}H_n(x){d}x  
=\frac{\sqrt{\pi}}{2 \sqrt{2}}  \left( \sum_{n=0}^{\infty}  \frac{(-1)^n  t^{2n}}{2^n n!}  \right)  \left(1 - \frac{2}{\sqrt{\pi}} \sum_{n=0}^{\infty}  \frac{(-1)^n t^{2n+1}}{ \sqrt{2} \, 2^n n! (2n+1)}   \right)  \;.
\ee
Identifying the coefficient of the term $\propto t^{2n+1}$ on both sides of this identity yields 
\bea 
\frac{1 }{ (2n+1)!} \int_{-\infty}^{0} e^{-2x^2}H_{2n+1}(x) {d} x  &=&  - \sum_{m=0}^{n}  \frac{(-1)^{n-m} }{(n-m)! 2^{n-m}  } \frac{(-1)^m }{ 2^{m+1} (2m+1) m!} \\
&=& \frac{(-1)^{n+1} }{2^{n+1} } \sum_{m=0}^{n}  \frac{1}{ (2m+1) \ m! (n-m)!   }
\eea
Finally the sum over $m$ can be expressed in terms of a hypergeometric function, which gives finally 
\bea \label{eval_2F1} 
\int_{-\infty}^{0} e^{-2x^2}H_{2n+1}(x) {d}x  = \frac{(-1)^{n+1} (2n+1)! }{2^{n+1} n!  } \ {}_2F_1(\frac{1}{2}, -n , \frac{3}{2}, -1) \;,
\eea
\end{itemize}
as given in the first line of Eq. (\ref{eq:GenFuncIntegral2}).

\section{Dyson Brownian Motion and Non-crossing Brownian paths}
\label{app:DBM-Noncrossing}

In this appendix we derive the relation given in Eq. \eqref{relGreg2} of the text.

\subsection{Relation between propagators}

As in the main text, we call $  { P }_N^{ \mathrm {DBM } } ( \vec { x } , t | \vec { x }_0 , t_0 ) $ the propagator of the Dyson Brownian motion with Dyson index $\beta =2$, and ${ P }^{Br}_N ( \vec { x } , t | \vec {x }_0 , t_0 )$ the propagator for independent Brownian Motions with boundary condition ${ P }^{Br}_N ( \vec { x } , t | \vec {x }_0 , t_0 ) = 0 $ whenever $x_i = x_j$. Let us first show the following relation 
 \be \label{rel1} 
  { P }_N^{ \mathrm {DBM } } ( \vec { x } , t | \vec { x }_0 , t_0 ) = \frac { \prod _ { i < j } \left( x _ { j } - x_ { i } \right) } { \prod _ { i < j } \left( x_ { 0j } - x_ { 0i } \right) } \   { P }^{Br}_N ( \vec { x } , t | \vec {x }_0 , t_0 ) \;.
\ee
We follow the argument of \cite{SchehrRambeau}, and consider a general $\beta$ for now. The propagator ${ P }^{Br}_N$ satisfies the diffusion equation 
\begin{equation}
\frac { \partial } { \partial t  }  { P }^{Br}_N \left(   \vec { x } , t | \vec {x }_0 , t_0 \right) 
= \frac { 1 } { 2 } \sum _ { i = 1 } ^ { N } \frac { \partial ^ { 2 } } { \partial x _ { i } ^ { 2 } } { P }^{Br}_N \left(   \vec { x } , t | \vec {x }_0 , t_0 \right) 
\end{equation}
together with the non-crossing condition, i.e. 
${ P }^{Br}_N \left(   \vec { x } , t | \vec {x }_0 , t_0 \right) = 0 \text { if } x _ { i } = x _ { j } 
$.

On the other hand, the Dyson Brownian motion  propagator $  { P }_N^{ \mathrm {DBM } } $ satisfies (see e.g. \cite{Katori_book})
\begin{equation} \label{DBMeq} 
 \frac { \partial } { \partial t }   { P }_N^{ \mathrm {DBM } }
 =
  \frac { 1 } { 2 } \sum _ { i = 1 } ^ { N } \frac { \partial ^ { 2 } } { \partial x _ { i } ^ { 2 } } { P }_N^{ \mathrm {DBM } }
  - 
  \frac { \beta } { 2 } \sum _ { i = 1 } ^ { N } \frac { \partial } { \partial x _ { i } } \left[ \sum _ { 1 \leq j \neq i \leq N } \frac { 1 } { x _ { i } - x _ { j } } { P }_N^{ \mathrm {DBM } } \right]
\end{equation}

Applying the transform:
\begin{equation}
  { P }_N^{ \mathrm {DBM } } ( \vec { x } , t | \vec { x }_0 , t_0 ) 
   = \frac { \exp \left[ \frac { \beta } { 2 } \sum _ { 1 \leq i < j \leq N } \log \left( x _ { j } - x _ { i } \right) \right] } { \exp \left[ \frac { \beta } { 2 } \sum _ { 1 \leq i < j \leq N } \log \left( x _ {0 j } - x _ { 0 i } \right) \right] } \times 
   { W }_N^{ \mathrm {DBM } } ( \vec { x } , t | \vec { x }_0 , t_0 ) 
\end{equation}
we obtain: 
\begin{equation}
\frac { \partial } { \partial t }  { W }_N^{ \mathrm {DBM } }  
=
 \frac { 1 } { 2 } \sum _ { i = 1 } ^ { N } \frac { \partial ^ { 2 } } { \partial x _ { i } ^ { 2 } }  { W }_N^{ \mathrm {DBM } } - \frac { \beta } { 8 } ( \beta - 2 ) \sum _ { i = 1 } ^ { N } \sum _ { 1 \leq j \neq i \leq N } \frac { 1 } { \left( x _ { j } - x _ { i } \right) ^ { 2 } }  { W }_N^{ \mathrm {DBM } } 
\end{equation}

For $\beta = 2$,  $ { W }_N^{ \mathrm {DBM } } $ verifies the same equation as $ { P }^{Br}_N $.  It also verifies the annihilating condition, since $  { P }_N^{ \mathrm {DBM } } ( \vec { x } , t | \vec { x }_0 , t_0 ) \sim \left( x _ { j } - x _ { i } \right) ^ { \beta } , x _ { i } \rightarrow x _ { j }$. We conclude $ { W }_N^{ \mathrm {DBM } }  = { P }^{Br}_N $   by unicity of the solution of this linear PDE, and thus we obtain Eq. \eqref{rel1}.

\subsection{Equivalence of the two processes}

Assuming known final positions $x_i$ at time $t$ and initial positions $\vec{x}_0$ at time $t_0$, the probability  to be in $\vec{y}$ at some intermediate time $\tau$ is the same in the $DBM$ and Brownian cases, by telescoping the extra factor:

\be 
\frac{
  { P }_N^{ \mathrm {DBM } } ( \vec { y } ,\tau | \vec { x }_0 , t_0 )   \ { P }_N^{ \mathrm {DBM } }   ( \vec { x } , t | \vec { y } , \tau ) }{    { P }_N^{ \mathrm {DBM } }  ( \vec { x } , t | \vec { x }_0 , t_0 )  }
  =
  \frac{
{ P }^{Br}_N ( \vec { y } ,\tau | \vec { x }_0 , t_0 ) \  { P }^{Br}_N  ( \vec { x } , t | \vec { y } , \tau ) }{  { P }^{Br}_N ( \vec { x } , t | \vec { x }_0 , t_0 ) }
\ee

More generally, the finite-dimensional distributions are equal for the two processes. Assuming fixed final and initial positions, the probability to be in $(y_1, \cdots, y_m)$ at times $t_0 < \tau_1 < \cdots < \tau_m < t$:

\be 
\frac{
  { P }_N^{ \mathrm {DBM } } ( \vec { y }_1 ,\tau _1| \vec { x }_0 , t_0 )  \cdots   { P }_N^{ \mathrm {DBM } }   ( \vec { x } , t | \vec { y }_m , \tau_m ) }{    { P }_N^{ \mathrm {DBM } }  ( \vec { x } , t | \vec { x }_0 , t_0 )  }
  =
  \frac{
{ P }^{Br}_N ( \vec { y }_1 ,\tau_1 | \vec { x }_0 , t_0 ) \cdots   { P }^{Br}_N  ( \vec { x } , t | \vec { y }_m , \tau_m ) }{  { P }^{Br}_N ( \vec { x } , t | \vec { x }_0 , t_0 ) }
\ee

From this equivalence we obtain that, conditioning on fixed final positions, the probability to stay below a deterministic moving barrier $g(t)$ is the same for the two processes. The relation between the propagators is thus still correct when adding a moving barrier: 
 \bea
 && { P }_N^{ \mathrm {DBM } } ( \vec { x } , t | \vec { x }_0 , t_0 ; g(t) ) =
  { P }_N^{ \mathrm {DBM } } ( \vec { x } , t | \vec { x }_0 , t_0  )
  \times {\rm Pr}(\text{DBM remains below the barrier $g(t)$}| \vec { x } , t ; \vec { x }_0 , t_0) \nonumber \\
&&  = { P }_N^{ \mathrm {DBM } } ( \vec { x } , t | \vec { x }_0 , t_0  )
\times  {\rm Pr}(\text{Brownian motions remains below the barrier $g(t)$ and do not cross}| \vec { x } , t ; \vec { x }_0 , t_0)
   \nonumber \\
  && =
   \frac { \prod _ { i < j } \left( x _ { j } - x_ { i } \right) } { \prod _ { i < j } \left( x_ { 0j } - x_ { 0i } \right) } \   { P }^{Br}_N ( \vec { x } , t | \vec {x }_0 , t_0 ; g(t) ) \;, \label{greg3} 
\eea
which shows the Eq. \eqref{relGreg2} of the text.

\subsection{Alternative derivation of Eq.~\eqref{greg3}}
 
 We present here another derivation of the relation between the propagators in the presence of a moving barrier.
 
\subsubsection{Constant barrier $g(t) = 0$}
 
If the barrier is fixed at $g(t)=0$, the propagator of the DBM in the presence of the barrier is the solution of Eq. 
\eqref{DBMeq} which vanishes at coinciding arguments and furthermore 
satisfies the additional condition
\be
 { P }_N^{ \mathrm {DBM } } ( \vec { x } , t | \vec { x }_0 , t_0 ;g(t)=0) = 0 \quad , \quad \text{if any} \, \, x_i=0 
\ee 
Since $ P_N^{ \mathrm {Br } } ( \vec { x } , t | \vec { x }_0 , t_0 ;g(t)=0)$ satisfies the same additional condition, the relation  \eqref{greg3} is still valid in this case.

\subsubsection{Moving barrier $g(t)$}

For a moving barrier $g(t)$, we consider the processes $z_i(t)=x_i(t) - g(t)$. For these processes
the absorbing boundary condition is fixed at $z_i=0$. The Langevin equation satisfied by a shifted processes
$z_i(t)$ reads
\be
\frac{dz_i(t)}{dt} = \frac{dx_i(t)}{dt} - \frac{dg(t)}{dt} = \frac{\beta}{2} \sum_{j \neq i} \frac{1}{z_i(t) - z_j(t)} 
- g'(t) + \xi_i(t) 
\ee
This Langevin equation is identical to the original one up to an additional drift term $- g'(t)$. 
The corresponding Fokker-Planck equation is identical to Eq. \eqref{DBMeq} 
with $x_i \to z_i$ together with the additional term from the drift (the arguments of all the 
functions are now the $z_i$)
\begin{align} 
- \sum_i \frac{\partial}{\partial z_i} (- g'(t)   { P }_N^{ \mathrm {DBM } })  =&  g'(t) \sum_i \frac{\partial   { P }_N^{ \mathrm {DBM } }}{\partial z_i} \\
=& \frac { \prod _ { i < j } \left( z_ { j } - z_ { i } \right) } { \prod _ { i < j } \left( z_ { 0j } - z_ { 0i } \right) } g'(t) \sum_i \left( \frac{\partial   { W }_N^{ \mathrm {DBM } }}{ \partial z_i} + \frac{\beta}{2} \sum_{j\neq i} \frac{1}{z_i - z_j}   { W }_N^{ \mathrm {DBM } }\right) 
\end{align}
However, we note that, by symmetry:
\be
\sum_i \sum_{j \neq i} \frac{1}{z_i - z_j} = 0
\ee
Such that the additional term is: 
\be 
\label{eq:Additional}
- \sum_i \frac{\partial}{\partial z_i} (- g'(t)  { P }_N^{ \mathrm {DBM } })  =- \frac { \prod _ { i < j } \left( z_ { j } - z_ { i } \right) } { \prod _ { i < j } \left( z_ { 0j } - z_ { 0i } \right) }
 \sum_i \frac{\partial  }{ \partial z_i}  ( - g'(t)   { W}_N^{ \mathrm {DBM } } )
\ee

We see that, in the translated frame, the PDE verified by the translated ${ W}_N^{ \mathrm {DBM } }$ is exactly the same as that of ${ P}_N^{ Br}$. As a consequence, the relation \eqref{greg3} between the two propagators still
holds for an arbitrary $g(t)$.

\section{Bulk density of the Dyson Brownian motion with a boundary at $W=0$}\label{App:Bueckner}

In this section, we derive the large $N$ limit of the density for the Dyson Brownian motion with a boundary at $W=0$. The starting point of our computations is the joint PDF of the positions $x_i<0$ given, at large time $t$, by \eqref{eq:JointPDBMzero}, which we write here with an overall prefactor $K_N$ including all terms that do not depend on $\vec{x}$:
\bea  \label{starting}
P_N^{DBM}(\vec{x},t|\vec{x_0},t_0;0) &\approx& K_N  \prod_{i=1}^N x_i \prod_{i<j} \left[(x_j - x_i) ( x_j^2 - x_i^2)\right] \  e^{- \frac{1}{2t} \sum_{i=1}^N x_i^2} \\
&\approx& K_N  \prod_{i=1}^N x_i \prod_{i<j} \left[(x_j - x_i)^2 ( x_j + x_j)\right] \  e^{- \frac{1}{2t} \sum_{i=1}^N x_i^2} \;.
\eea
Note that this joint PDF is very similar to the one encountered in the so called $O(n)$ matrix model \cite{Kostov,Kostov2}, with the value $n=-1$ in this case. To compute the density in the limit of large $N$, we will follow the method exposed in \cite{TheseNadal2012,BorotNadal2012}, which is based on a method developed by Bueckner \cite{Bue66}. We first perform a change of variables 
\bea \label{def_yi}
y_i = - \frac{x_i}{\sqrt{2Nt}} \;,
\eea
such that the joint PDF of the $y_i$'s reads, 
\bea \label{joint_yi}
P_{\rm joint}(y_1, \cdots, y_N) = \frac{1}{Z'_N}  e^{-E_N ( \vec{y}  ) }
\eea
where 
\bea \label{EN_y} 
 E_N ( \vec{y}  ) =   N \sum_i y_i^2 - \frac{1}{2} \sum_{i \neq j} \ln \abs{y_i + y_j } -  \sum_{i \neq j} \ln \abs{ y_j - y_i} - \frac{1}{2} \sum_i \ln \abs{y_i} \;.
\eea 
Let us introduce the average bulk density $\tilde r(y)$ 
\bea \label{rtilde_app}
\tilde r(y) = \frac{1}{N}  \sum_i \langle \delta(y - y_i) \rangle \;,
\eea
where the average is computed with respect to the joint PDF in (\ref{joint_yi}). In the limit of large $N$, the density can be computed using a standard 
Coulomb gas method and one finds that $\tilde r(y)$ is given by the solution of the following integral equation 
\be 
\label{eq:Bueck}
y= \dashint_0 ^\infty dy' \frac{\tilde r(y')}{y-y'} + \frac{1}{2} \int_0^\infty dy' \frac{\tilde r(y)'}{y+y'} \;,
\ee
which holds for $y$ inside the support of $\tilde r$, together with the normalisation condition $\int_0^\infty \tilde r(y) dy = 1$. It turns out that $\tilde r(y)$ has a finite support $[0,L]$, and the solution of (\ref{eq:Bueck}) can be obtained explicitly along the lines explained in Ref.~\cite{TheseNadal2012} (see Section 6.3).

Let us introduce the resolvent 
\be \label{def:W}
W(z) = \int_0^\infty dy \frac{\tilde r(y)}{z-y}, 
\ee
which is defined on the complex plane with a cut on $[0,L]$. Equation \eqref{eq:Bueck} gives the following constraint on the resolvent, for $y \in [0,L]$:
\be 
\label{eq:Weqresolvent}
2 y = W(y+i 0^+) + W(y-i 0^+) - W(-y) \;.
\ee
Hence, $W$ is the solution of the following Riemann-Hilbert problem \cite{BorotNadal2012,TheseNadal2012}:
\begin{enumerate}
\item $ W$ is analytic everywhere except on the cut $[0,L]$,
\item $ W(z) \sim  \frac{1}{z} $ as $\abs{z} \to \infty$, which follows from its definition (\ref{def:W}) together with the normalization of $\tilde r$,
\item $ W(z) \in \mathbb{R}$ for $z \in [L,+\infty]$,
\item $W$ satisfies \eqref{eq:Weqresolvent} \;.
\end{enumerate}
Note the last condition can also be written as $y = \mathrm{Re} [W(y)] - \frac{1}{2} W(-y) $, see equation (6.60) in \cite{TheseNadal2012}, and $W$ has a jump as it approaches the cut, i.e. $ W(y \pm i 0^+)=\operatorname{Re}[W(y)] \mp i \pi \tilde r(y)$.

The solution of this Riemann-Hilbert problem can be found as $W(z) = h(z) + \widetilde{W}(z)$ with a particular solution of (\ref{eq:Weqresolvent}) given by 
$ h(z) = \frac{2}{3}z$ (see Eq. (6.62) of \cite{TheseNadal2012}) while the homogeneous solution $\widetilde{W}(z)$ reads 
\be 
\widetilde{W}(z) = P_0(z^2) \phi_0(z) + P_1(z^2) \phi_1(z)
\ee
where $P_{0,1}(x)$ are polynomials while the functions $\phi_{0,1}(z)$ are given by 
\be 
\left\{
    \begin{array}{ll}
        \phi_0(z)  = 2 \frac{\cos( \frac{\omega}{3} + \frac{\pi}{6} )  }{ \sqrt{3} } \\
        \\
        \phi_1(z) = \frac{\sin( \frac{\omega}{3} + \frac{\pi}{6} ) }{ \tan( \omega) }
    \end{array}
\right. \;, \quad \quad {\rm with} \quad
\frac{L}{z}=\sin \omega \;.
\ee
The polynomials $P_{0,1}$ as well as the edge of the support $L$ are then obtained by imposing that $W(z) \sim \frac{1}{z} $ as $\abs{z} \to \infty$. Using the asymptotic behaviours for large~$z$~\cite{TheseNadal2012}
\bea
\phi_{0}(z) &=& 1-\frac{1}{3} \tan \left[\frac{\pi}{6}\right] \frac{L}{z} + \mathcal{O}\left( \frac{1}{z^2}\right)
= 1 - \frac{1}{3 \sqrt{3}}\frac{L}{z} + \mathcal{O}( \frac{1}{z^2}) \;, \\
\phi_{1}(z) &=&\sin \left[\frac{\pi}{6}\right] \frac{z}{L}+\frac{1}{3} \cos \left[\frac{\pi}{6}\right]-\left(\frac{1/9+1}{2}\right) \sin \left[\frac{\pi}{6}\right] \frac{L}{z} + \mathcal{O}\left(\frac{1}{z^2}\right) \\
&=& \frac{z}{2L} + \frac{1}{2\sqrt{3}} - \frac{5}{18} \frac{L}{z}+ \mathcal{O}\left(\frac{1}{z^2}\right) \;,
\eea
one obtains 
\be 
\left\{
    \begin{array}{ll}
        A(x)= A = \frac{2}{3 \sqrt{3}} L  = \frac{1}{\sqrt{2}} \;, \\
        \\
        B(x) = B = - \frac{4}{3} L = - \sqrt{6} \;, \\
        \\
        L = \left( \frac{3}{2} \right) ^{3/2} \;.
    \end{array}
\right.
\ee
Finally, the resolvent is given by (we recall that $\frac{L}{z}=\sin \omega$) 
\bea \label{resolvent_final}
W(z) = \frac{2}{3} z + \sqrt{\frac{2}{3}} \cos ( \frac{\omega}{3} + \frac{\pi}{6} ) - \sqrt{6} \frac{\sin( \frac{\omega}{3} + \frac{\pi}{6} ) }{ \tan( \omega) } \;,
\eea
from which one obtains the density $\tilde r(y)$ using the relation 
\be 
\tilde r( y) = - \frac{1}{\pi } \mathrm{Im} ( W(y +i 0^+) ) \;.
\ee  
Since $z=y \in [0,L]$ corresponds to $\omega=\frac{\pi}{2}-i \eta$ with $\eta>0$ such that $\frac{L}{y}=\sin \omega=\cosh \eta$ the density is given by \cite{TheseNadal2012}
\be 
\tilde r(y) =
\frac{1}{2 \pi \sqrt{2} }  \left( (e^{-\eta /3} -  e^{ \eta /3})  + 3  \frac{y}{2 L}( e^{ \eta } -  e^{ - \eta } )   ( e^{ \eta /3} +  e^{ -  \eta /3} ) \right) \;,
\ee
with $e^{\pm \eta} = \frac{L}{y} \pm \sqrt{\frac{L^2}{y^2} -1 }$ and $L = \left( \frac{3}{2} \right) ^{3/2}$, which eventually yields the expression given in Eq. (\ref{rtilde}). 

As stated in the text, this is in accordance with Proposition 2.5 of \cite{KuijlaarsMolag}. Because of the change of variables we have applied in this paper, the relation between the variable $s$ from this work and our variable $y$ is the following
\begin{equation}
s=y^2 \;,
\end{equation}
such that the density $\frac{d \mu_{V, \frac{1}{2}}^{*}(s)}{d s}$ obtained in \cite{KuijlaarsMolag} is related to $\tilde{r}(y)$ through :
\begin{equation}
\label{eq:G20}
\frac{d \mu_{V, \frac{1}{2}}^{*}}{d s} (y^2) = \frac{\tilde{r}(y)}{2y} \;.
\end{equation}
This relation between the two formulas can be proved by changing variables to $\omega = \frac{1-\sqrt{1-\frac{y^{2}}{L^{2}}}}{1+\sqrt{1-\frac{y^{2}}{L^{2}}}}$. Indeed, replacing $y$ by $\omega$ in both sides of (\ref{eq:G20}) through $y =L \sqrt{1-\left(\frac{1-\omega}{1+\omega}\right)^{2}} $, one shows that :
\begin{equation}
\frac{d \mu_{V, \frac{1}{2}}^{*}}{d s} (y^2) = \frac{\tilde{r}(y)}{2y} =  
\frac{\left(1- w^{1/3}\right) \left(w^{1/3}+1\right)^3}{6 \sqrt{3} \pi  w^{2/3}} \;.
\end{equation}

\end{appendix}


\begin{thebibliography}{100}

\bibitem{Doob1949} J. L. Doob, {\it Heuristic approach to the Kolmogorov Smirnov Theorems}, Ann. Math. Stat. {\bf 20}, 393 (1949).

\bibitem{Brei1966} L. Breiman, {\it First Exit Times from a Square Root Boundary}, 5th Berkeley Symp. {\bf 2}, 9 (1966).

\bibitem{Uch1980} K. Uchiyama, {\it Brownian first exit from and sojourn over one sided moving boundary and application}, Z. Wahrscheinlichkeit {\bf 54}, 75 (1980).

\bibitem{Sal1988} P. Salminen, {\it On the First Hitting Time and the Last Exit Time for a Brownian Motion to/from a Moving Boundary.}, Adv. Appl. Probab. {\bf 20}, 411 (1988).

\bibitem{Nov1981} A. A. Novikov, {\it On Estimates and the Asymptotic Behavior of Nonexit Probabilities of a Wiener Process To a Moving Boundary}, Math. USSR-Sbornik {\bf 38}, 495 (1981).

\bibitem{AK13} F. Aurzada, T. Kramm, {\it First exit of Brownian motion from a one-sided moving boundary}, Prog. Probab. {\bf VI}, 213 (2013).

\bibitem{AS15} F. Aurzada, T. Simon, {\it Persistence probabilities and exponents}, L{\'e}vy Matters {\bf V}, 183 (2015).

\bibitem{Kolm1933} A.N. Kolmogorov,  {\it On the Empirical Determination of a Distribution}, Giorn. Inst. Ital. Attuari {\bf 4}, 83 (1933).

\bibitem{Smir1936} N. V. Smirnov, {\it Sur la distribution de w2}, Comptes Rendus (Paris), {\bf 202}, 449 (1936).

\bibitem{ChiBou2012} R. Chicheportiche, J.-P. Bouchaud, {\it Weighted Kolmogorov-Smirnov test: Accounting for the tails}, Phys. Rev. E {\bf 86}, 1 (2012).

\bibitem{Chandra1943} S. Chandrasekhar, {\it Stochastic problems in Physics and Astronomy}, Rev. Mod. Phys. {\bf 15}, 1 (1943).

\bibitem{Maj1999} S. N. Majumdar, {\it Persistence in nonequilibrium systems}, Curr. Sci. {\bf 77}, 370 (1999). 

\bibitem{Maj2005} S. N. Majumdar, {\it Brownian functionals in Physics and Computer Science}, Curr. Sci. {\bf 89}, 2076 (2005).

\bibitem{BrayMajSchehr2013} A. J. Bray, S. N. Majumdar, G. Schehr, {\it Persistence and First-Passage Properties in Non-equilibrium Systems}, Adv. Phys. {\bf 62}, 225 (2013).

\bibitem{Red2001} S. Redner, {\it A Guide to First-Passage Processes}, Cambridge University Press (2001).

\bibitem{BraySmith2007} A. J. Bray, R. Smith, {\it Survival of a diffusing particle in an expanding cage}, J. Phys. A Math. Theor. {\bf 40}, 10965 (2007).

\bibitem{BraySmith2007deux} A. J. Bray, R. Smith, {\it The survival probability of a diffusing particle constrained by two moving, absorbing boundaries}, J. Phys. A Math. Theor. {\bf 40}, 1 (2007).

\bibitem{RedKrap1996} P. L. Krapivsky, S. Redner, {\it Life and death in an expanding cage and at the edge of a receding cliff}, Am. J. Phys. {\bf 64}, 546 (1996).

\bibitem{Tur1992} L. Turban, {\it Anisotropic critical phenomena in parabolic geometries: The directed self-avoiding walk}, J. Phys. A. Math. Gen. {\bf 25}, 127 (1992).

\bibitem{Fish1984} M. E. Fisher, {\it Walks, walls, wetting, and melting}, J. Stat. Phys. {\bf 34}, 667 (1984).

\bibitem{HF84} D. A. Huse, M. E. Fisher, {\it Commensurate melting, domain walls, and dislocations}, Phys. Rev. B {\bf 29}, 239 (1984).

\bibitem{KratGutVien2000} C. Krattenthaler, A. J. Guttmann, X. G. Viennot, {\it Vicious walkers, friendly walkers and Young tableaux: II. With a wall}, J. Phys. A Math. Gen. {\bf 33}(48), 8835 (2000).

\bibitem{BraWin2004} A. J. Bray, K. Winkler, {\it Vicious walkers in a potential}, J. Phys. A Gen. Phys. {\bf 37}, 2 (2004).

\bibitem{Katori2002} M. Katori, H. Tanemura, {\it Scaling limit of vicious walks and two-matrix model}, Phys. Rev. E  {\bf 66}, 011105 (2002).

\bibitem{KrajLacroix}
A.~Krajenbrink, B. Lacroix-A-Chez-Toine, P.~Le~Doussal, {\it Distribution of Brownian coincidences}, \href{https://arxiv.org/abs/1903.06511}{arXiv:1903.06511}, (2019).

\bibitem{KIK08} M. Katori , M. Izumi, N. Kobayashi, {\it Two Bessel bridges conditioned never to collide, double Dirichlet series, and Jacobi theta function}, J. Stat. Phys. {\bf 131}, 1067 (2008).

\bibitem{SMCR08} G. Schehr, S. N. Majumdar, A. Comtet, J. Randon-Furling, {\it Exact distribution of the maximal height of p vicious walkers}, Phys. Rev. Lett. {\bf 101}, 150601 (2008).

\bibitem{KIK08b} N. Kobayashi, M. Izumi, M. Katori, {\it Maximum distributions of bridges of noncolliding Brownian paths}, Phys. Rev. E {\bf 78}, 051102 (2008).

\bibitem{RS11} J. Rambeau, G. Schehr, {\it Extremal statistics of curved growing interfaces in 1+1 dimensions}, Europhys. Lett. {\bf 91}, 60006 (2010).

\bibitem{FMS11} P. J. Forrester, S. N. Majumdar, G. Schehr, {\it Non-intersecting Brownian walkers and Yang-Mills theory on the sphere}, Nucl. Phys. B 
{\bf 844}, 500 (2011).

\bibitem{Lie12} K. Liechty, {\it Nonintersecting Brownian motions on the half-line and discrete Gaussian orthogonal polynomials}, J. Stat. Phys. {\bf 147}, 582 (2012).

\bibitem{S12} G. Schehr, {\it Extremes of N Vicious Walkers for Large $N$: Application to the Directed Polymer and KPZ Interfaces}, J. Stat. Phys. {\bf 149}, 385 (2012).

\bibitem{SMCF13} G. Schehr, S.N. Majumdar, A. Comtet, P. J. Forrester, {\it Reunion Probability of N Vicious Walkers: Typical and Large Fluctuations for Large $N$}, J. Stat. Phys. {\bf 150}, 491 (2013).

\bibitem{Rem2017a} G. B. Nguyen, D. Remenik, {\it Non-intersecting Brownian bridges and the Laguerre Orthogonal Ensemble}, Ann. I. H. Poincar\'e B {\bf 53}, 2005 (2017).

\bibitem{Rem2017b} G. B. Nguyen, D. Remenik, {\it Extreme statistics of non-intersecting Brownian paths}, Electron. J. Probab. {\bf 22}, 1 (2017).

\bibitem{Borodin2009} A. Borodin, P. L. Ferrari, M. Prahofer, T. Sasamoto, J. Warren, {\it Maximum of Dyson Brownian motion and non-colliding systems with a boundary}, Electron. Commun. Probab. {\bf 14}, 486 (2009).

\bibitem{Dys62} F. J. Dyson, {\it A Brownian-motion model for the eigenvalues of a random matrix}, J. Math. Phys. {\bf 3}, 1191 (1962).

\bibitem{Tao2012} T. Tao, {\it Topics in Random Matrix Theory}, Amer. Math. Soc. (2012).

\bibitem{Muttalib} K.A. Muttalib, {\it Random matrix models with additional interactions}, J. Phys. A Math. Gen. {\bf 28}(5), L159 (1995).

\bibitem{Bor1998} A. Borodin, {\it Biorthogonal ensembles}, Nucl. Phys. B. {\bf 536}, 704 (1998).

\bibitem{BorotNadal2012} G. Borot, C. Nadal, {\it Purity distribution for generalized random Bures mixed states}, J. Phys. A Math. Theor. {\bf 45}, 075209 (2012).

\bibitem{ClaeysBiorthogonal} T. Claeys, S. Romano, {\it Biorthogonal ensembles with two-particle interactions}, Nonlinearity. {\bf 27}(10), 2419 (2014).

\bibitem{PLDMajSch2018} P. Le Doussal, S. N. Majumdar, G. Schehr, {\it Periodic Airy process and equilibrium dynamics of edge fermions in a trap}, Ann. Phys. (N. Y.) {\bf 383}, 312 (2018).

\bibitem{fermions_review}
D.S. Dean, P. Le Doussal, S.N. Majumdar, G. Schehr, {\it Non-interacting fermions at finite temperature in a d-dimensional trap: universal correlations}, Phys. Rev. A {\bf 94}, 063622 (2016).

\bibitem{Lam1972} J. Lamperti, {\it Semi-Stable Markov Processes I}, Probab. Theory Rel. {\bf 22}(3), 205 (1972).

\bibitem{BorAmbFla2005} P. Borgnat, P. O. Amblard, P. Flandrin, {\it Scale invariances and Lamperti transformations for stochastic processes}, J. Phys. A. Math. Gen. {\bf 38}, 2081 (2005).

\bibitem{MajSirBraCor1996} S. N. Majumdar, C. Sire, A. J. Bray, S. J. Cornell, {\it Nontrivial exponent for simple diffusion}, Phys. Rev. Lett. {\bf 77}, 2867 (1996).

\bibitem{DerHakZei1996} B. Derrida, V. Hakim, R. Zeitak, {\it Persistent spins in the linear diffusion approximation of phase ordering and zeros of stationary Gaussian processes}, Phys. Rev. Lett. {\bf 77}, 2871 (1996).

\bibitem{MS96}
S. N. Majumdar, C. Sire, {\it Survival Probability of a Gaussian Non-Markovian Process: Application to the $T=0$ Dynamics of Ising Model}, Phys. Rev. Lett., {\bf 77}, 1420 (1996).

\bibitem{Risken} H. Risken, {\it  The Fokker-Planck Equation}, Springer (Berlin) (1996).

\bibitem{KarMcGreg1959} S. Karlin, J. McGregor, {\it Coincidence probabilities}, Pacific J. Math. {\bf 9}, 1141 (1959).

\bibitem{KatTan2002} M. Katori, H. Tanemura, {\it Scaling limit of vicious walks and two-matrix model}, Phys. Rev. E {\bf 66}, 1 (2002).

\bibitem{MajCom2005} S. N. Majumdar, A. Comtet, {\it Airy distribution function: from the area under a Brownian excursion to the maximal height of fluctuating interfaces}, J. Stat. Phys, {\bf 119}, 777 (2005).

\bibitem{DLMF} See the {\it Digital Library of Mathematical Functions} at https://dlmf.nist.gov/ formula 13.9.16	.

\bibitem{mehta}
M. L. Mehta,  {\it Random Matrices}, New York: Academic (1991).

\bibitem{forrester} P. J. Forrester, {\it Log-Gases and Random Matrices},
London Math. Soc. monographs (2010). 

\bibitem{Katori2003} M. Katori, H. Tanemura, T. Nagao, N. Komatsuda, {\it Vicious walks with a wall, noncolliding meanders, and chiral and Bogoliubov-de Gennes random matrices}, Phys. Rev. E {\bf 68}, 16 (2003).

\bibitem{Bruijn} N. De Bruijn, {\it On some multiple integrals involving determinants}, J. Indian Math. Soc, 133 (1955).

\bibitem{SchehrRambeau} J. Rambeau, G. Schehr, {\it Distribution of the time at which $N$ vicious walkers reach their maximal height},
Phys. Rev. E {\bf 83}, 061146 (2011).

\bibitem{OConnell2005} P. Biane, P. Bougerol, N. O'Connell, {\it Littelmann paths and Brownian paths}, Duke Math. J. {\bf 130}, 127 (2005).

\bibitem{Tierz2007} Y. Dolivet, M. Tierz, {\it Chern-Simons matrix models and Stieltjes-Wigert polynomials}, J. Math. Phys. {\bf 48}(2), 023507 (2007).
%
\bibitem{Sommers2003} H.-J. Sommers, K. Zyczkowski, {\it Bures volume of the set of mixed quantum states}, J. Phys. A Math. Gen. {\bf 36}, 10083 (2003).

\bibitem{Sommers2004} H.-J. Sommers, K. Zyczkowski, {\it Statistical properties of random density matrices}, J. Phys. A Math. Gen. {\bf 37}, 8457 (2004).

\bibitem{Kostov} I. K. Kostov, {\it O(n) vector model on a planar random lattice: spectrum of anomalous dimensions}, Mod. Phys. Lett. A {\bf 4}, 217 (1989).

\bibitem{Kostov2} I. K. Kostov, M. Staudacher, {\it Multicritical phases of the O(n) model on a random lattice}, Nucl. Phys. B {\bf 384}(3), 459 (1992).

\bibitem{Kostov3} I. K. Kostov, {\it Loop amplitudes for nonrational string theories}, Phys. Lett. B {\bf 266}, 317 (1991).

\bibitem{TheseNadal2012} C. Nadal, {\it Th\`ese : Matrices al\'eatoires et leurs applications {\`a} la physique statistique et physique quantique}, Universit{\'e} Paris-Sud XI (2011).

\bibitem{KuijlaarsMolag} A. B. J. Kuijlaars, L. D. Molag, {\it The local universality of Muttalib-Borodin biorthogonal ensembles with parameter $\theta = \frac{1}{2}$}, Nonlinearity {\bf 32}, 3023 (2019).

\bibitem{Wright} E. M. Wright, {\it The asymptotic expansion of the generalized Bessel function}, Proc. London Math. Soc. {\bf 38}, 257 (1934).

\bibitem{ManYor2008} R. Mansuy, M. Yor, {\it Aspects of Brownian Motion}, Springer (Berlin) (2008).


\bibitem{MO15}
S. N. Majumdar, H. Orland, {\it Effective Langevin equations for constrained stochastic processes},
J. Stat. Mech. P06039 (2015).

\bibitem{UsHardBox}
B. Lacroix-A-Chez-Toine, P. Le Doussal, S. N. Majumdar, G. Schehr, {\it Statistics of fermions in a d-dimensional box near a hard wall}, EPL {\bf 120}, 
10006 (2017). 

\bibitem{UsHardBoxLong} B. Lacroix-A-Chez-Toine, P. Le Doussal, S. N. Majumdar, G. Schehr, {\it Non-interacting fermions in hard-edge potentials}, J. Stat. Mech. 123103 (2018).

\bibitem{FerrariSpohn} P. L. Ferrari, H. Spohn, {\it Constrained Brownian motion: fluctuations away from circular and parabolic barriers}, Ann. Probab. {\bf  33}(4), 1302 (2005).

\bibitem{Nechaev} S. Nechaev, K. Polovnikov, S. Shlosman, A. Valov, A. Vladimirov, {\it Anomalous one-dimensional fluctuations of a simple two-dimensional random walk in a large-deviation regime}, Phys. Rev. E {\bf 99}(1), 012110 (2019).

\bibitem{Smith} N. R. Smith, B. Meerson, {\it Geometrical optics of constrained Brownian excursion: from the KPZ scaling to dynamical phase transitions}, J. Stat. Mech. {\bf 2019}(2), 023205 (2019).

\bibitem{DalBasd2002} J.-L. Basdevant, J. Dalibard, {\it M{\'e}canique Quantique}, Editions de l'Ecole polytechnique (2002).

\bibitem{Katori_book} M. Katori, {\it Bessel processes, Schramm-Loewner evolution, and the Dyson model}, Springer (Singapore) (2015).

\bibitem{Bue66} H. F. Bueckner, {\it On a class of singular integral equations}, J. Math. Anal. Appl. {\bf 14}, 392 (1966).

\end{thebibliography}
 \end{document}